\documentclass[final,3p,times]{elsarticle}

\usepackage{bm}
\usepackage{lineno,hyperref}
\usepackage{color}
\usepackage{amsmath}
\usepackage{graphicx}
\usepackage{subcaption}
\usepackage{amssymb}
\usepackage{url}

\newcommand{\be}{\begin{equation}}
\newcommand{\ee}{\end{equation}}

\newcommand{\bx}{{\bm{x}}}

\newcommand{\bX}{{\bm{X}}}
\newcommand{\obx}{\overline{\bm{x}}}
\newcommand{\bn}{\widehat{\bm{n}}}

\newcommand{\bU}{{\bm{U}}}

\begin{document}

\begin{frontmatter}

\title{Turbulence and fire-spotting effects into wild-land fire simulators}

\author[BCAM]{Inderpreet KAUR\corref{mycorrespondingauthor}}
\ead{ikaur@bcamath.org}
\author[UNIBO,BCAM]{Andrea MENTRELLI}
\author[SPE]{Fr\'ed\'eric BOSSEUR}
\author[SPE]{Jean-Baptiste FILIPPI}
\author[BCAM,IKER]{Gianni PAGNINI}

\cortext[mycorrespondingauthor]{Corresponding author}

\address[BCAM]{
BCAM--Basque Center for Applied Mathematics
Alameda de Mazarredo 14, E-48009 Bilbao, Basque Country -- Spain}
\address[UNIBO]{
Department of Mathematics \& (AM)$^2$--Alma Mater Research Center on Applied Mathematics\\
University of Bologna, via Saragozza 8, I-40123 Bologna, Italy}
\address[SPE]{Laboratoire SPE--Sciences Pour l'Environnement CNRS - UMR6134
University of Corsica, BP 52, F-20250 Corte, Corsica -- France}
\address[IKER]{
Ikerbasque--Basque Foundation for Science Calle de Mar\'a D\'iaz de Haro 3, E-48013 Bilbao, 
Basque Country -- Spain}

\begin{abstract}
This paper presents a mathematical approach to model the effects and the role of phenomena with random nature such as turbulence and fire-spotting 
into the existing wildfire simulators. The formulation proposes that the propagation of the fire-front is the sum of a drifting component 
(obtained from an existing wildfire simulator without turbulence and fire-spotting) and a random fluctuating component. 
The modelling of the random effects is embodied in a probability density function accounting for the fluctuations around the 
fire perimeter which is given by the drifting component. 
In past, this formulation has been applied to include these random effects into a wildfire simulator based on an Eulerian moving interface method,
namely the Level Set Method (LSM), 
but in this paper the same formulation is adapted for a wildfire simulator based 
on a Lagrangian front tracking technique, namely the Discrete Event System Specification (DEVS). The main highlight of the present study is the comparison of the performance of a Lagrangian and an Eulerian moving interface method when applied to wild-land fire propagation.
Simple idealised numerical experiments are used to investigate the potential applicability of the proposed formulation to DEVS and to compare its behaviour with respect to the LSM. The results show that DEVS based wildfire propagation model qualitatively improves its performance 
(e.g., reproducing flank and back fire, 
increase in fire spread due to pre-heating of the fuel by hot air and firebrands,
fire propagation across no fuel zones, secondary fire generation, \dots)
when random effects are included according to the present formulation. The performance of DEVS and LSM based wildfire models is comparable and the only differences which arise among the two are due to the differences in the geometrical construction of the direction of propagation.
Though the results presented here are devoid of any validation exercise and provide only a proof of concept, they show a strong inclination towards an intended operational use. 
The existing LSM or DEVS based operational simulators like WRF-SFIRE and ForeFire respectively can serve as an ideal basis for the same.
\end{abstract}


\end{frontmatter}


\section{Introduction and motivations}
Modelling of wild-land fire propagation serves as a crucial tool to combat the high economic and environmental damage associated with the wildfires. 
An effective and swift modelling can aid in building up an efficient plan for fire suppression and restrict the life and property damage to a minimum. Modelling the propagation of wild-land fire is a complex, multi-scale, multi-physics and multi-discipline process, 
and the crux of this simulation lies in delivering either a versatile or a specialised model which is easy to implement and is capable of providing timely information. With the availability of higher computational power, various new and improved modelling approaches have been developed over the last decade. An extensive review by Sullivan \cite{sullivan-ijwf-2009a,sullivan-ijwf-2009b,sullivan-ijwf-2009c} provides a comprehensive overview of the wide spectrum of physical, 
empirical, statistical and mathematical analogue models available in literature.

\smallskip

Most of the statistical and mathematical models available in literature comprise of a Rate of Spread (ROS) formulation and a moving interface technique. The analytical formulation of the ROS is developed independent of the moving interface scheme and can be characterised in terms of the wind speed, slope, fuel characterisation, combustion properties, along with other experimental data. Various formulations for the ROS are available in literature \cite{rothermel-1972,balbi_etal-cf-2009,mallet_etal-cma-2009,finney-cjfr-2002,finney-ijwf-2003}, and they are usually versatile enough to be used with most moving interface schemes. Among the different moving interface methods available, Eulerian Level Set Method (LSM) is extensively used in wildfires to represent the propagating fire-line \cite{sethian_etal-arfm-2003,mallet_etal-cma-2009,mandel_etal-gmd-2011}. LSM is a scheme which represents a moving interface on a simple Cartesian grid. This method is particularly appropriate for wildfire simulations, as it permits an accurate computation of the front normal vector which is used to propagate the front. Other approaches available in literature which focus on wildfire propagation modelling include the Lagrangian Discrete Event System Specification (DEVS) \cite{karimabadi_etal-jcp-2005,omelchenko_etal-jcp-2012,filippi_etal-s-2009,filippi_etal-nhess-2014} 
and Huygens' principle \cite{anderson_etal-jams-1982,richards-ijnme-1990,richards-cst-1993,richards-ijwf-1995,glasa_etal-mcs-2008,rios_etal-nhess-2014}. The Lagrangian DEVS approach simulates the propagation of the wildfire without defining any underlying mesh to represent the burning state. DEVS being an event driven simulation considers time as a continuous variable and permits faster simulations over higher resolution. 
On the other hand, wildfire models based on Huygens' principle utilise an elliptical spread at each point of the fire-front and in some cases also benefit from some analytical results \cite{anderson_etal-jams-1982,richards-ijnme-1990,richards-cst-1993,richards-ijwf-1995,glasa_etal-mcs-2008,rios_etal-nhess-2014}.

\smallskip
Wild-land fire propagation involves the heat transport to the surroundings in accordance with the atmospheric wind, 
and other factors like fuel distribution, elevation, and orography. 
Different statistical and mathematical models utilized for wild-fire propagation in operational simulators
include a part of the physical processes based on the experimental data 
or its mathematical analogues, but are usually limited to the average value without taking into account the random fluctuations. 
These models accommodate the variability introduced by the changes in fuel type, 
wind conditions or orography but fail in quantifying a sudden erratic output of the fire due to random effects,
e.g., due to turbulence or fire-spotting. 
The generation of heat introduces a turbulent flow in the vicinity of 
the fire which in turn allows a wider contact of the hot air with the unburned fuel 
\cite{clark_etal-jam-1996,potter-ijwf-2002,potter-ijwf-2012a,potter-ijwf-2012b,clements_etal-jgr-2008,mandel_etal-gmd-2011}. 
This convective and radiative transfer of heat causes a rapid ignition of the unburned fuel bed and leads to a faster spread of the fire. 
Turbulence can also facilitate the generation of spot fires when the firebrands are transported away from the main fire due to 
advection \cite{sardoy_etal-cf-2007,sardoy_etal-cf-2008,kortas_etal-fsj-2009,perryman-2009,perryman_etal-ijwf-2013}. 
The effect of such random phenomena can range from subdued to catastrophic depending upon the concurrent weather conditions and fuel characteristics. 
Existing operational wildfire simulators are limited by their formulation to decipher these random phenomena, 
and require implementation of additional schemes/variables to represent such processes.

\smallskip
In past, few approaches have been proposed to model different random processes within the wildfire propagation 
\cite{favier-pla-2004,hunt-c-2007,boychuk_etal-ees-2009,perryman_etal-ijwf-2013} in order to provide a more realistic fire spread. 
In Ref. \cite{boychuk_etal-ees-2009}, the authors incorporate the stochastic fire-spotting phenomenon based on a continuous-time Markov chain on a lattice. 
The state of the lattice site is controlled according to the local transition rate functions, where the physical processes like - fire spread, 
spotting and burnout compete against each other. In Ref. \cite{perryman_etal-ijwf-2013},
the authors integrate the mathematical models with a cellular automata 
scheme to demonstrate the variability in the landing patterns of firebrands. 
In Ref. \cite{ntaimo_etal-sim-2004}, the authors discuss a DEVS based model which incorporates the uncertainty 
by sampling certain variables from arbitrary probability distributions. 

\smallskip
The present authors in a number of their recent works 
\cite{pagnini_etal-crs4-2012,pagnini_etal-mfbr-2012,pagnini-cedyacma-2013,pagnini-semasimai-2014,pagnini_etal-nhess-2014,
pagnini_etal-ecmi-2014} 
have proposed a new approach to include the effects of random processes in some operational wildfire propagation models. 
They derive a formulation to modify the modelled position of the fire-front to include random processes. 
The motion of the propagating fire-front is randomised by the addition of a random displacement
distributed according to a probability density function (PDF) corresponding to the turbulent transport of heat and to the fire-spotting landing distance.  
The driving equation of the resulting averaged process is analogous to an evolution equation of the reaction-diffusion type, 
where the ROS controls the source term. If the motion is considered without such random processes, 
the diffusive part disappears, and the spread of the fire-line is identical to the one given by the chosen operational simulator and driven only by the ROS. The proposed formulation to include random processes holds true for any method of determination of the ROS and the statistical spread is determined by the PDF of displacements of the random contour points marked as active burning points. The ensemble averaging of the displacements smoothens out of the noisy fluctuations and results in a smooth fire-line contour. This formulation can be implemented into existing wildfire models as a \textit{post-processing} scheme at each time step without calling for any major changes in the original framework. The efficacy of this formulation by using a LSM based fire propagation model has already been shown 
\cite{pagnini_etal-crs4-2012,pagnini_etal-mfbr-2012,pagnini-semasimai-2014,pagnini_etal-nhess-2014,pagnini_etal-ecmi-2014}.

\smallskip
In this study, the authors proceed with their preliminary attempt \cite{kaur_etal-praga-2015} to include the proposed formalism for random processes (turbulence and fire-spotting), already discussed and implemented into an Eulerian LSM based wild-land fire simulator 
\cite{pagnini_etal-crs4-2012,pagnini_etal-mfbr-2012,pagnini-cedyacma-2013,pagnini-semasimai-2014,pagnini_etal-nhess-2014,
pagnini_etal-ecmi-2014}, 
into a Lagrangian DEVS based wild-land fire simulation model: ForeFire (\url {https://github.com/forefireAPI/firefront}) \cite{filippi_etal-s-2009}. 
ForeFire is an open-source wildfire simulator developed by University of Corsica, France \cite{filippi_etal-affr-2014} 
to serve as a basis for an operational simulator and  provides flexibility to include new physics options. 
ForeFire is based on a Lagrangian technique and integrates the so-called Balbi model for the ROS \cite{balbi_etal-cf-2009} 
with a DEVS \cite{zeigler-ths-2000} 
based front tracking method to simulate the wildfire propagation. 
Coupling the front tracking method with DEVS permits the utilisation of time as a continuous variable; and hence removes the limitation of establishing 
a trade off between the temporal resolution and the scale of the simulation according to the Courant--Friedrichs--Lewy (CFL) condition. ForeFire model is developed devoid of any description of random processes such as turbulent heat transfer and fire-spotting and therefore hinders the evolution of the fire across fuel break zones. 

\smallskip

The aim of this article is the implementation of the mathematical formulation for turbulence and fire-spotting in a DEVS based wildfire model (ForeFire) to improve its efficiency. Operational utilisation of ForeFire for active wildfire propagation is the main driving factor for improving its output specially while dealing with realistic situations. This article tries to provide a proof of concept through a series of idealised numerical simulations to show the potential efficacy of the method in adapting the fire-line under the effect of turbulence and fire-spotting. Besides this, the investigation also provides a unique scope to compare the performance of Eulerian and Lagrangian approach based moving interface methods. Although a number of studies on fluid flows focus on the comparison of performance of Eulerian and Lagrangian methods for particle transport  \cite{carvalho_etal-amm-2007,zhang_etal-ae-2007,saidi_etal-ae-2014}, none of the studies concentrate on their comparison in context of wildfire propagation. In the present article, various test cases are described to compare the performance of these two schemes in application to wildfires.

\smallskip
The remaining paper is organised as follows. 
The salient features of DEVS and LSM based wildfire simulators are described in section \ref{overview}. 
The description of the formulation for random effects is provided in section \ref{formulation} and
the numerical set-up to include this formulation into the DEVS based simulator is described in section \ref{devs}.
Section \ref{setup} describes the different simulation set-ups for evaluation. A discussion of the numerical behaviour of the two methods and a detailed comparison of their performance is provided in section \ref{Discussion}. A brief sensitivity analysis of the spatial distribution of the fire-brand landing patterns is also presented in section \ref{firebrand}. 
Concluding remarks are provided in section \ref{conclusions}. 

\section{Overview of wildfire simulators}
\label{overview}
Modelling of wildfire spread focusses on the development of a phenomenological and theoretical formula for a suitable representation of the ROS of fire and the simulation of the propagation of the fire perimeter to reconstruct the shape of fire over time. Various ROS models available in literature \cite{rothermel-1972,balbi_etal-cf-2009,mallet_etal-cma-2009,finney-cjfr-2002,finney-ijwf-2003} 
provide an analytical formulation of the ROS for given wind conditions, slope and fuel characteristics. 
The available ROS models are usually developed independent of the moving interface methods and 
user discretion is advised in selecting one  according to the requirements. 
Since this study aims at studying the effect of the random processes using wildfire simulators based over different moving interface methods, an analysis of different ROS models is not presented in this article. 

\smallskip
Wildfire simulators based on the DEVS and the LSM techniques are inherently different, but the concept of Lagrangian markers and the constant perimeter resolution in the DEVS front tracking method can be considered analogous to the concept of active fuel points and constant action arc length in LSM tracking method as discussed in Ref. \cite{pagnini_etal-nhess-2014}. This section provides a brief overview of the salient features of the DEVS and LSM based wildfire simulators.
\subsection{DEVS based wildfire simulator}

The temporal scheme used to simulate the fire perimeter in wild-land fire simulator ForeFire is based on the DEVS \cite{zeigler-s-1987}. 
DEVS handles the time advancement in terms of the increment of physical quantities instead of a discrete time step. 
The resulting front is a polygon whose marker points have real (i.e.. non-discrete) coordinates instead of being located on nodes of a regular mesh or grid. 
This computationally in-expensive Lagrangian technique simulates the evolution of an interface without any underlying grid 
representing the state of the system. Each fire-front is discretised into a series of markers connected to the next marker by a piecewise linear segment.
The fire-break zones like roads and rivers are also represented as arbitrary shaped polygons. Each marker is associated with a propagation speed and direction, and each time a marker moves, 
the intersection with the neighbourhood is checked to take care of collision and topological changes. 

The propagation speed of the marker is defined by the ROS, while the propagation direction is defined by a front normal function. While spline interpolation may be used to estimate this normal, the bisector angle made by the marker with its immediate left and right neighbours is used as a computationally effective approximation of this outward normal. 

\smallskip
In DEVS, time is treated as a continuous parameter and each marker evolves according to its own independent time step. 
The time advancement of the markers is event based and all markers do not share the same time step. 
Due to different time advances for each marker, the CFL condition applies only locally and the markers move asynchronously. All events, triggering marker movement, are time sorted in an event list and processed by a scheduler.
This self adjusting temporal resolution in Lagrangian formulation is computationally efficient and provides an efficient way to simulate 
the spatially in-homogeneous problem. 
The simulation in DEVS advances as new events are generated; the generation of new events is managed by the following two 
criteria:
\begin{description}
\item[Collision criterion:]  
A collision happens when a marker moves into a different area, 
e.g., from an unburned to a burned area or from an active fire to a fuel break. Each collision generates a modification of the shape.
\item[Quantum distance criterion:]  
Quantum distance $\Delta q$ is defined as the maximum distance each marker is allowed to cover during advancement. 
The actual resolution of the simulation is limited by this parameter and details smaller than the quantum distance 
$\Delta q$ may not be accounted for.  
\end{description}

\smallskip
Three type of events can be defined to control the front propagation: {\it decomposition}, {\it regeneration} and {\it coalescence}. 
The {\it decomposition} function is activated when a marker enters a different area (e.g., while approaching a fuel break zone). 
As soon as the marker enters a new area, two new markers are created on the boundary of the new area. 
In {\it regeneration}, the markers are redistributed to refine the shape; if two markers are separated by a distance greater 
than the perimeter resolution $\Delta c$, a new marker is generated between the two markers.  
The {\it coalescence} function reconstructs the fire-perimeter by merging the markers. All markers with separation less than $\Delta r$ are merged together. For stability and to avoid cross over of two markers, 
$\Delta r = \Delta c/2$ is assumed and the condition $\Delta c \geq 2\Delta q$ should be respected. 
The precision of the method is highly dependent on the choice of $\Delta q$ and $\Delta c$. Quantum distance $\Delta q$, should be of a much higher resolution 
than the wind data for minimal error. A detailed description of the DEVS front tracking method can be found in Ref. \cite{filippi_etal-s-2009}.

\smallskip
ForeFire library is developed with C/C++/Java and Fortran bindings for a UNIX compatible environment and compiled using a SWIG (\url{ http://swig.org/}) build platform. 
NetCDF library with legacy C++ interface is required to build ForeFire, and SCons (\url{http://www.scons.org/}) python tool is used to make the library. 
A python based interface is used to launch the simulations with software calls to the main code. 
A command line mode interpreter is also available to run simulations.

\subsection{LSM based wildfire simulator}
%

LSM is one of the widely used and successful tools for tracking fronts in any dimension \cite{osher_fedkiw-2003,sethian_etal-arfm-2003}. 
It is particularly suitable for problems where the speed of the evolving interface is dependent on the interface properties 
and the boundary conditions at the interface. 
Wild-land fire propagation is one of problems where LSM finds a frequent application to represent the evolving fire-fronts
\cite{mallet_etal-cma-2009,rehm_etal-nist-2009,mandel_etal-gmd-2011}. 
In particular, it is the basis for the operational software WRF-SFIRE (\url {http://github.com/jbeezley/wrf-fire/}).

\smallskip
Let \(\Gamma\) be a simple closed curve representing the fire-front interface in two-dimensions, 
and let $\Omega$ be the region bounded by the fire-front $\Gamma$. If the interface is made up of more than one closed surface, 
the domain $\Omega$ is not simply connected and represents more than one independently evolving bounded areas.
Let $\gamma : S \times [0,+ \infty[\rightarrow R$ be an {\it implicit} function defined on the domain of interest $S \subseteq R^{2}$ 
such that the level set $\gamma(x,t) = \gamma_{*}$ coincides with the evolving front, i.e., $\Gamma(t) = \{x \in S \mid \gamma(x,t) = \gamma_{*}\}$. 
If $\Gamma$ is an ensemble of $n$ surfaces, the ensemble of the $n$ interfaces is considered as an {\it interface}. 
The evolution of the isocontour $\gamma = \gamma_*$ in time is governed by 
\be
\frac{D\gamma}{Dt}=\frac{\partial\gamma}{\partial t} + \frac{d \bx }{dt}\cdot \nabla\gamma = \frac{D\gamma_*}{Dt}= 0 \,.
\label{hamilton_jacobi}
\ee

\smallskip
If the motion of the interface is assumed to be directed towards the front normal $\bn$
\be
\bn = - \frac{\nabla\gamma}{\parallel\nabla\gamma\parallel} \,,
\ee 
then, Eq. \eqref{hamilton_jacobi} reduces to 
\be
\frac{\partial\gamma}{\partial t} = \mathcal{V}(\bx,t)\parallel\nabla\gamma\parallel \,.
\ee

\smallskip
In application to wild-land fire modelling, an indicator function $\phi(\bx,t)$ is introduced to allow a convenient identification 
of the burned area and the fire perimeter
\be
\phi(\bx,t)=
\begin{cases}
1, & \bx \in \Omega(t) \,, \\
0, & \bx \not\in \Omega(t) \,.
\end{cases} 
\label{indicatorphi}
\ee
The boundary of $\Omega$ can be identified as the front-line contour of the wildfire
and $\mathcal{V}$ denotes the ROS of the fire-front. 

\smallskip
The LSM code makes use of a general-purpose library which aims at providing a robust and efficient tool for studying the evolution 
of co-dimensional fronts propagating in one-, two- and three-dimensional system. 
The library, written in Fortran2008/OpenMP, 
along with standard algorithms useful for the calculation of the front evolution by means of the classical LSM, 
includes Fast Marching Method algorithms.
All the results presented in this paper are obtained by making use of the above-mentioned software; 
Matlab \url{(www.mathworks.com/products/matlab/)} 
and open source softwares such as \texttt{SciPy} \cite{SciPy} and \texttt{Matplotlib} \cite{Matplotlib} in the IPython framework \cite{IPython} 
have been used for visualisation purposes. 

\section{Random effects formulation}
\label{formulation}
The proposed approach is based on the idea to split the motion of the front position into a drifting part and a fluctuating part
\cite{pagnini_etal-crs4-2012,pagnini_etal-mfbr-2012,pagnini-cedyacma-2013,pagnini-semasimai-2014,pagnini_etal-nhess-2014,
pagnini_etal-ecmi-2014,mentrelli_etal-caim-2015,mentrelli_etal-jcp-2015}. 
This splitting allows specific numerical and physical choices which can be useful to improve the algorithms and the models. 
In particular, the drifting part can be related to existing methods for moving interfaces, 
for example the Eulerian LSM \cite{sethian_etal-arfm-2003} or the Lagrangian DEVS \cite{karimabadi_etal-jcp-2005,omelchenko_etal-jcp-2012}. 
The fluctuating part is the result of a comprehensive statistical description of the system which includes the random effects in agreement 
with the physical properties of the system. 
As a consequence, the fluctuating part can have a non-zero mean, implying that the drifting part does not correspond to the average motion. 
Due to this fact, the present splitting is distinct from the well-known Reynolds decomposition frequently adopted in turbulent flows. 

\smallskip
The motion of the each burning point can be random due to the effect of turbulence and/or fire-spotting. 
Let $\bX^{\omega}(t,\obx_0)$ be the $\omega$ realisation of the random trajectory of an active burning point at time $t$ 
with the initial condition $\bX^{\omega}(0,\obx_0)=\obx_0$. The trajectory of a single interface particle can be described by 
the one-particle PDF $f^{\omega}(\bx;t)=\delta(\bx-\bX^{\omega}(t,\obx_0))$, where $\delta(\bx)$ is the Dirac delta function. 
The random front contour can be derived by using the sifting property of the Dirac delta function peaking at a random position. 
This random position represents the stochastic trajectory which describes the random motion of the front line and includes both 
drifting and fluctuating part of the front position.  
In the $\omega$ realisation, the evolution in time of the function $\gamma^{\omega}(\bx,t)$, 
which embeds the random fire-line $\Gamma^{\omega}$ is given by
\cite{pagnini_etal-crs4-2012,pagnini_etal-nhess-2014,mentrelli_etal-caim-2015,mentrelli_etal-jcp-2015} 
\be
\gamma^{\omega}(\bx,t) = \int_S \gamma(\obx,t) \delta(\bx-\bX^{\omega}(t,\obx) \, d\obx \,. 
\ee  
The effective indicator of the burned area enclosed by the random front $\phi_{e}(\bx,t):S \times [0,+\infty[ \rightarrow [0,1]$ 
may be defined as 
\begin{eqnarray} 
\phi_{e}(\bx,t) &=& \left\langle  \int_{\Omega(t)}\delta(\bx-\bX^{\omega}(t,\obx)) \, d\obx \right\rangle \nonumber \\
&=& \int_{\Omega(t)} \langle \delta(\bx-\bX^{\omega}(t,\obx)) \rangle \, d\obx \nonumber \\
&=& \int_{\Omega(t)}f(\bx;t|\obx) \, d\obx \\
&=& \int_S \phi(\obx,t) f(\bx;t|\obx) \, d\obx \,,
\label{phi_e}
\end{eqnarray}
where $ f(\bx;t|\obx) = \left\langle \delta(\bx-\bX^{\omega}(t,\obx))\right\rangle$ is the PDF of the displacement 
of the active burning points around the position $\obx$ which is obtained from existing wildfire simulators. 

An arbitrary threshold value $\phi_{e}^{th}$ is chosen to serve as the criterion to mark the region as burned or unburned, 
i.e., $\Omega_{e}(\bx,t)=\left\lbrace \bx \in S \mid \phi_{e}(\bx,t) > \phi_{e}^{th}\right\rbrace$.

\smallskip
The evolution of the effective indicator $\phi_{e}(\bx,t)$ can be obtained by using the Reynold transport theorem to Eq. (\ref{phi_e}) 
\cite{pagnini_etal-prl-2011,pagnini_etal-crs4-2012}
\be
\frac{\partial \phi_{e}}{\partial t} = \int_{\Omega(t)} \frac{\partial f}{\partial t} d \obx 
+ \int_{\Omega(t)} \nabla_{\obx} \cdot [V(\obx,t) f(\bx;t|\obx)] \, d\obx \,. 
\label{evolution_phi_e}
\ee

 Since $f(\bx;t|\obx)$ is a PDF, its evolution in time is here assumed to be governed by an equation of the type
\be
\frac{\partial f}{\partial t}=\varepsilon f \,,
\label{epsilon}
\ee
where $\varepsilon$ is an operator acting on space variable $\bx$, and not on $\obx$ and $t$. For example, the operator $\varepsilon$ is the Laplacian when $f(\bx;t|\obx)$ is the Gaussian density function. 
Using (\ref{epsilon}), equation (\ref{evolution_phi_e}) can be written as  
\be
\frac{\partial \phi_{e}}{\partial t} = \varepsilon \phi_{e} + \int_{\Omega(t)} \nabla_{\obx} \cdot 
[V(\obx,t) f(\bx;t|\obx)] \, d\obx \,. 
\ee 
The velocity of the fire-line propagation is controlled by the ROS $\mathcal{V}$ and defined through $V = \mathcal{V}(\obx, t) \bn$, 
while turbulence and fire-spotting phenomena are modelled by modifying the PDF function. 
This method is compatible with the estimation of the ROS by any technique. 

\smallskip 
The modelling of the effects of the random processes is handled by the PDF $f(\bx;t|\obx)$, 
accounting for the two independent random variables representing turbulence and fire-spotting respectively. 
It should be noted that for brevity, here fire-spotting is assumed to be an independent downwind phenomenon; 
hence the effect of fire-spotting is accounted only for the leeward part of the fire-line. 
Taking into account these assumptions the PDF for the random processes can be defined as
\be
f(\bx;t|\obx) = 
\begin{cases}
\displaystyle
\int_{0}^{\infty} G(\bx-\obx- \ell \, \bn_{U};t) q(\ell;t) \, d\ell \,, &  \bn \cdot \bn_{U} \geq 0 \,,\\
\\
G(\bx-\obx;t), & otherwise \,,
\end{cases}
\label{pdf}
\ee
where $\bn_U$ is the unit vector aligned with the mean wind direction.

\smallskip
Turbulent diffusion is assumed to be isotropic and modelled by a bi-variate Gaussian PDF
\be
G(\bx-\obx;t)=\frac{1}{4\pi D t}
\exp{\left\lbrace \frac{(x-{\overline{x}})^{2}+(y-{\overline{y}})^{2}}{4 D t}\right\rbrace} \,, 
\ee
where $D$ is the turbulent diffusion coefficient  
such that $\left\langle (x-{\overline{x}})^{2}\right\rangle = 
\left\langle (y-{\overline{y}})^{2}\right\rangle = 2Dt$. 
In the present model, the whole effect of the turbulent processes over different scales is assumed to be parametrised only by the turbulent diffusion coefficient. Also, only the turbulent fluctuations with respect to the mean transport are considered. Hence, the characterisation of turbulence is independent of the mean characteristics of the motion, and in particular the estimation of the turbulent diffusion coefficient $D$ is independent of the mean wind. Moreover, since the simulations are performed with a flat terrain and without boundaries in the spatial domain, horizontal isotropy is also assumed. Within this framework, the mean wind does not break the isotropy, because only the fluctuations are considered, but a non-trivial orography can indeed break the isotropy, e.g., valleys or slopes. Even if an exact estimation of $D$ is out of the scope of the present study, $D$ is one of the necessary parameters for simulations and a quantitative estimation is required.
The desired diffusion coefficient $D$ corresponds to the turbulent heat convection generated by the fire. In literature, the ratio between the total heat transfer and the molecular conduction of heat is known 
as the Nusselt number: ${\rm Nu}=(D+\chi)/\chi$. Where, $\chi$ is the thermal diffusivity of the air at ambient temperature and is well-known in literature:
$\chi=2 \times 10^{-5} \, {\rm m^2 s^{-1}}$. It is also experimentally known that the relation between the Nusselt number ${\rm Nu}$ and 
the Rayleigh number ${\rm Ra}$ is given by ${\rm Nu} \simeq 0.1 \, {\rm Ra}^{1/3}$ \cite{niemela_etal-jfm-2006}. The Rayleigh number is the ratio between the convection and the conduction of heat and it is defined 
as ${\rm Ra}=\gamma \, \Delta T \, g \, h^3/(\nu \chi)$, where $\gamma$ is the thermal expansion coefficient, $\Delta T$ the temperature difference between the bottom and the top of the convective cell, $h$ the dimension of the convection cell, $g$ the acceleration gravity and $\nu$ the kinematic viscosity. From above,  $D$ can be computed by the formula:
\be 
D \simeq 0.1 \, \chi \, [\gamma \, \Delta T \, g \, h^3/(\nu \chi)]^{1/3} - \chi ,
\label{diffusion_coeff}
\ee

where values $\gamma=3.4 \times 10^{-3} \, {\rm K}^{-1}$, $g=9.8 \, {\rm m s^{-2}}$ and 
$\nu=1.5 \times 10^{-5} \, {\rm m^2 s^{-1}}$ are taken from literature. Since the heat transfer is considered in the horizontal plane, perpendicular to the vertical ``heating wall'' embodied by the fire, the length scale of the convective cell is assumed to be $h=100 \, {\rm m}$ with a temperature difference 
$\Delta T= 100 \, {\rm K}$. Considering these values, the scale of the turbulent diffusion coefficient turns out to be approximately $10^{4}$ times the thermal diffusivity of air at ambient temperature ($\chi$).

\smallskip
The jump length of the firebrands varies according to the meteorological conditions, 
the intensity of the fire and the fuel characteristics \cite{albini-usafs-1983, morris-usafs-1987, catchpole-2002}. 
A precise formulation of the landing distributions is difficult due to the limited small scale experimental results \cite{houssami_etal-ft-2015} 
and the complexities involved in characterising the detailed aspect of firebrand generation and landing. The firebrands land only in the positive direction and their landing distribution increases with distance to reach a maximum before dropping to zero \cite{hage-jam-1961,manzello_etal-fsj-2008}. In past, researchers have utilized different statistical distributions to fit/reproduce the distribution pattern of particles deposited on ground from a given source, e.g., the lognormal distribution is used to depict the distribution of the short-range firebrands (up to $4500 \, \rm{m}$) \cite{sardoy_etal-cf-2008} 
while other studies also indicate the use of a Rayleigh distribution \cite{wang-ft-2011} or a Weibull distribution \cite{kortas_etal-fsj-2009} to describe the landing distributions. 

\smallskip
In this study, a lognormal distribution is used to describe the downwind distribution of the firebrands: 
\be
q(\ell;t) = \frac{1}{\sqrt{2\pi} \, \sigma(t) \, \ell}\exp\left\lbrace -\frac{(\ln \ell/\ell_0 -\mu(t))^{2}}{2 \, \sigma(t)^{2}} \right\rbrace \,, 
\label{lognormal}
\ee
where $\mu(t) = \left\langle \ln \ell/\ell_0\right\rangle $ and $\sigma(t)=\left\langle (\ln \ell/\ell_0 -\mu(t))^{2}\right\rangle$ 
are the mean and the standard deviation of $\ln \ell/\ell_0$ respectively, and $\ell_0$
is a unit reference length.

\smallskip
Since the fuel ignition due to hot air and firebrands is not an instantaneous process, a suitable criterion related to an ignition delay 
is introduced for application to wildfire modelling. This ignition delay was previously considered as a heating-before-burning mechanism due to the hot air 
 \cite{pagnini_etal-crs4-2012,pagnini_etal-mfbr-2012} and generalised to include fire-spotting  \cite{pagnini_etal-nhess-2014}.
In particular, since the fuel can burn because of two pathways, i.e., hot-air heating and firebrand landing, 
the resistance analogy suggests that the resulting ignition delay can be approximately computed as resistances acting in parallel. 
Hence, representing $\tau_h$ and $\tau_f$ as the ignition delay due to hot air and firebrands respectively, the joint ignition delay $\tau$ is
\be
\frac{1}{\tau}=\frac{1}{\tau_h} + \frac{1}{\tau_f} = \frac{\tau_h + \tau_f}{\tau_h \tau_f} \,.
\ee

\smallskip
In this study, the delay effect is portrayed as a heating-before-burning mechanism and described by an accumulative process of the effective fire-front over time, i.e.,
\be
\psi(x,t) =\int_{0}^{t} \phi_{e}(x,\eta)\frac{d\eta}{\tau} \,, 
\label{psi}
\ee
where, $\psi(x,0)=0$ corresponds to the initial unburned fuel. 
This accumulation can be understood as an accumulation of heat that results in an increase of the fuel temperature $T(\bx,t)$. 
Finally, the function $\psi(\bx,t)$ may be related to the fuel temperature $T(\bx,t)$ as follows:
\be
\psi(\bx,t)\propto\frac{T(\bx,t)-T_a(\bx)}{T_{ign}-T_a(\bx)} \,,
\label{psi_temp} 
\ee
where $T_{ign}$ is the ignition temperature, $T((\bx),0) = T_a(\bx)$  and $T(\bx,t)\leq T_{ign}$. 

\smallskip
Let $\Delta t$ be the ignition delay due to the effects of the random processes, then after the elapse of the ignition delay time  
$T(\bx,\Delta t) = T_{ign}$, and with the assumptions $T_a \ll T_{ign}$ and the constant of proportionality equal to $1$,
Eq. \eqref{psi_temp} reduces to 
\be
\psi(\bx,\Delta t) = 1 \,. 
\ee 
Hence, as a cumulative effect of hot air and firebrands on the unburned fuel, an ignition point $\bx$ at time $t$ is created when $\psi(\bx,t) = 1$, and the condition $\phi(\bx, t)=1$ is also imposed.

\section{Turbulence and fire-spotting effects into DEVS based wildfire simulator}
\label{devs}

ForeFire is a DEVS based wild-land fire model developed bereft of any additional terms or parameters to model turbulence or fire-spotting. As described earlier, DEVS is a Lagrangian method where each marker (active burning point in the fire-line) advances by the quantum distance and the advancement of each marker is managed by  \textit{decomposition}, \textit{regeneration} or \textit{coalescence} functions. When one such marker arrives to a new region, the marker decomposes to generate new markers at the interface between the two regions. If this new region belongs to a fuel break zone, the new markers are assigned null propagation velocity and their further evolution is inhibited. But additional information about turbulence and fire spotting can allow the fire-line to cross the fuel breaks. In this article, the DEVS based ForeFire model is extended to include the effects of turbulence and fire-spotting by \textit{post-processing} the output obtained from the model at each time step. The present method does not call for any major changes in the original framework of DEVS, and can be versatilely used to reconstruct its output to include these two processes. The detailed steps of the numerical procedure to re-construct the output of DEVS based ForeFire are described as follows:

\begin{enumerate}[Step 1.]
\item Beginning with an initial fire-line, the DEVS front tracking method is used to estimate the propagation of the markers to build up a new fire perimeter for the next time step. This output is modified to include the effects of turbulence and fire spotting by \textit{post-processing numerical enrichment}. This post-processing step is independent of the definition of ROS. 

\item The fire perimeter obtained from the active burning points is used to construct the indicator function $\phi(\bx,t)$ (\ref{indicatorphi}). The spatial information contained in  $\phi(\bx,t)$ is sufficient to modify the fire-line  with respect to turbulence and fire-spotting and serves as the input to the \textit{post-processing} step. Indicator function $\phi(\bx,t)$ has a value $0$ over the region of the available fuel and $1$ over the burned area. 

\item The spatial coordinates of the fire perimeter obtained from DEVS  are discrete values spread all over the domain and cannot be represented by a specific mesh/grid. For brevity, the effective indicator function $\phi_e(\bx,t)$ (\ref{phi_e}) is generated over a Cartesian grid to facilitate the computation of the function $\psi(\bx,t)$ (\ref{psi}) over the same grid. \textit{Point in polygon} is used to generate the indicator function $\phi(\bx,t)$ from the locus of points representing the fire perimeter in DEVS. It is remarked that $\phi(\bx,t)$, and $\phi_e(\bx,t)$ can also be computed from the real coordinates representing the fire perimeter in DEVS, but the inherent construction of the function $\psi(\bx,t)$ mandates its computation over a regular grid.

\item The value of the effective indicator $\phi_{e}(\bx,t)$ is computed through the numerical integration of the product of 
the indicator function $\phi(\bx,t)$ and the PDF of fluctuations according to Eq. (\ref{phi_e}). The effect of turbulence or fire-spotting is included by choosing the corresponding PDF (\ref{pdf}). 

\item The function $\psi(\bx,t)$ is updated for each grid point by integration in time with the current value of $\phi_{e}(\bx,t)$ (\ref{psi}). 

\item All points which satisfy the condition $\psi(\bx,t) \geq 1$ are labelled as new ignition points. The \textit{post-processing} enrichment of the input fire-front is completed at this step and the new markers describing the enriched fire-front are used to define a new front valid at the next time step in DEVS framework. With the inclusion of heating-before-burning mechanism, the new fire-front perimeter represents a larger area than the input fire perimeter. 

\item At the next time step, the new markers progress according to the DEVS front tracking method, and the updated perimeter is again subjected to the \textit{post-processing} to enrich the fire front with the random fluctuations pertaining to turbulence and fire-spotting. The sequence is repeated till the final ``event time" step or till the fire reaches the end of the domain.

\item The DEVS models time as a continuous parameter and updates the system in accordance with the events. The system is updated at user defined ``event times". In between the consecutive events, no change in the system occurs and a list of pending events is maintained in chronological order. 
The event list is updated at the next ``event time". Since, DEVS provides freedom to select an ``event time", it is chosen to coincide with the maximum allowable time step according to the CFL criterion. It is remarked  that this choice of ``event time" limits the efficiency of DEVS computation (increase in the computation time due to excessive computations), but is indispensable in order to introduce the heating-before-burning mechanism (\ref{psi}) in the current formulation. 
\end{enumerate}
 
\smallskip
It is remarked that the direction of the propagation vector for DEVS is decided by the bisector of the angle formed between 
a marker and its immediate neighbouring markers 
on left and right. This direction vector is a weak approximation of the normal and can be very different from the actual normal direction in the case of a non-circular profile.

\section{Simulation set-up}
\label{setup}
To estimate the performance of the DEVS front tracking method with the inclusion of turbulence and fire-spotting, a series of numerical experiments are performed. Similar set of experiments are also carried out with the LSM based simulator to enable a fair comparison between the two techniques. 
A formulation developed in Ref. \cite{mallet_etal-cma-2009} and Ref. \cite{mandel_etal-gmd-2011} is followed for LSM,  
while for the DEVS method, 
ForeFire fire simulator \cite{filippi_etal-s-2009} is used. 
To allow a direct comparison between the two, both models are parametrised in an identical set-up. 

\smallskip
In the present study, for brevity, no particular type of vegetation is defined and simulations are carried out with a pre-defined constant value of ROS. It is assumed that the ROS remains constant for a particular domain. The parametrisation of the numerical simulations is oversimplified and the test cases are chosen with the purpose of highlighting the potential applicability of the mathematical formulation. It is remarked that the chosen parametrisations for the numerical simulations do not reproduce any of the forest-fire or prescribed burn experiments but to emphasise the applicability of the formulation, the values of various parameters are chosen to approximately lie in the valid range. The present scope of this work is to provide a first look into the investigation of the performance of LSM and DEVS based fire simulators in the context of inclusion of the effects due to turbulence and fire-spotting.

\smallskip
A flat area of hypothetical homogeneous vegetation spread over a domain size of $5000\, \rm m \times 5000\, \rm m $ is selected for simulations. 
Different values of the ROS are utilised for different test cases. The ROS is assumed to be $0.05 \, \rm ms^{-1}$ along the normal direction $\bn$
in no wind conditions, while in the presence of wind, it is estimated by the $3\%$ model \cite{filippi_etal-nhess-2014}, i.e.,
\be
ROS = 0.03 \, \bU \cdot \bn \,, 
\label{three_percent} 
\ee
where $\bU$ is the mean wind velocity. Since, $3\%$ model limits the propagation only towards the mean wind direction, 
a ROS formulation provided by Mallet {\it et al.} \cite{mallet_etal-cma-2009} is followed to prescribe non-zero spread velocities for the flank and rear fires:
\be 
ROS(U,\theta)=
\begin{cases}
\varepsilon_{o}+a\sqrt{U\cos^{n}\theta}, & \text{if  } |\theta|\leq \frac{\pi}{2}\,, \\
\varepsilon_{o}(\alpha+(1-\alpha)|\sin\theta|),& \text{if  } |\theta| > \frac{\pi}{2}\,,
\end{cases}
\label{Mallet}
\ee
where $U^2=\bU \cdot \bU$, $\varepsilon_{o}$ is the flank velocity,  
($\varepsilon_{o}\alpha$) is the rear velocity with $\alpha\in[0,1]$, 
and $\theta$ is defined as the angle between the normal to the front and the wind direction. 
In the present set-up, the parameters are defined as 
$\alpha = 0.8$, $n = 3$, $a=0.5 \, \rm m^{1/2}s^{-1/2}$, $\varepsilon_{o}=0.2 \, \rm ms^{-1}$, and $U = 3\, \rm ms^{-1}$ \cite{mallet_etal-cma-2009}. 

\smallskip
In the case of the LSM based simulator, 
the domain is discretised with a Cartesian grid of $20\, \rm m$ both in $x$ and $y$ directions, 
while for DEVS the resolution of the simulation is established in the terms of quantum distance $\Delta q$ and perimeter resolution $\Delta c$ \cite{filippi_etal-s-2009}. Quantum distance $\Delta q$ is defined as the maximum allowable distance to be covered by a particle at each advance, 
while a measure of $\Delta c$ is used to decompose/regenerate/coalesce two particles on propagation. 
The choice of $\Delta q$ and $\Delta c$ is {\it dependent on the type of problem}, and in the present study, two sets of values are utilized. 
The simulations are performed with $\Delta q = 4\, \rm m $, $\Delta c = 18\, \rm m $ for zero wind conditions, and $\Delta q = 2\,\rm m $, $\Delta c = 8\,\rm m$ in the presence of wind. In DEVS based simulator, the locus of the points constituting the initial front are wind up in the clockwise direction to represent an expanding front (anticlockwise order represents a contracting front). For each numerical simulation presented here, the initial fire-front is represented by 200 markers. A brief study focussing on the numerical behaviour of DEVS and LSM based techniques is also presented in the next section.

\smallskip
The mean wind, wherever used, is assumed to be constant in magnitude and direction. The turbulent diffusion coefficient $D$, and ignition delays corresponding to hot air and firebrands heating are also assumed to be constant throughout the simulations. In this study, the ignition delay corresponding to the hot air $\tau_h$ and to the firebrand $\tau_f$ are assumed to be $10 \, \rm min$ and $1\,\rm min$ respectively. Physically, a measure of the ignition delay can be inferred from the fire intensity, fuel properties and the length scale at which the fuel is exposed to the heat source \cite{anderson-ft-1970}. According to Ref. \cite{anderson-ft-1970}, for the same fuel characteristics and low fire intensities the ignition delay due to hot air is  higher than the firebrand landing, but in the cases of high fire intensities, the ignition delay at different length scales can be comparable due to the functional relationship between the ability of the fuel surface to absorb energy and the ability of the fuel to conduct heat inward. Flammability classification of different natural fuels available in literature \cite{dimitrakopoulos_etal-ft-2001, dickinson_etal-jb-1985,bowman_etal-jb-1988} can aid in the selection of an appropriate value for ignition delay.

\smallskip
Appropriate values of $D$ are chosen by using formula \ref{diffusion_coeff} to assess the effect of increasing level of turbulence and to highlight its role in comparison to the firebrand landing. In this study, three different values of $D$ are utilised: $D= 0.30 \, {\rm m^2 s^{-1}}$, $D=0.15 \, {\rm m^2 s^{-1}}$, $D=0.075 \, {\rm m^2 s^{-1}}$.

Also, with reference to the lognormal distribution (\ref{lognormal}), 
the unit reference length $\ell_0$ in feet ($1\,\rm ft = 0.30\,\rm m)$.

\smallskip
Multiple idealised simulation tests are performed both in the presence and the absence of wind by neglecting and considering the effects of the random phenomena.
The {\it first} case evaluates an isotropic growth of the fire-line in the absence of wind by neglecting all the random processes. A fire-break zone 60 $\rm m$ wide is also considered in this case.
In the {\it second} test, the spread of fire-line for different ROS in different directions is studied. 
The {\it third} test discusses the propagation of the fire-line with wind when the ROS is defined according to the $3\%$ of the mean wind,
formula (\ref{three_percent}), and when it is defined according to formula (\ref{Mallet}) . 
The random processes are neglected for the first three test cases. The pure LSM and pure DEVS based wildfire models fail while managing the realistic cases of fire propagation across the fire-break zones, 
hence the {\it fourth} test evaluates the performance of DEVS when turbulent processes are also considered in the model. The effect of turbulent processes is studied both in the presence and the absence of the wind. 
The {\it fifth} test evaluates the performance of the two simulators when fire-spotting is also included along with turbulence. 
Fire-break zones are also introduced in the last two tests (the fourth and the fifth) to observe the propagation of the fire-line while encountering areas of null fuel. 
To simplify the simulation and maintain equivalence between the two simulators, the region across and behind the centre of the initial fire-line is demarcated as the leeward side and the windward side respectively. 

\smallskip
As an extension of the study, a brief analysis of the firebrand landing distribution is also made. 
With reference to the lognormal distribution (\ref{lognormal}) for firebrand landing,
different values of $\mu$ and $\sigma$ are chosen motivated by a possible correlation with some physical aspects of wildfire like changing fire intensity and meteorological conditions.  
For these simulations, only the LSM based simulator is used and the wind speed is assumed to be $3\,\rm ms^{-1}$, while the ROS is given by the Rothermel model
\cite{rothermel-1972,mandel_etal-gmd-2011}. 

\smallskip
All the simulations have been performed at BCAM--Basque Center for Applied Mathematics (\url {www.bcamath.org}) in Bilbao, Basque Country -- Spain.
\section{Discussion}  
\label{Discussion}
\subsection{Numerical behaviour of DEVS and LSM}
\begin{figure}[!thp]
	\centering
    \includegraphics[trim = 0.5cm 6cm 2cm 6cm, clip=true, width =6.5cm, height =7cm]{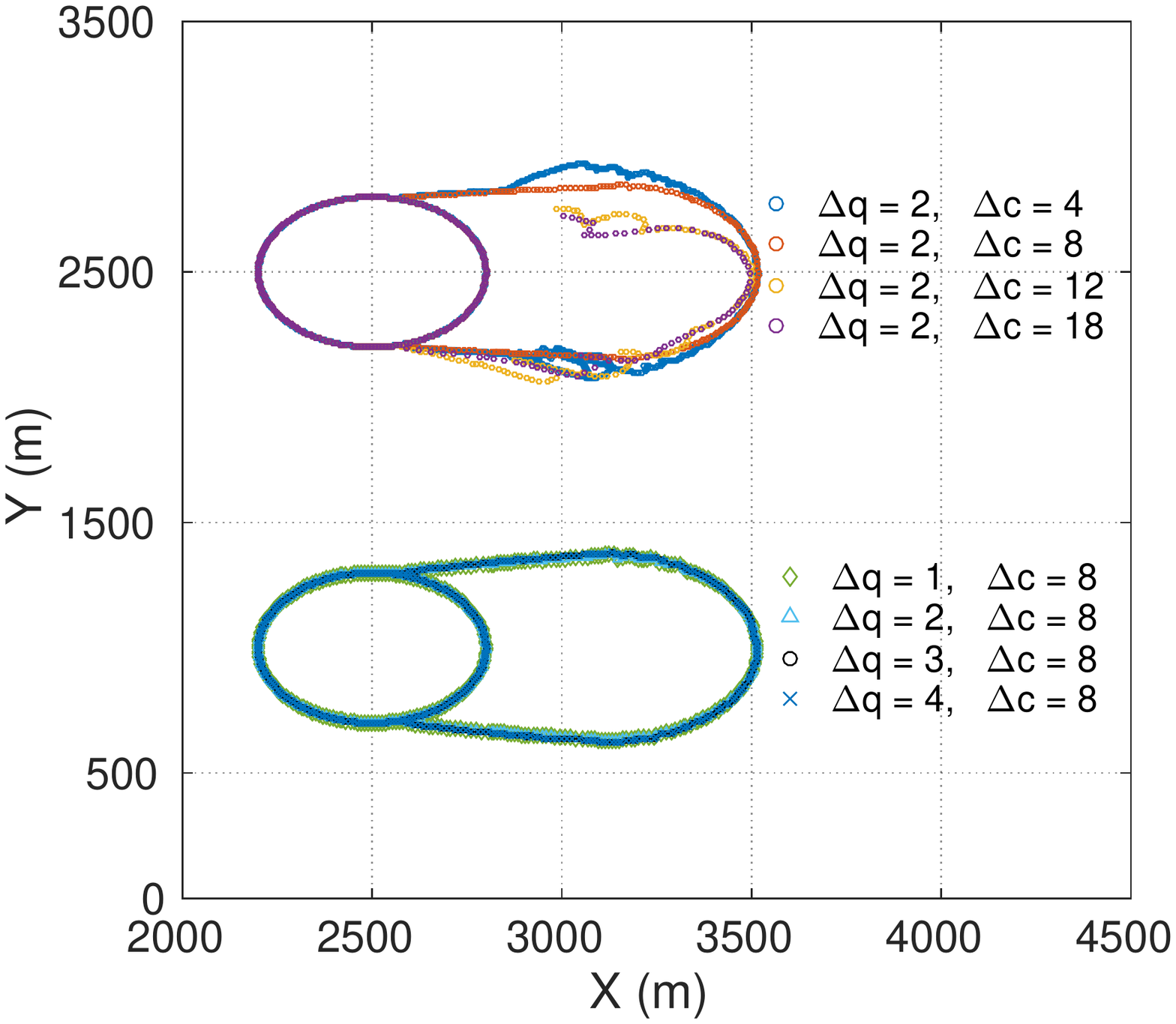}
   \caption{The fire-lines showing the numerical behaviour of DEVS for different pairs of $\Delta c$  and $\Delta q$. Each contour represents the fire-line at 134 min. The initial fire is a circle of radius 300 m and represented by 200 markers in the model.}
   \label{fig:deltaq_deltac}
\end{figure}

The stability of DEVS method relies on the choice of quantum distance $\Delta q$ and the perimeter resolution $\Delta c$ but the numerical simulations involving a uniform ROS in all directions (e.g., a constant 0.05 $\rm m s^{-1}$ irrespective of wind or direction) are unaffected by the choice of these two parameters. In a homogenous domain, a uniform ROS forces all the markers to advance with an identical speed at a same time step and leads to generation of synchronous events. In this case, all the markers are driven by the same CFL condition, and hence, the motion of the  markers is analogous to an Eulerian technique. On the contrary, markers with different speed generate different number of discrete events and the resolution of $\Delta q$ and $\Delta c$ comes into picture.  In context of the idealised test cases involving wind, an evaluation of performance of the method concerning  different values of $\Delta q$ and $\Delta c$ is demonstrated in Fig.~\ref{fig:deltaq_deltac}. The numerical simulations show the evolution of the fire-line in the presence of an east wind of 3 $\rm ms^{-1}$ with ROS as 3\% of the normal wind, for different pairs of $\Delta q$ and $\Delta c$. Each contour represents the fire-front at 134 $\rm min$. The perimeter resolution should be at least twice of the quantum distance ($\Delta c \geq 2\Delta q$) to ensure that two markers do not cross-over \cite{filippi_etal-james-2009}. Taking into account this restriction, the upper plot of Fig.~\ref{fig:deltaq_deltac} highlights the differences that arise in the evolution of the fire-front with respect to different values of $\Delta c$. DEVS employs a neighbourhood check for each marker to identify its two immediate neighbours and utilises these neighbours to regulate the shape of the front by merging and generating new markers. A large value of $\Delta c$ can consider disconnected markers as neighbours and can lead to a undesired generation or merging of markers. The loss of information by unsought merging can cause the front to collapse. On the other hand, decreasing the perimeter resolution $\Delta c$ without increasing the resolution of the marker jumps causes the \textit{regeneration} function to be evoked frequently and leads to creation of a large number of markers with small separations. Over time, too many markers at a coarse spatial scale leads to excessive unnecessary computations and small numerical errors introduced by the computation can escalate quickly over time. A finer scale of $\Delta q$ is desired with small values of $\Delta c$ for a complete optimisation of the numerical simulation.
The lower plot in Fig.~\ref{fig:deltaq_deltac} shows the variation of the fire-front with respect to increasing resolution of $\Delta q$. Though a finer resolution of $\Delta q$ resolves the fire map at a better resolution, it has no significant impact on the simulation results; on the other hand, it increases the computation cost of the simulation. In the present scenario, the homogeneous vegetation and topography of the present domain causes the Lagrangian movement of markers to be driven predominantly by \textit{regeneration} and the fine scale information about the domain is redundant. But on the other hand, in an inhomogeneous domain (involving diverse vegetation, irregular topography, etc), where both \textit{decomposition} and \textit{coalescence} play an important role, a finer marker jump length is necessary to account for the finer details of the domain. Usually, the value of $\Delta q$ is taken to be around 2-3 $\rm m$ to ensure a correct  representation of the variable vegetation, changing topography and fire break zones like roads, rivers etc. Concerning a pure quantitative analysis of the total error in DEVS, the interested readers are referred to \cite{filippi_etal-james-2009} and \cite{kofman-s-2002}. 

\smallskip

\begin{figure}[!thp]
	\centering
    \includegraphics[ width =9cm, height =6cm]{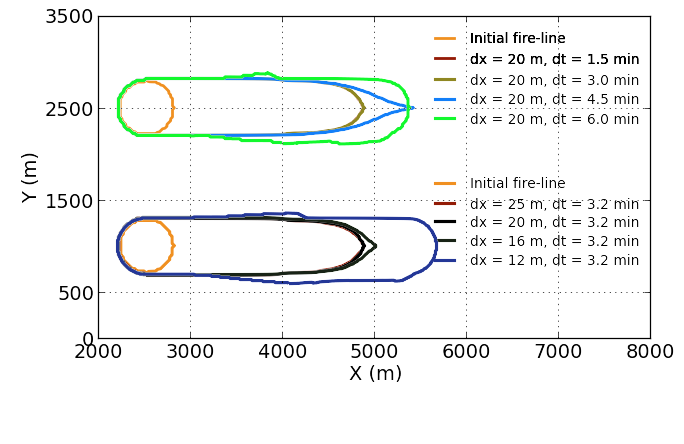}
   \caption{The fire-lines showing the numerical behaviour of LSM for  different pairs of $dt$ and $dx$. Each contour represents the fire-line at 380 min and the initial fire is a circle of radius 300 m.}
   \label{fig:dt_dx}
\end{figure}

The numerical behaviour of the LSM is dependent on the time step ($dt$), the grid resolution ($dx,dy$) and the maximum absolute speed of all points on the grid. The iterative time step involved while solving the differential equations needs to satisfy the CFL criteria for a converging solution. The CFL condition defines the maximum limit of the time step in order to limit the  evolution of the contour to at most one grid cell at each time step. This upper limit on the time step is important for the stability of the system and a converging solution. In one-dimension, the CFL condition can be described as:
\be
dt \leq k \, \frac{\min(dx,dy)}{v_{max}}
\label{cfl}
\ee
where, $v_{max}$ is the maximum absolute speed, and  $0<k<1$ is a constant. To demonstrate the numerical behaviour of LSM, similar numerical simulations with 3\% model and an east wind of  $3\rm\,ms^{-1}$ are repeated for different values of $dt$ and $dx$ . The upper figure in Fig.~\ref{fig:dt_dx} shows the fire-line contours at 380 min for various values of $dt$, and when $dx =dy = 20 \rm \,m$. With values of $dt$ smaller than the threshold defined by CFL condition, the fire-line evolves without any instabilities. Similar results are achieved with further decrease in the time step but with an increase in the computation time. On the other hand, when for the same speed function, $dt$ is chosen to violate the CFL criteria, the fire-line evolves with an undesirable behaviour and this obscurity magnifies over time. In the same way, CFL condition can also be interpreted as the constraint on the grid resolution. The lower figure in Fig.~\ref{fig:dt_dx} shows the evolution of the fire-line with respect to different values of $dx$. The increase in the grid resolution over an identical time step introduces numerical instability and causes the  fire-lines to evolve erroneously. For the same velocity field, shorter time steps are required over small grid spacings in order to respect the CFL condition, though it leads to an increase in the computation costs. Hence, accounting for the numerical behaviour of the LSM and to avoid the excessive computation, in the present study, $dt$ is defined according to (\ref{cfl}), with $k = 0.5$.

\smallskip
It is remarked here that since  the mathematical formulation for turbulence and fire-spotting is implemented in DEVS and LSM based wildfire simulators as a \textit{post-processing} scheme, it preserves the original framework of the wildfire simulators and does not affect the numerical stability and convergence of both these methods. 

\subsection{Comparison of DEVS and LSM} 
\begin{figure}[p]
   \begin{subfigure}{0.5\textwidth}
	\centering
    \includegraphics[width = 5.2cm, height =5cm]{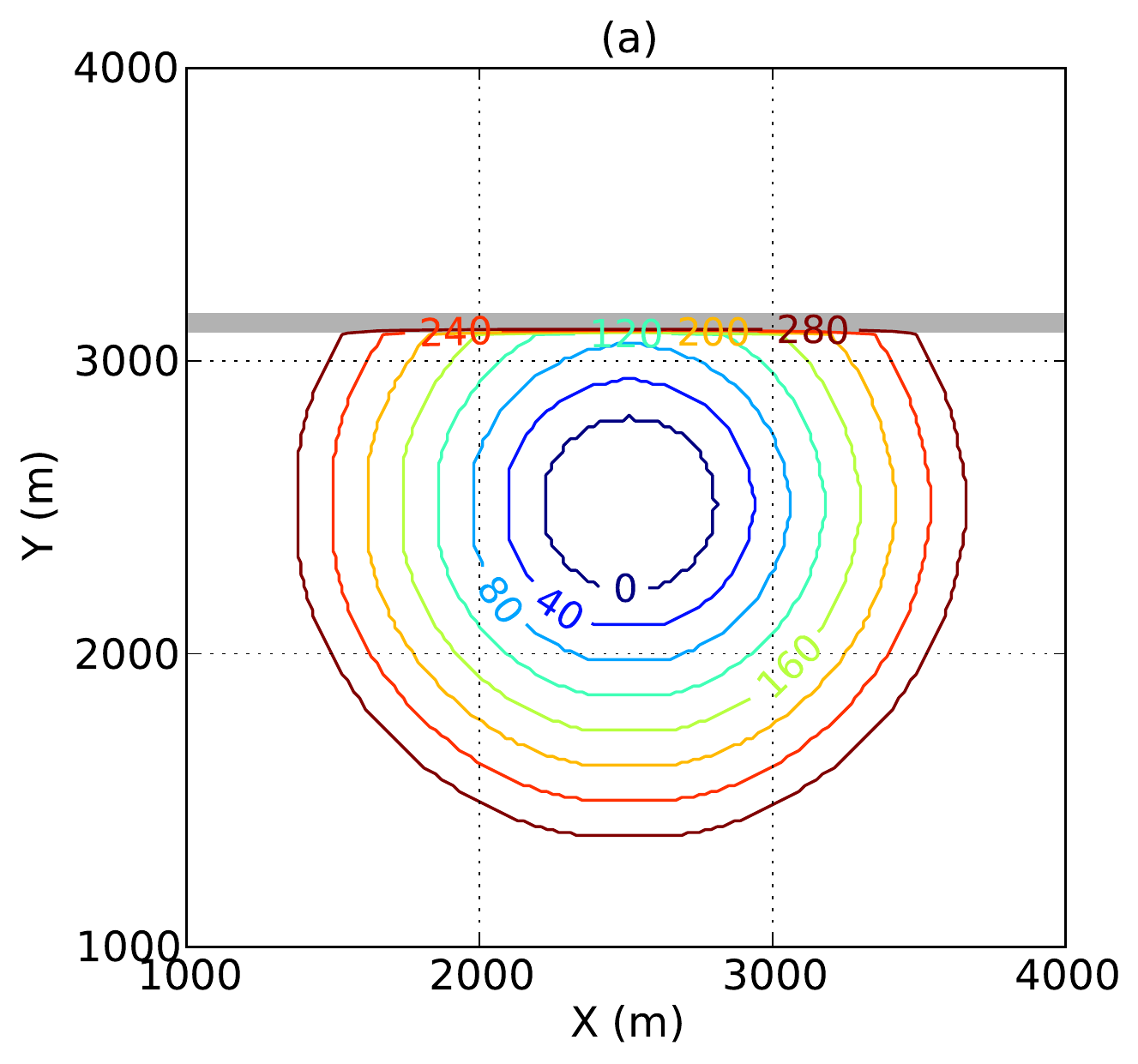}
       \end{subfigure}
    \begin{subfigure}{0.5\textwidth}
	\centering
    \includegraphics[trim = 2cm 8cm 4cm 8cm, clip=true, width = 8cm, height =6.3cm]{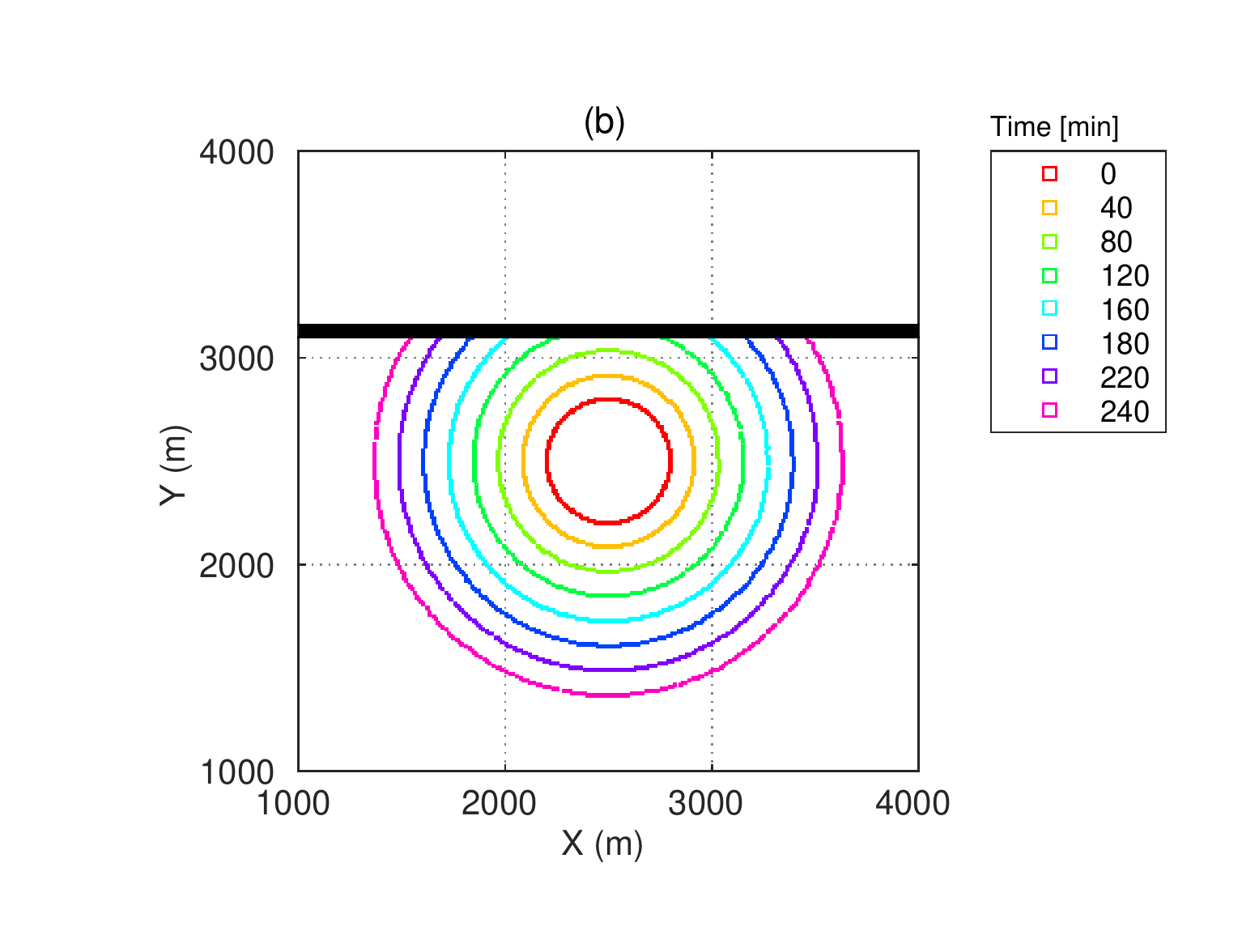}
    \end{subfigure}
         \caption{Evolution in time of the fire-line contour without random processes in absence of wind for the a) LSM and b) DEVS based simulators.
The initial fire-line is a circle of radius $300\,\rm m$. The fire-break zone is 60 m wide.}
   \label{fig:FF_LSM_no_wind_firebreak}
\end{figure}

In the first three test scenarios, the evolution of the fire-line without the impact of random processes in both LSM and DEVS approaches is analysed. Fig.~\ref{fig:FF_LSM_no_wind_firebreak} presents the propagation of the fire-line in no wind conditions for both the simulators. The advancement of the initial circular fire-line follows an isotropic growth and with identical initial conditions the two different moving interface schemes provide a similar evolution of the fire-line. But the fire-line fails to propagate across the fire-break zone in both the techniques.
%
The evolution is shown only up to $140\,\rm min$, but an extended run up to $250\,\rm min$ (not shown) indicates the limitation of the two approaches to reproduce the fire jump across the no fuel zone. 

\begin{figure}[p]
   \begin{subfigure}{0.5\textwidth}
	\centering
    \includegraphics[height = 5cm, width = 5cm]{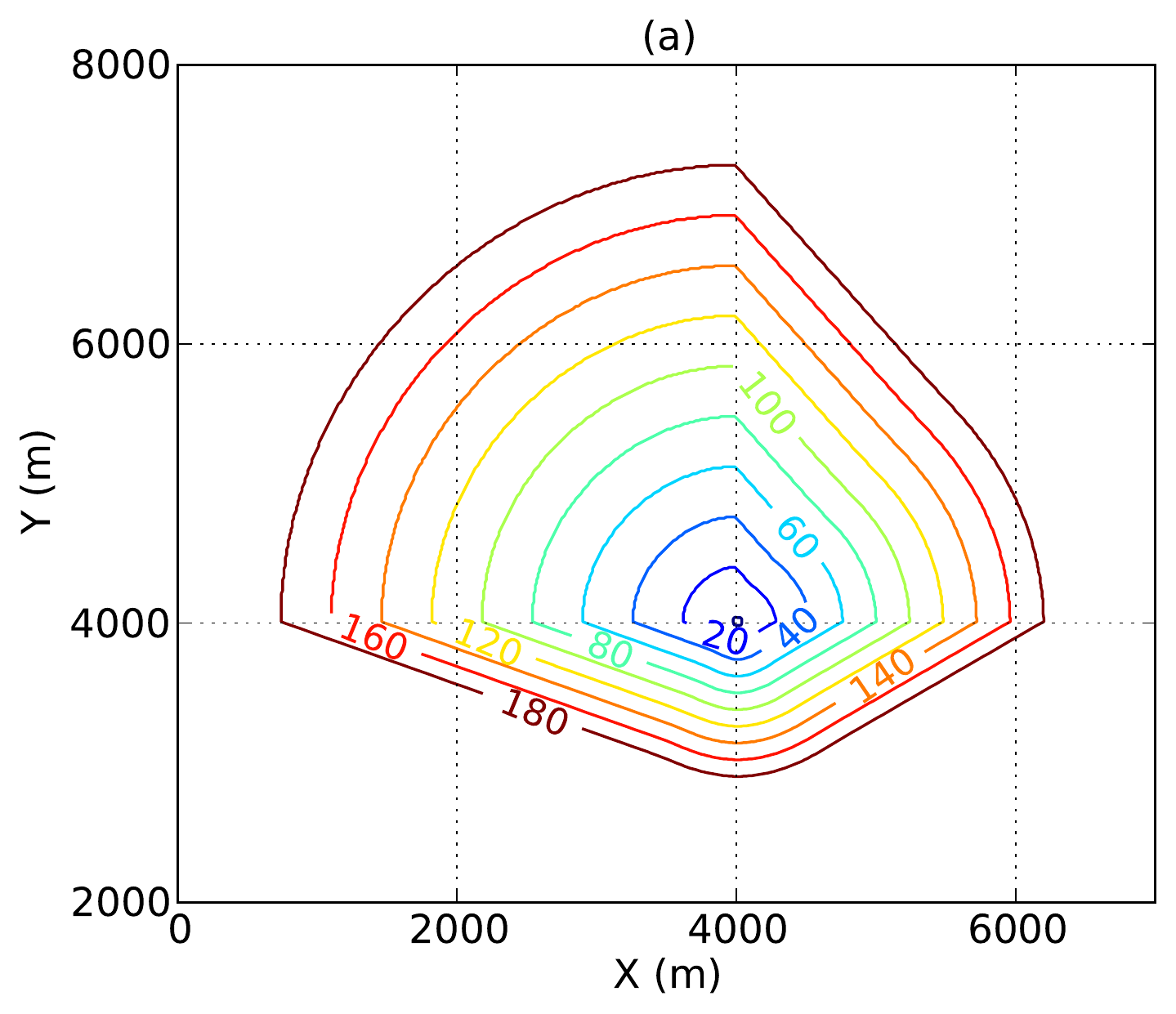}
       \end{subfigure}
    \begin{subfigure}{0.5\textwidth}
	\centering
    \includegraphics[trim = 2cm 8cm 4cm 8cm, clip=true, width = 8cm, height =6cm]{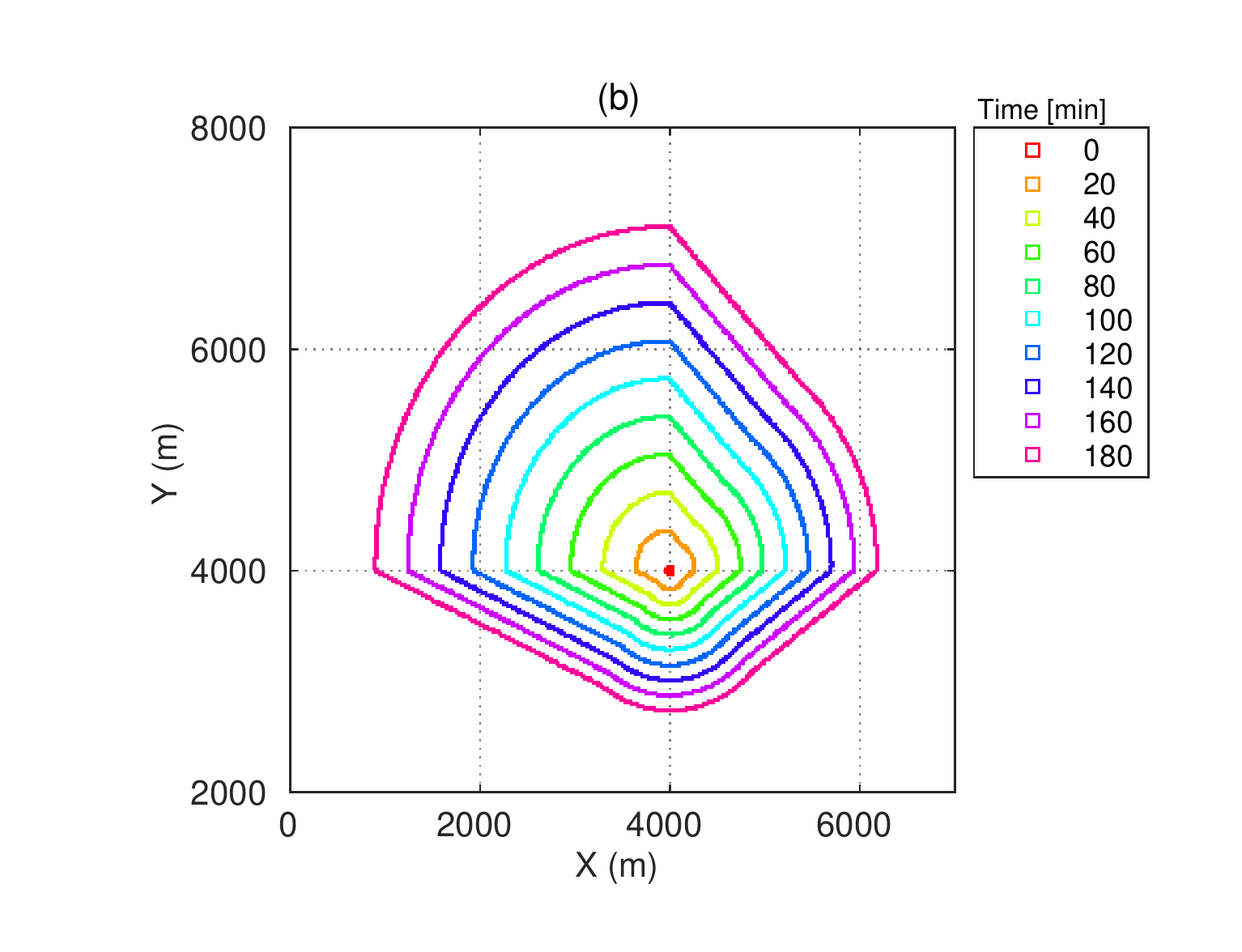}
    \end{subfigure}
     \caption{Evolution in time of the fire-line contour without random processes in absence of wind with a non-homogeneous ROS for a) LSM and b) DEVS based simulators.
The initial fire-line is a circle of radius $30\,\rm m$.}
   \label{fig:FF_LSM_nowind_inhomo}
\end{figure}

\smallskip 
Fig.~\ref{fig:FF_LSM_nowind_inhomo} presents the growth of an initial spot fire but with a non-homogeneous ROS in absence of any wind. 
The different values of the ROS can be attributed to different fuel types, but here no effort has been made to describe the fuel type. 
The fire-line propagates with different speed in the three different directions, but the isotropy is preserved. The results from these two test cases indicate that in the absence of wind, the two advancing schemes: LSM and DEVS, show an identical behaviour in simulating situations with constant ROS and hence, this observation provides a crucial background and scope for the comparison of situations with increasing variability and complexity.

\begin{figure}[p]
   \begin{subfigure}{0.5\textwidth}
	\centering
    \includegraphics[scale = 0.38]{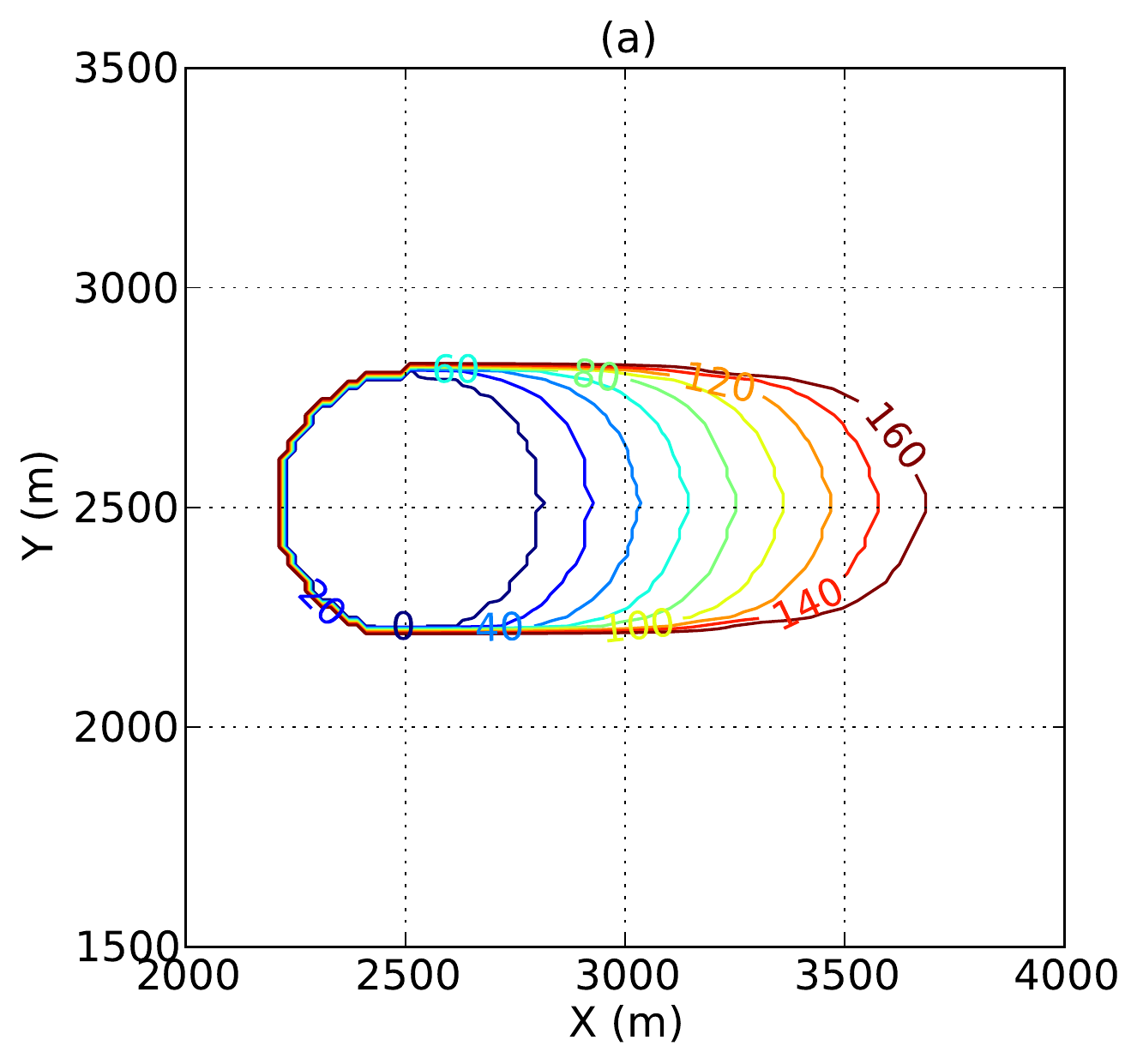}
       \end{subfigure}
    \begin{subfigure}{0.5\textwidth}
	\centering
    \includegraphics[trim = 1.5cm 8cm 4cm 8cm, clip=true, scale = 0.4]{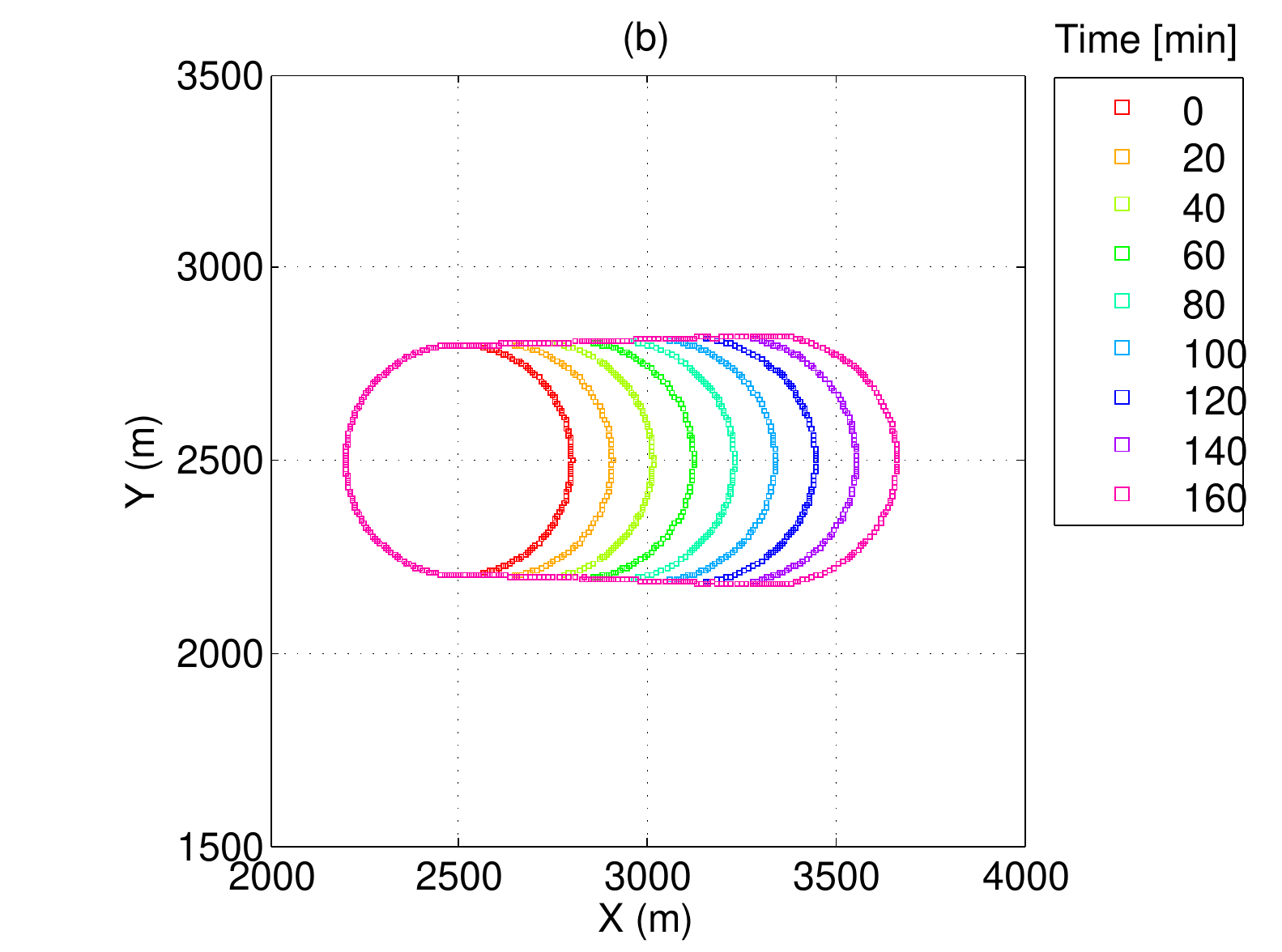}
    \end{subfigure}
            \caption{Evolution in time of the fire-line contour without random processes with an east wind of $3\, \rm m s^{-1}$ for a) LSM and b) DEVS
based simulators. The initial fire-line is a circle of radius $300\, \rm m$.}
   \label{fig:FF_LSM_wind_circle}
\end{figure}

\begin{figure}[p]
   \begin{subfigure}{0.5\textwidth}
	\centering
    \includegraphics[scale = 0.38]{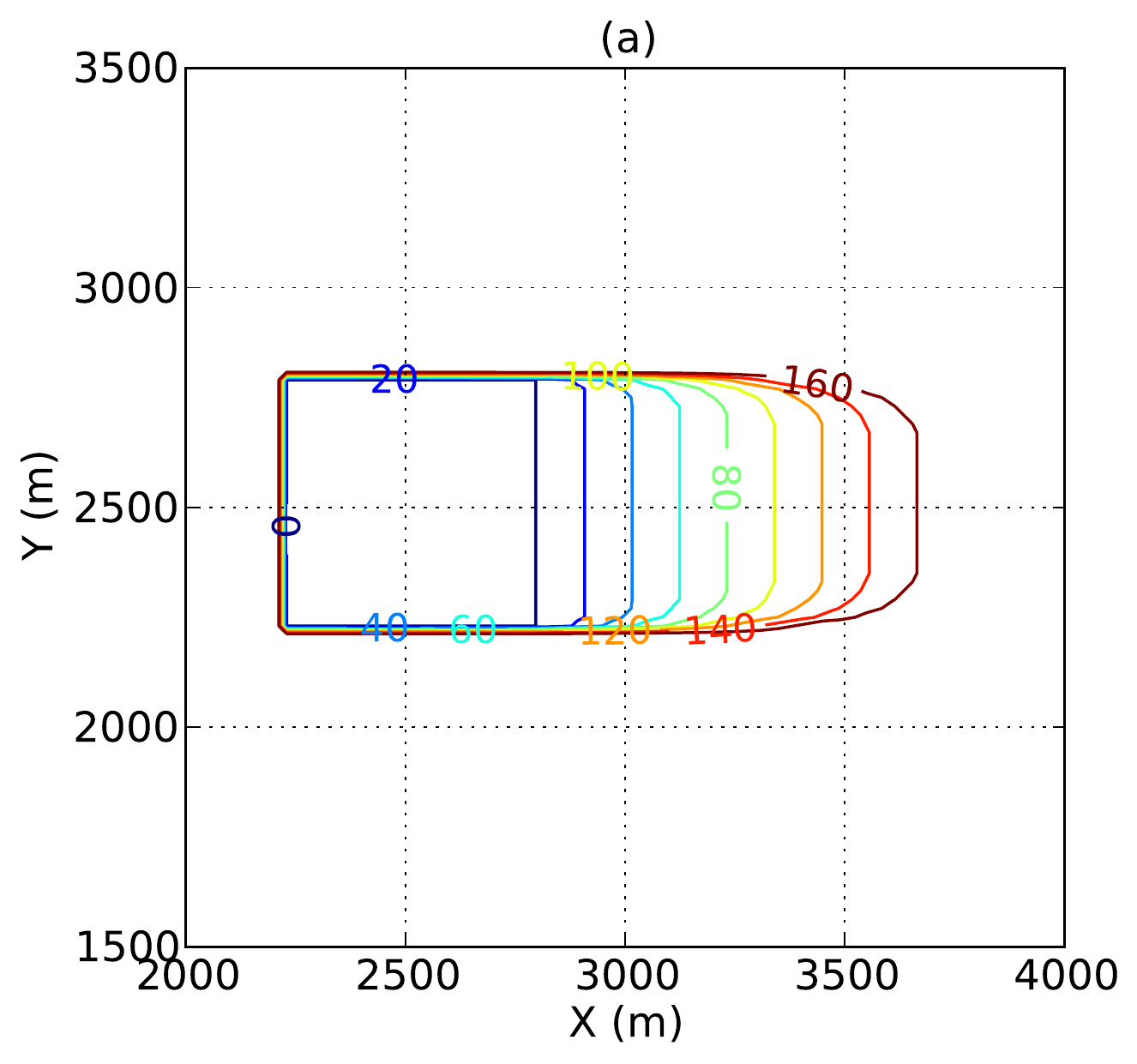}
       \end{subfigure}
    \begin{subfigure}{0.5\textwidth}
	\centering
    \includegraphics[trim = 1.5cm 8cm 4cm 8cm, clip=true, scale = 0.4]{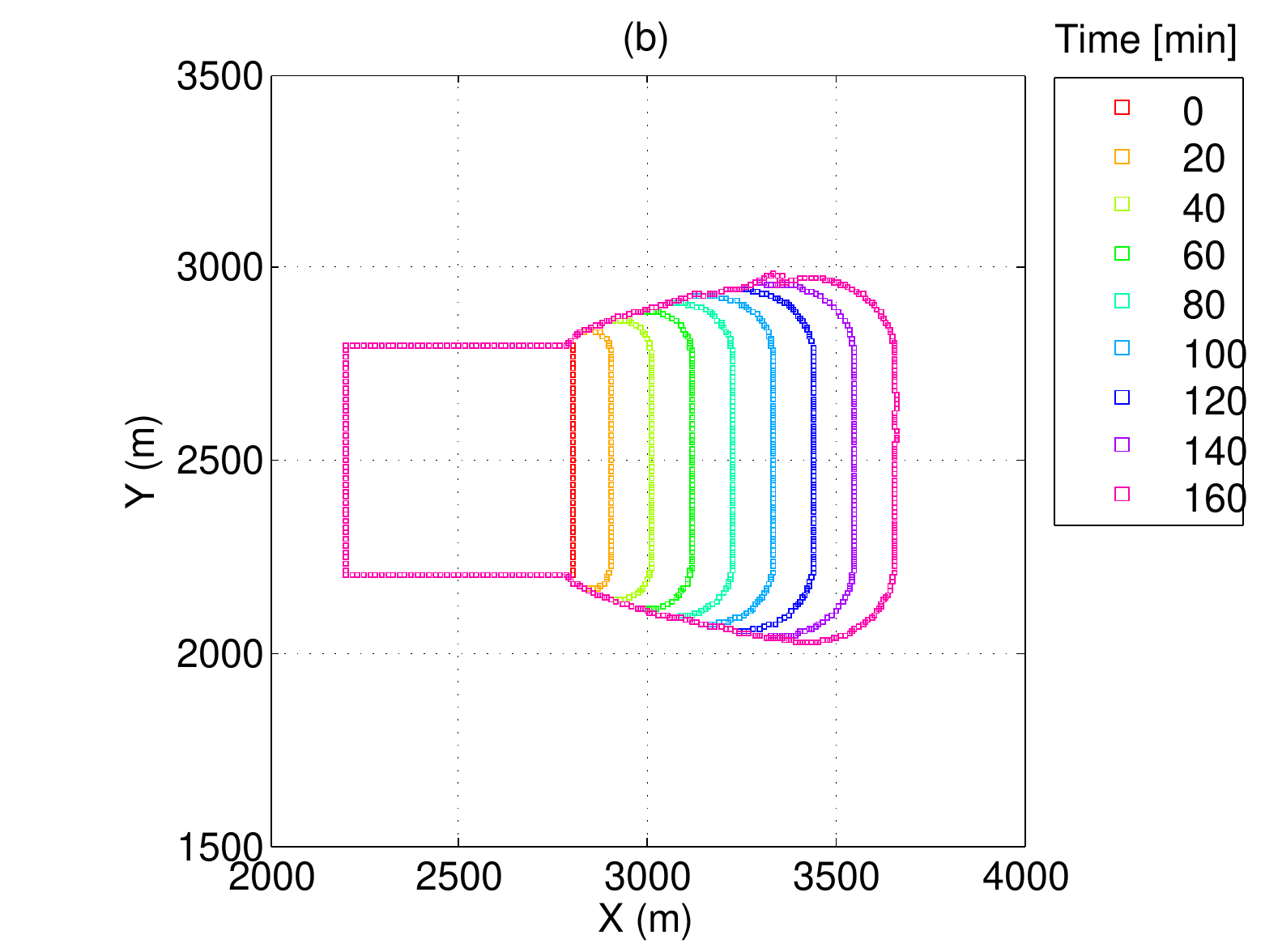}
    \end{subfigure}
     \caption{Same as Fig.~\ref{fig:FF_LSM_wind_circle}, but when the initial fire-line is a square of side $600\, \rm m$.}
   \label{fig:FF_LSM_wind_square}
\end{figure}

\smallskip 
Fig.~\ref{fig:FF_LSM_wind_circle} shows the  evolution of the fire-front of a circular initial profile with radius $300 \, \rm m$
in case of a weak wind of $3 \, \rm m s^{-1}$ 
directed in the east direction. The isochronous fronts are plotted at every $20\,\rm min$ and follow an oval shape for both the simulators. 
DEVS fire contours diverge slightly from the mean wind direction 
and develop into an increasing flanking fire over time. This divergence in the evolution of fire-front occurs due to differences in the computation of 
the normal between the two approaches. This fact can be very well appreciated when an initial square profile is considered.  Fig.~\ref{fig:FF_LSM_wind_square} shows the progress of the fire when the initial fire perimeter is considered to be a square with side $600 \,\rm m$. 
Under the effect of a constant zonal wind and restricted $3\%$ model (\ref{three_percent}), the evolution in LSM strictly follows the initial square shape, but in DEVS the active burning markers at the corners advance spuriously to provide an additional flanking spread. The DEVS markers at the corners are affected by the approximate normal and this leads to a larger flank fire. Whereas, the normal direction for markers pointing towards the direction of wind is identical for both the methods, hence the head fires arrive to same location in both the simulations. 
The flank fires generated by the approximate construction of the normal seem to be in a more qualitative agreement with a realistic case, but 
within the stated analytical formulation of the ROS (\ref{three_percent}), the lateral progress of the flank fires is an anomalous behaviour. 

\begin{figure}[th]
   \begin{subfigure}{0.5\textwidth}
	\centering
    \includegraphics[ width = 5cm, height =5cm]{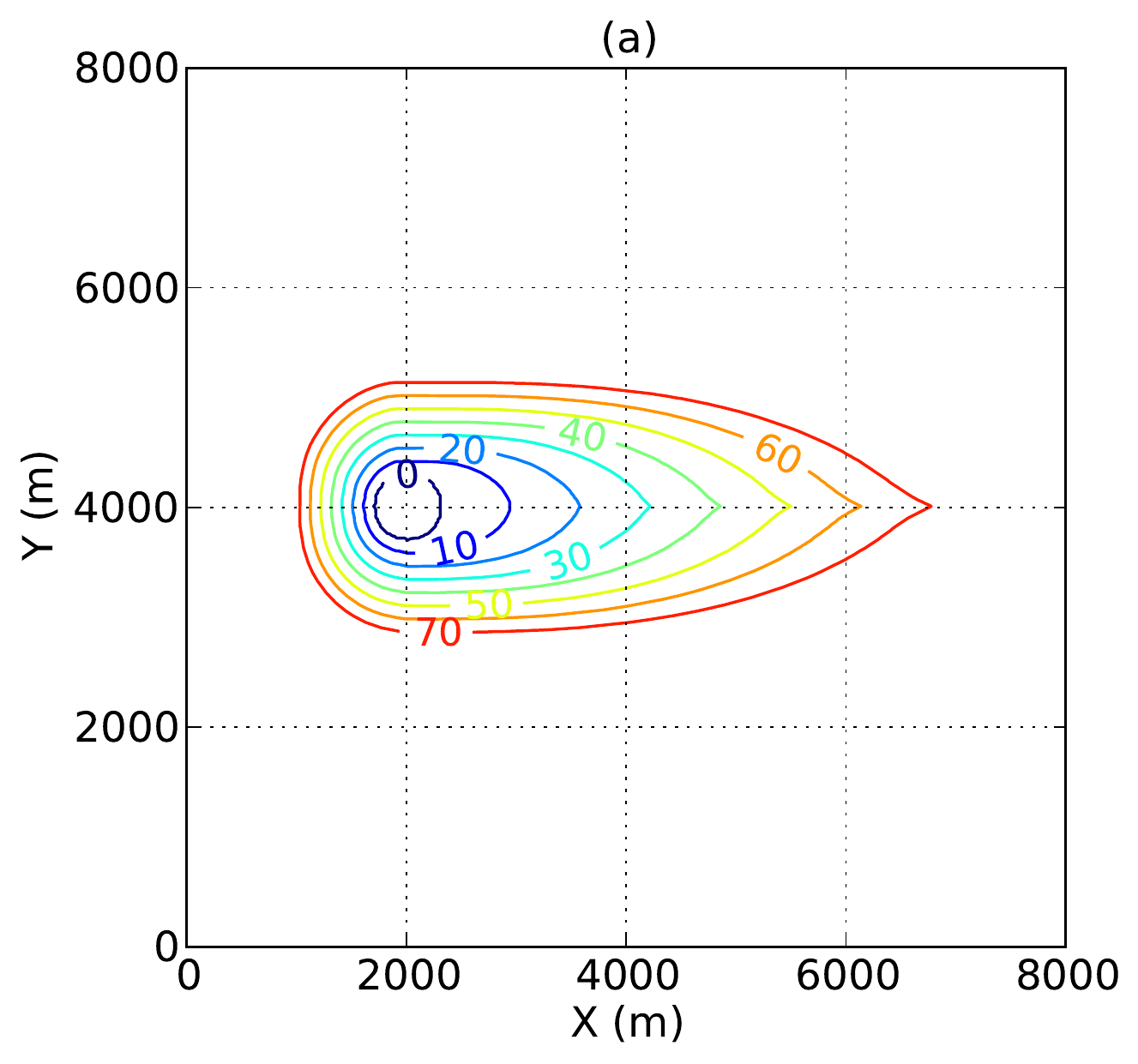}
       \end{subfigure}
    \begin{subfigure}{0.5\textwidth}
	\centering
    \includegraphics[trim = 2cm 8cm 4cm 8cm, clip=true, width = 8cm, height =6cm]{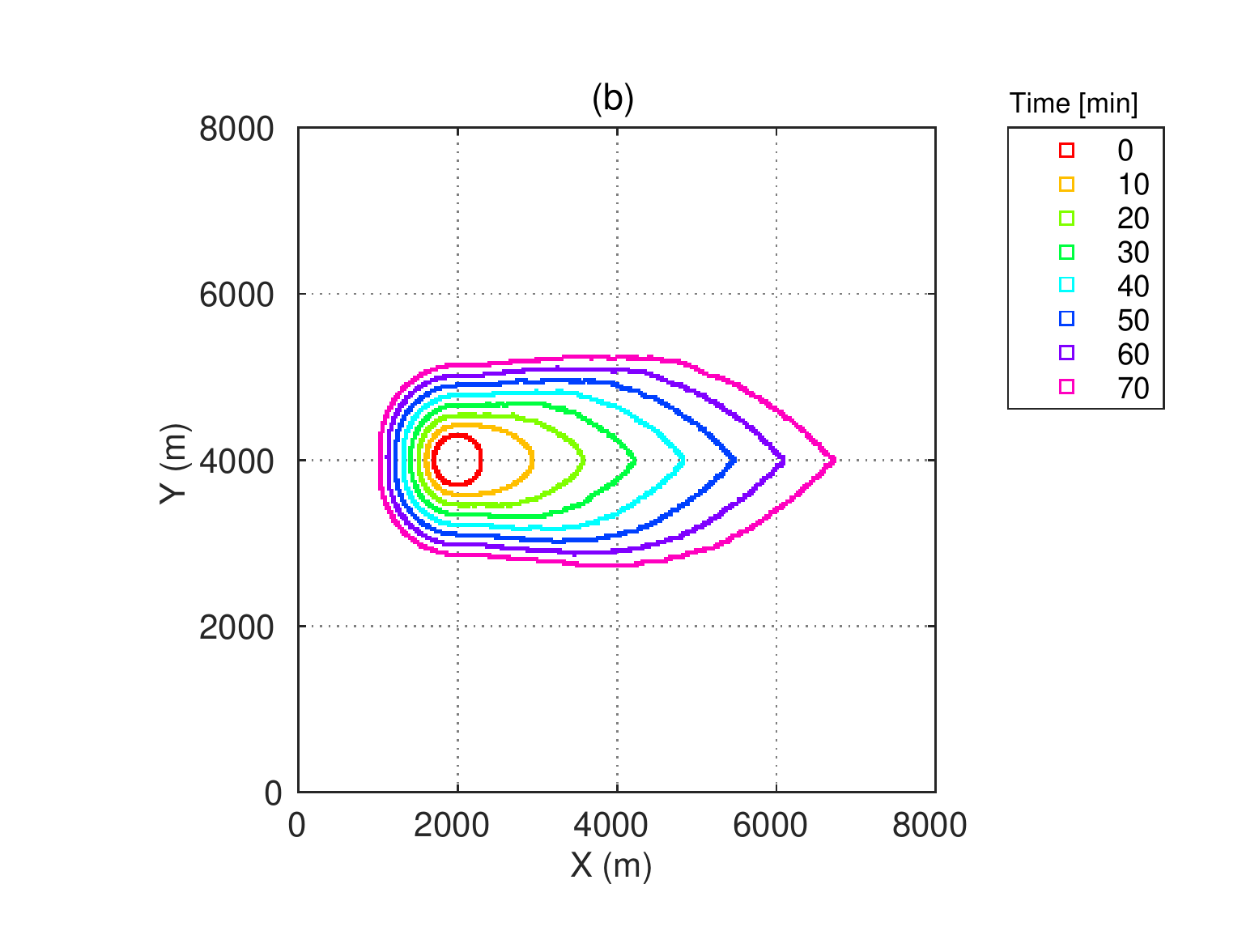}
    \end{subfigure}
      \caption{Evolution in time of the fire-line contour without random processes with ROS given by formula (\ref{Mallet}) 
and when $\theta$ is the angle between the outward normal in a contour point and the mean wind direction for a) LSM and b) DEVS based simulators. 
The mean wind velocity is $3 \, \rm ms^{-1}$ in the positive $x$-direction.}
      \label{fig:FF_LSM_Mallet_with_normal}
\end{figure}

\begin{figure}[th]
   \begin{subfigure}{0.5\textwidth}
	\centering
    \includegraphics[width = 5cm, height =5cm]{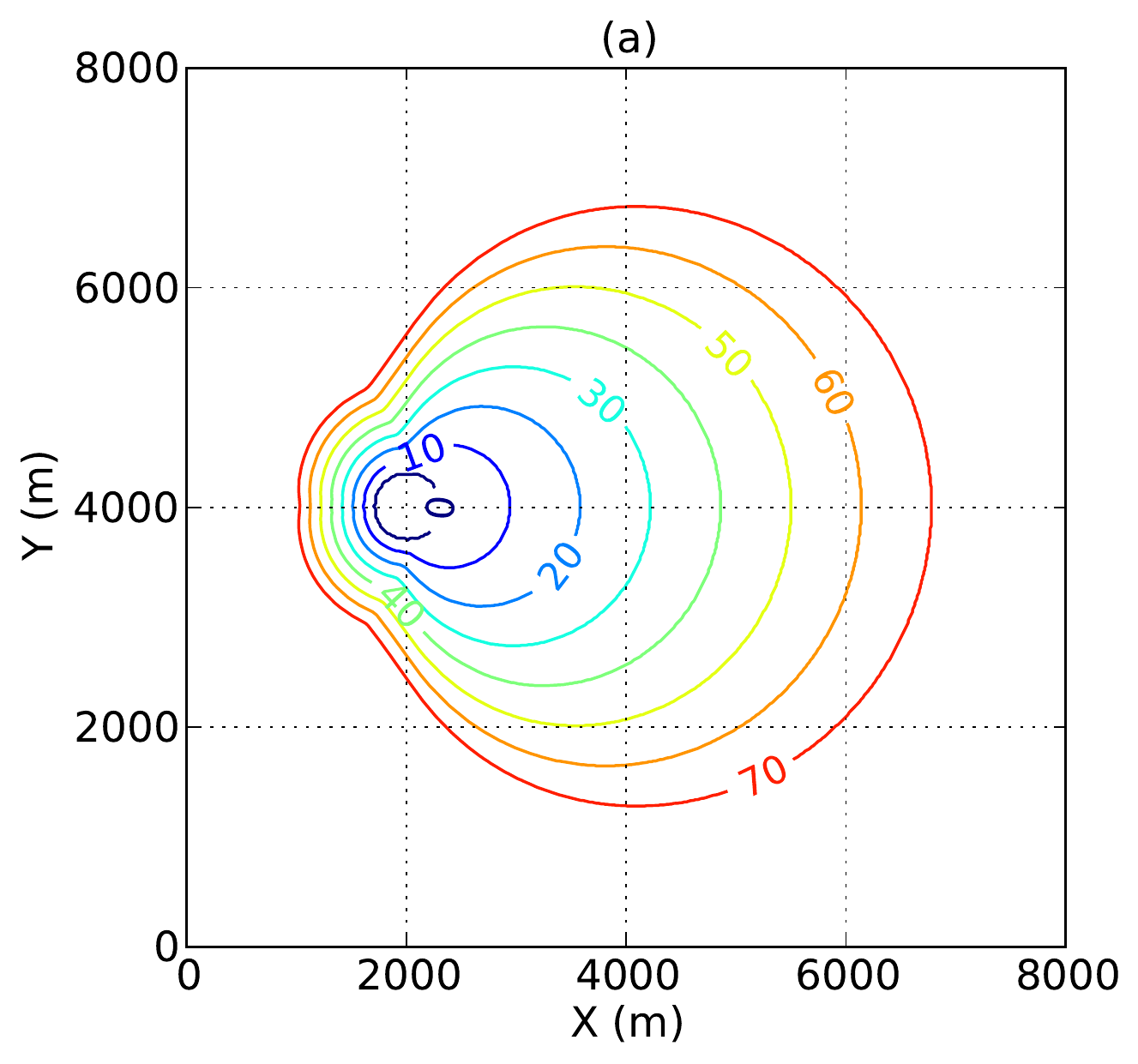}
       \end{subfigure}
    \begin{subfigure}{0.5\textwidth}
	\centering
    \includegraphics[trim = 2cm 8cm 4cm 8cm, clip=true, width = 8cm, height =6cm]{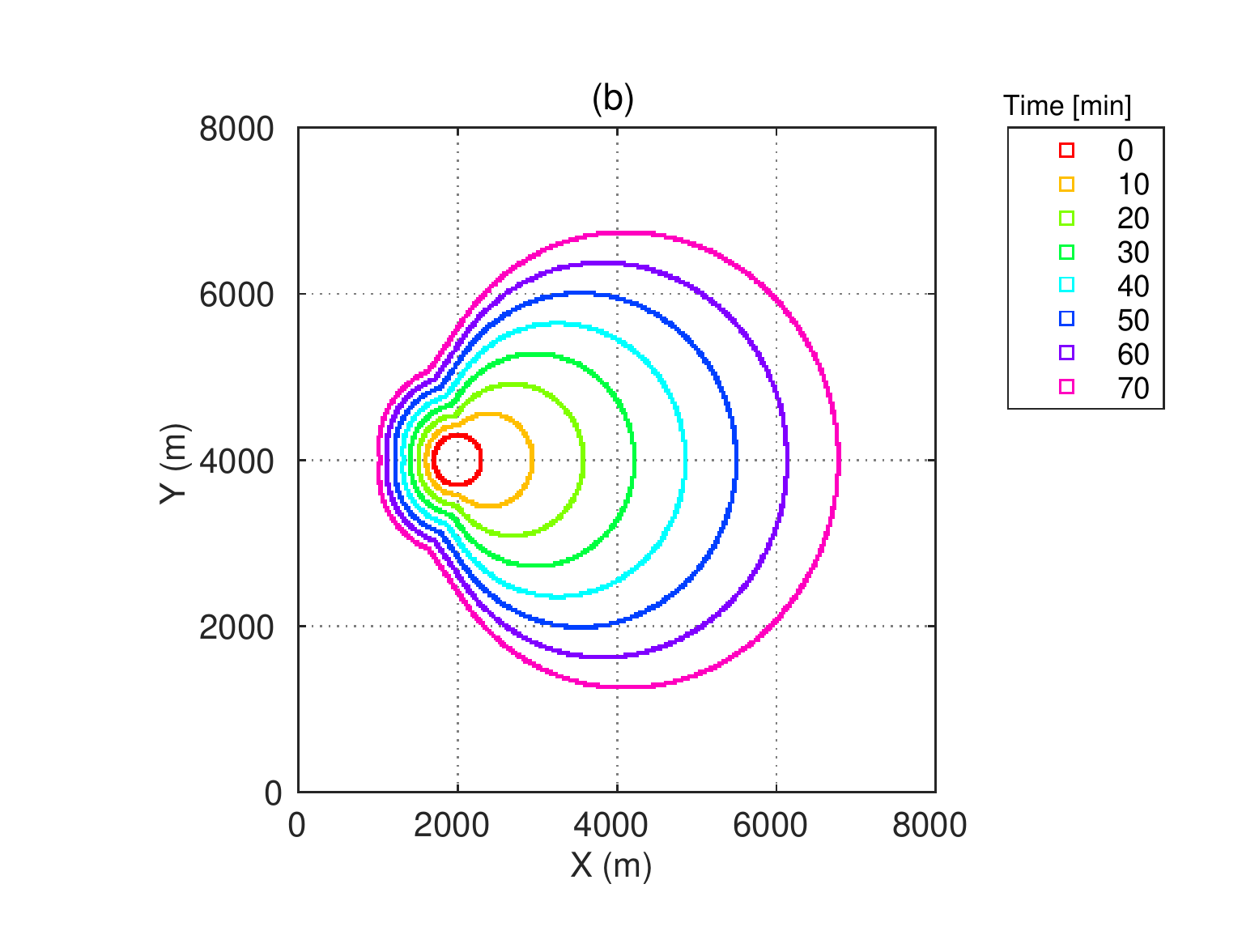}
    \end{subfigure}
      \caption{Same as Fig. \ref{fig:FF_LSM_Mallet_with_normal}, but when $\theta$ is the angle between the line joining a contour point and the mean wind direction.}
      \label{fig:FF_LSM_Mallet_no_normal}
\end{figure}

\smallskip
The ROS formulation developed by Mallet {\it et al.} \cite{mallet_etal-cma-2009},
here formula (\ref{Mallet}), provides a simple way to study the evolution of flank fire by introducing different ROS for 
the head, flank and rear directions. In the original paper, $\theta$ is defined as the angle between the normal to the front and the wind direction. 
Since, the normal computation in DEVS approach is approximate, two separate tests are performed to evaluate the effect of such
approximation on the spread: 
{\it firstly} when $\theta$ is computed according to the original definition \cite{mallet_etal-cma-2009}, 
and {\it secondly} when $\theta$ is assumed to be the angle between 
the line joining the front and the wind direction to enforce an identical definition of $\theta$ in both the methods. 
Fig.~\ref{fig:FF_LSM_Mallet_with_normal} shows the evolution of fire when $\theta$ is computed according to the original definition;
the advancement of the head and rear fires are identical, but the flank fire spread is spuriously larger in DEVS because of the approximate computation of the front normal direction. 
On the other hand, it is evident from Fig.~\ref{fig:FF_LSM_Mallet_no_normal} that in the second case, 
when the computation of $\theta$ is not affected by the approximate computation of the front normal in DEVS, the fire-line advancement is identical to LSM in all directions.

\begin{figure}[p]
   \begin{subfigure}{0.5\textwidth}
	\centering
    \includegraphics[trim = 2cm 8cm 4cm 8cm, clip=true, width = 8cm, height =6cm]{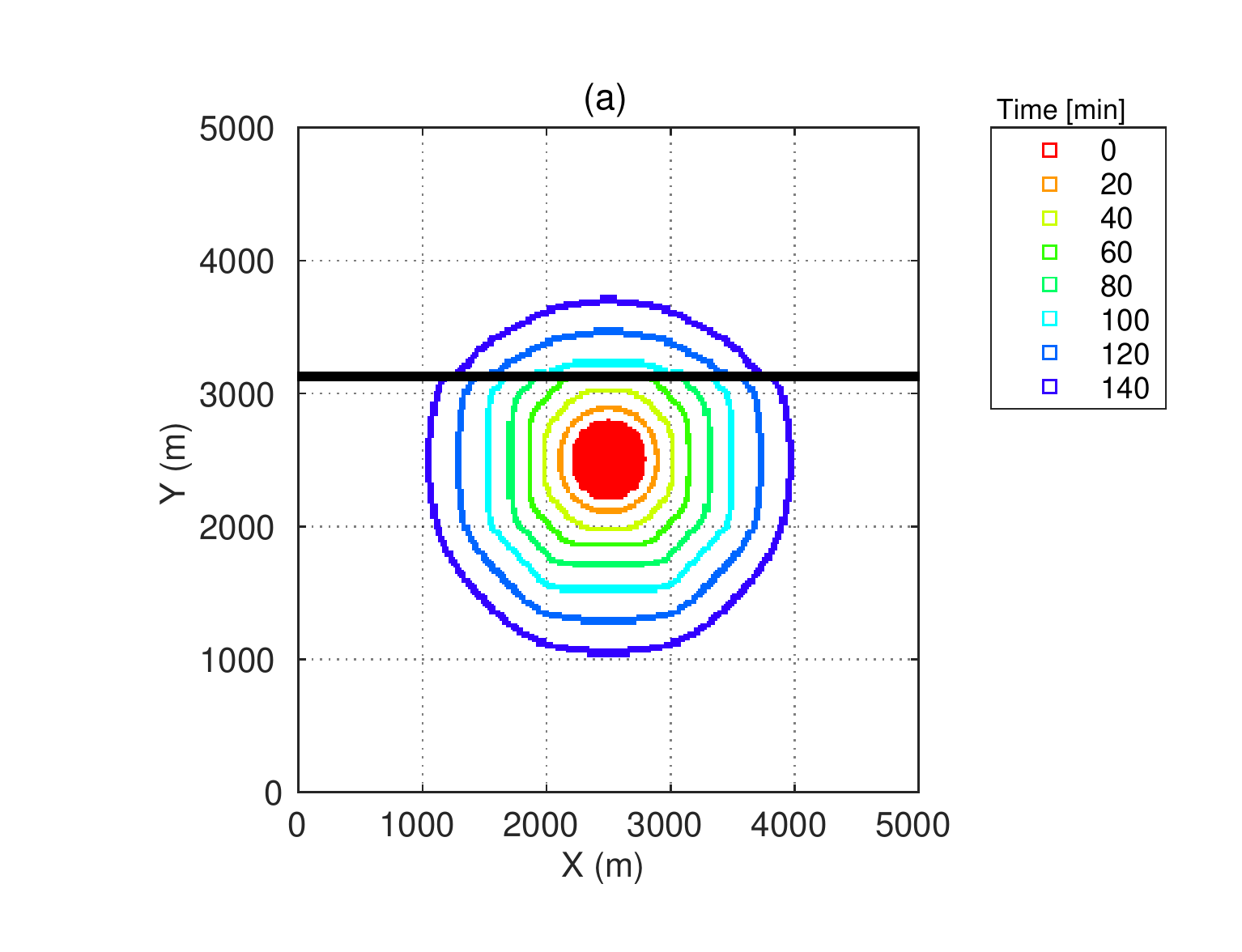}
       \end{subfigure}
    \begin{subfigure}{0.5\textwidth}
	\centering
    \includegraphics[trim = 2cm 8cm 4cm 8cm, clip=true, width = 8cm, height =6cm]{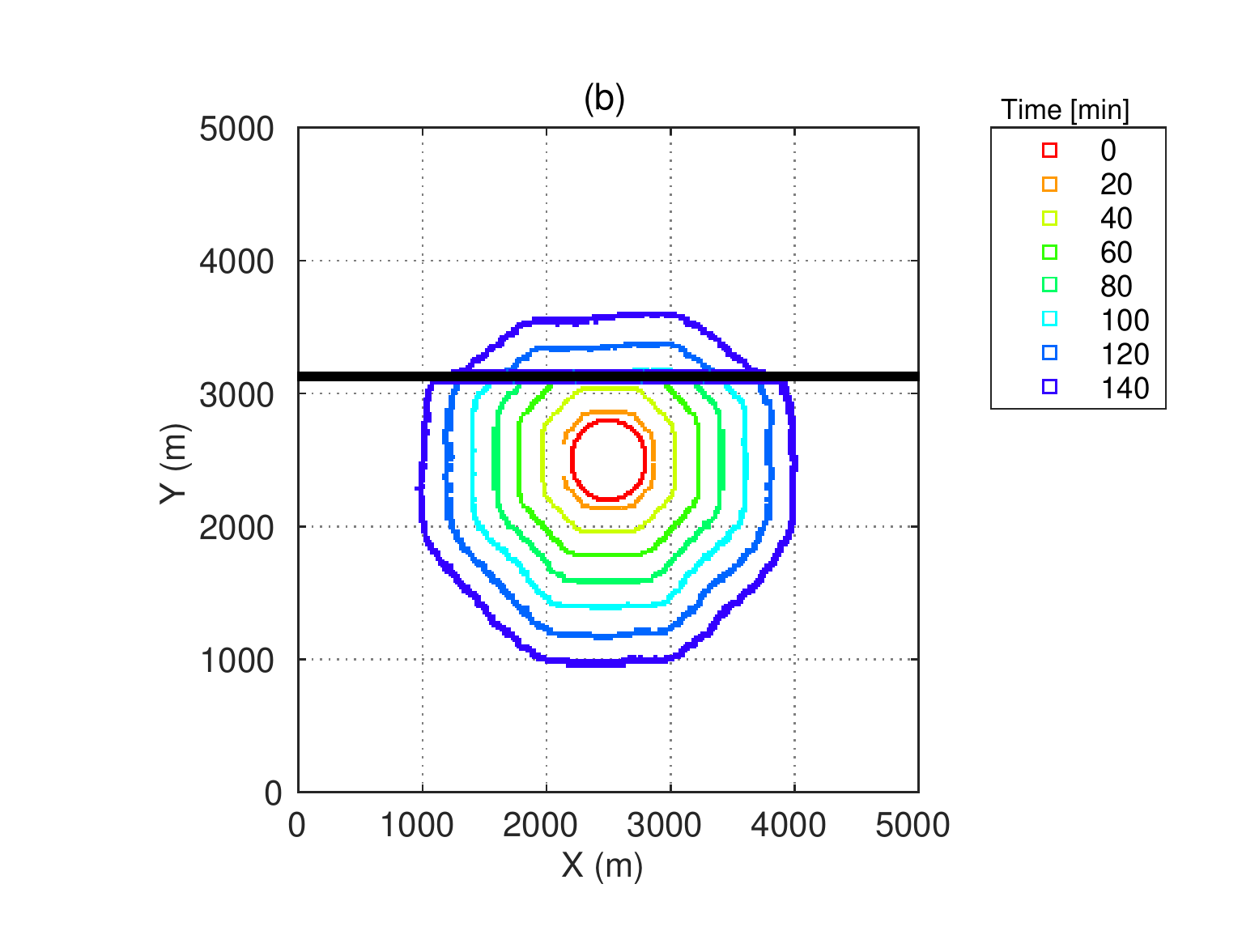}
    \end{subfigure}
         \caption{Evolution in time of the fire-line contour with turbulence in absence of wind for a) LSM and b) DEVS based simulators. 
The initial fire-line is a circle of radius $300\,\rm m$. The turbulent diffusion coefficient is $D=0.15 \,\rm m^{2} s^{-1}$.}
   \label{fig:FF_LSM_no_wind_turb_100}
\end{figure}

\begin{figure}[p]
	\centering
    \includegraphics[trim = 4cm 8cm 4cm 8cm, clip=true, scale = 0.5]{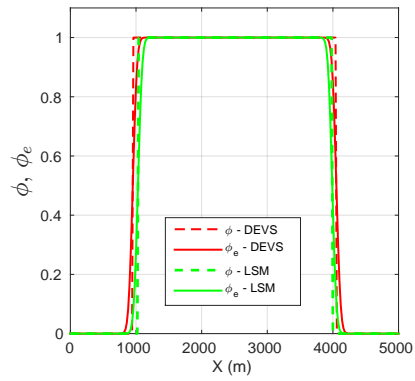}
    \caption{Cross section of the indicator function $\phi(\bx,t)$ and $\phi_{e}(\bx,t)$ at $y =2500\,\rm  m$ and $t = 140\,\rm min$, 
corresponding to the front evolution showed in Fig.~\ref{fig:FF_LSM_no_wind_turb_100}.}
    \label{fig:phi_phieff_turb_100}
\end{figure}

\smallskip 
As shown in Fig.~\ref{fig:FF_LSM_no_wind_firebreak}, in case of fire-break zones, pure LSM and DEVS based simulators are inherently 
unable to simulate the realistic situations of fire overcoming a fire break. 
But Fig.~\ref{fig:FF_LSM_no_wind_turb_100} shows that with the introduction of turbulence, the simulators can model the effect of hot air 
to overcome fire-break zones. Here the value of turbulent diffusion coefficient $D$ is assumed to be $0.15\,\rm m^{2}s^{-1}$. 
The evolution of the fire-line is almost similar for both the simulators, though a slight underestimation can be visibly observed in the DEVS approach.  
A cross section of $\phi(\bx,t)$ and $\phi_e(\bx,t)$ at $y=2500\,\rm m$ (Fig.~\ref{fig:phi_phieff_turb_100}) 
provides a more detailed illustration of the differences in the evolution of the fire-line between the two approaches. The introduction of marginal differences with the inclusion of turbulence can be explained by the contrast in the construction of the indicator function between the two simulators. In LSM, the indicator function is defined over a Cartesian grid by definition, while on the other hand, in DEVS, a gridded indicator function is constructed from the original real coordinates. The \textit{point in polygon} technique is used to transform the real coordinates to a Cartesian grid. This technique provides only an approximate measure of the same; small errors can be introduced by the improper classification of the points lying very close to the boundary \cite{hormann_etal-cgta-2001}, 
and can lead to a slight underestimation/overestimation of the original fire-line.

\begin{figure}[p]
   \begin{subfigure}{0.5\textwidth}
	\centering
    \includegraphics[trim = 2cm 8cm 4cm 8cm, clip=true, width = 8cm, height =6cm]{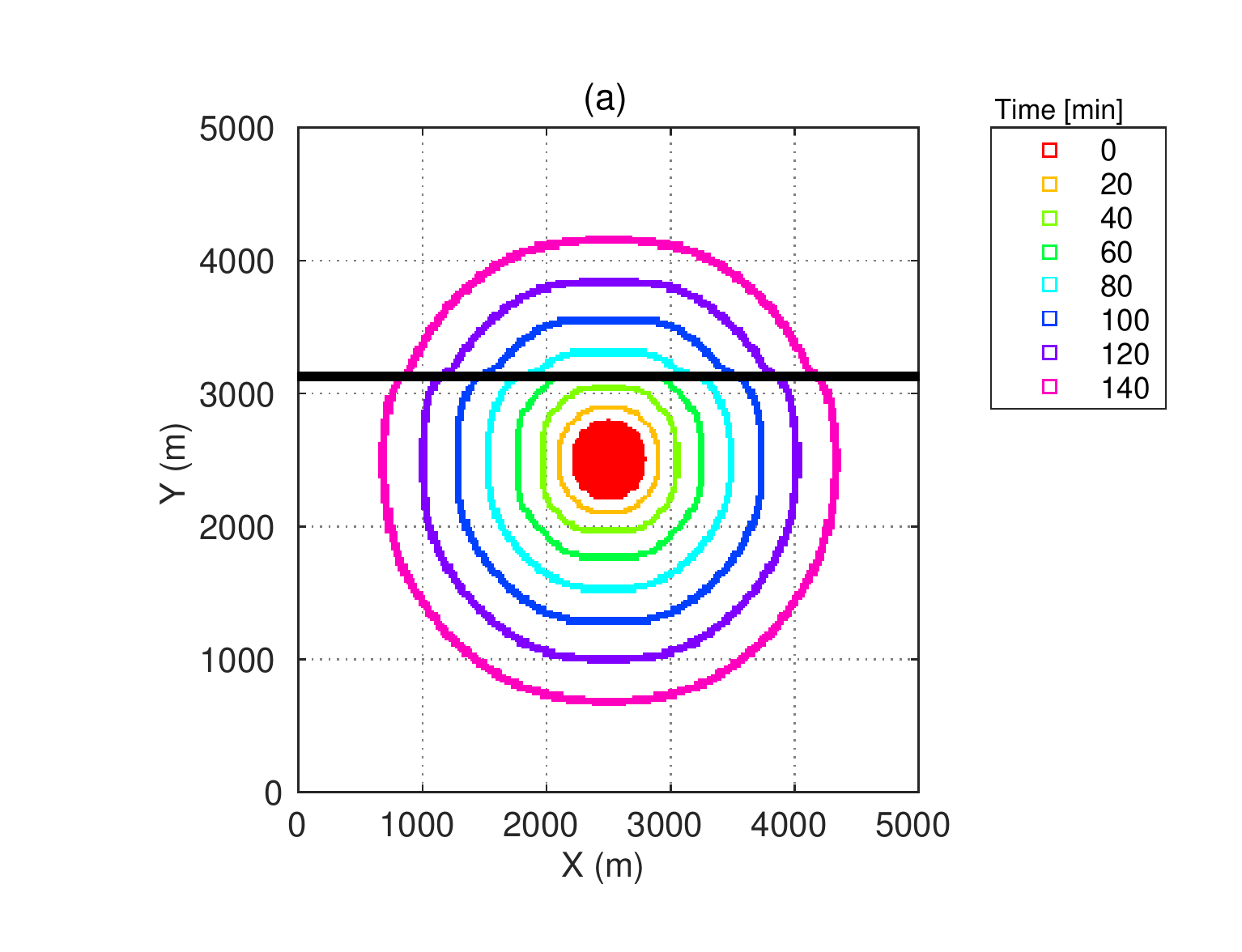}
       \end{subfigure}
    \begin{subfigure}{0.5\textwidth}
	\centering
    \includegraphics[trim = 2cm 8cm 4cm 8cm, clip=true, width = 8cm, height =6cm]{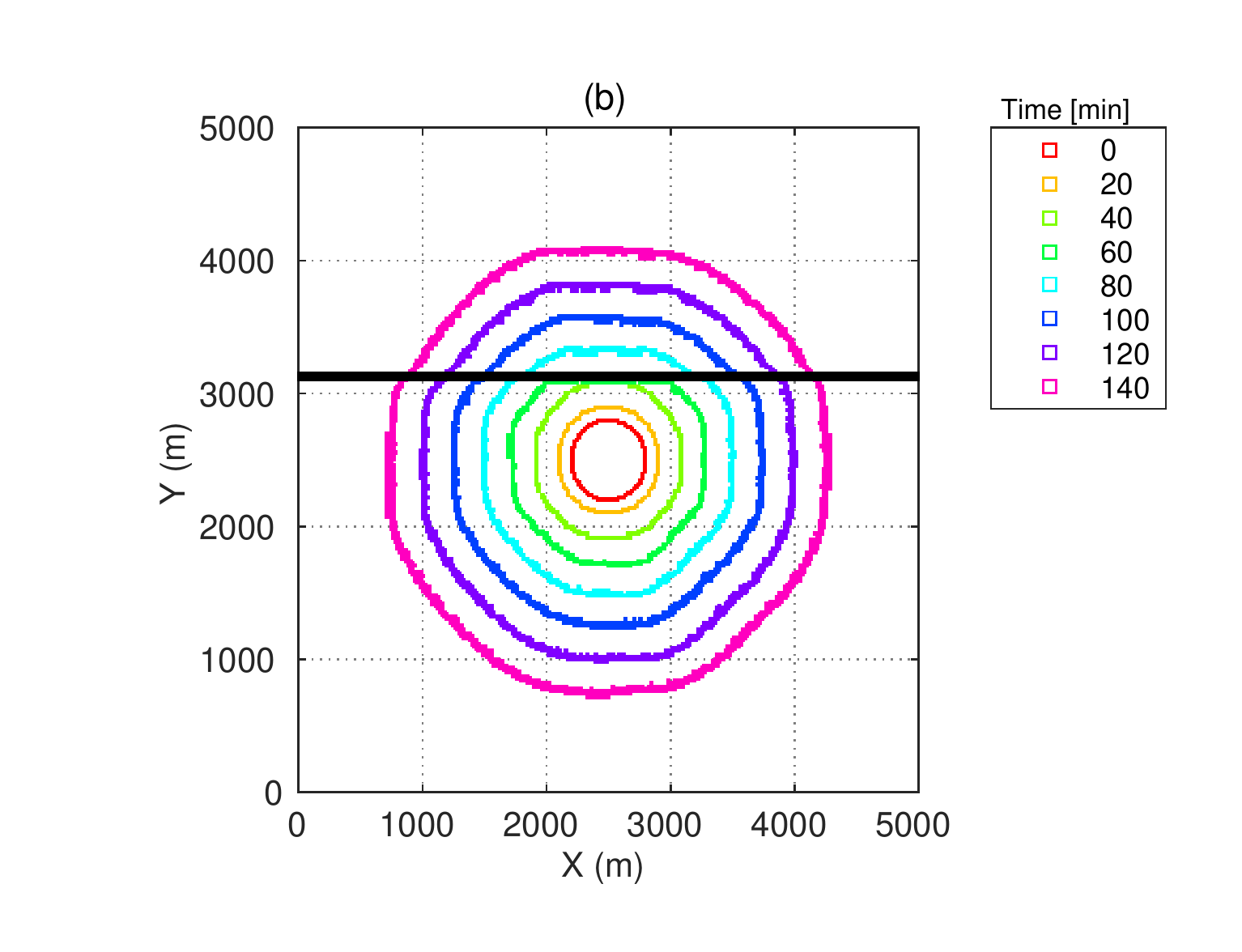}
    \end{subfigure}
     \caption{Same as Fig.~\ref{fig:FF_LSM_no_wind_turb_100}, but with turbulent diffusion coefficient $D=0.30 \,\rm m^{2} s^{-1}$.}
     \label{fig:FF_LSM_no_wind_turb_200}
\end{figure}

\smallskip
Fig.~\ref{fig:FF_LSM_no_wind_turb_200} shows the results for the two simulators when the turbulent diffusion coefficient is $D=0.30\,\rm m^{2} s^{-1}$. 
Stronger turbulence causes a rapid propagation of the fire-line and an earlier ignition across the fire-break zone. 
A detailed analysis of the effect of varying turbulence over {\it long-term propagation} with the LSM can be found in 
\cite{pagnini_etal-crs4-2012,pagnini_etal-mfbr-2012}. 

\begin{figure}[p]
   \begin{subfigure}{0.5\textwidth}
	\centering
    \includegraphics[trim = 2cm 8cm 4cm 8cm, clip=true, width = 8cm, height =6cm]{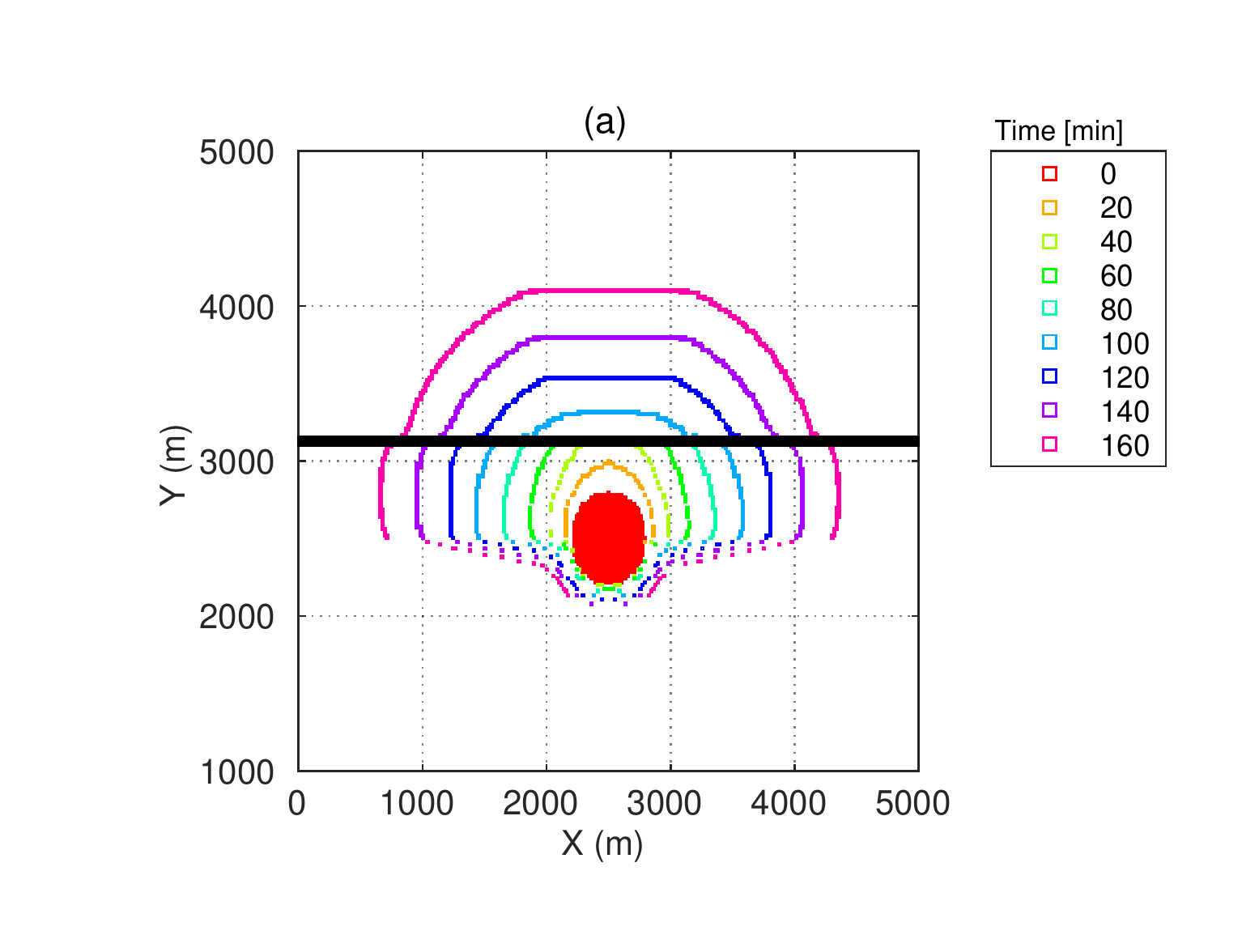}
       \end{subfigure}
    \begin{subfigure}{0.5\textwidth}
	\centering
    \includegraphics[trim = 2cm 8cm 4cm 8cm, clip=true, width = 8cm, height =6cm]{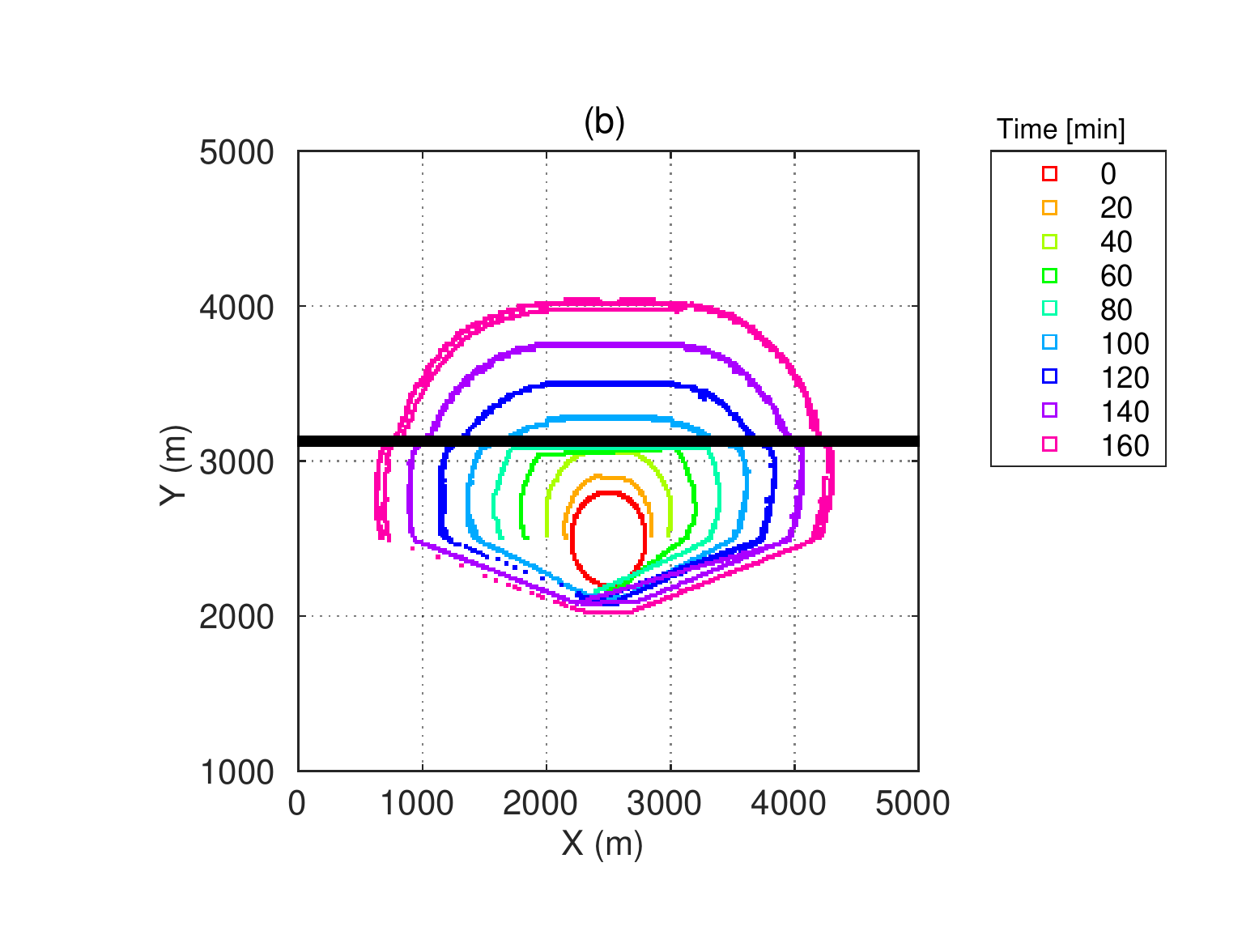}
    \end{subfigure}
     \caption{Evolution in time of the fire-line contour with turbulence and a north wind of $3\,\rm m s^{-1}$ for a) LSM and b) DEVS based simulators. 
The turbulent diffusion coefficient is $D=0.15\,\rm m^{2} s^{-1}$.}
     \label{fig:FF_LSM_with_wind_turb_100}
\end{figure}

\smallskip
Fig.~\ref{fig:FF_LSM_with_wind_turb_100} presents the effect of inclusion of turbulence with a non-zero wind profile  
and $3\%$ model (\ref{three_percent}). The effect of turbulence is most pronounced in the direction of the wind and it can be clearly observed that with turbulence, the spread over the head, flank and back fire-lines is faster and the head fire is also able to overcome the fuel break. The quantum of increase in the fire spread can also be appreciated in comparison to Fig.~\ref{fig:FF_LSM_wind_circle}, 
which presents the ideal case with the same initial conditions. Both simulators show almost similar characteristics in the spread of fire, though the flank fire has a slightly larger spread in DEVS.

\begin{figure}[p]
   \begin{subfigure}{0.5\textwidth}
	\centering
    \includegraphics[trim = 2cm 8cm 4cm 8cm, clip=true, width = 8cm, height =6cm]{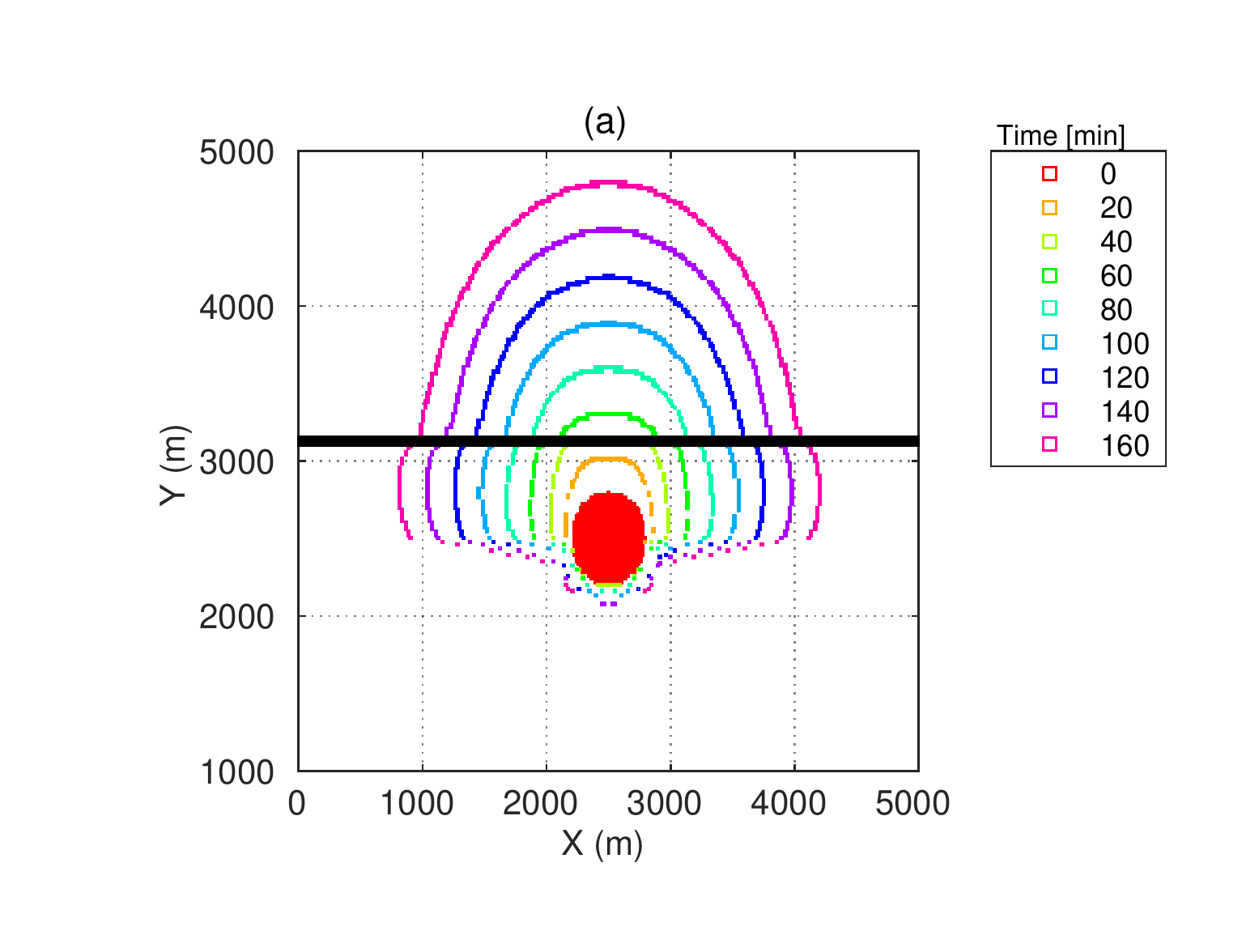}
       \end{subfigure}
    \begin{subfigure}{0.5\textwidth}
	\centering
    \includegraphics[trim = 2cm 8cm 4cm 8cm, clip=true, width = 8cm, height =6cm]{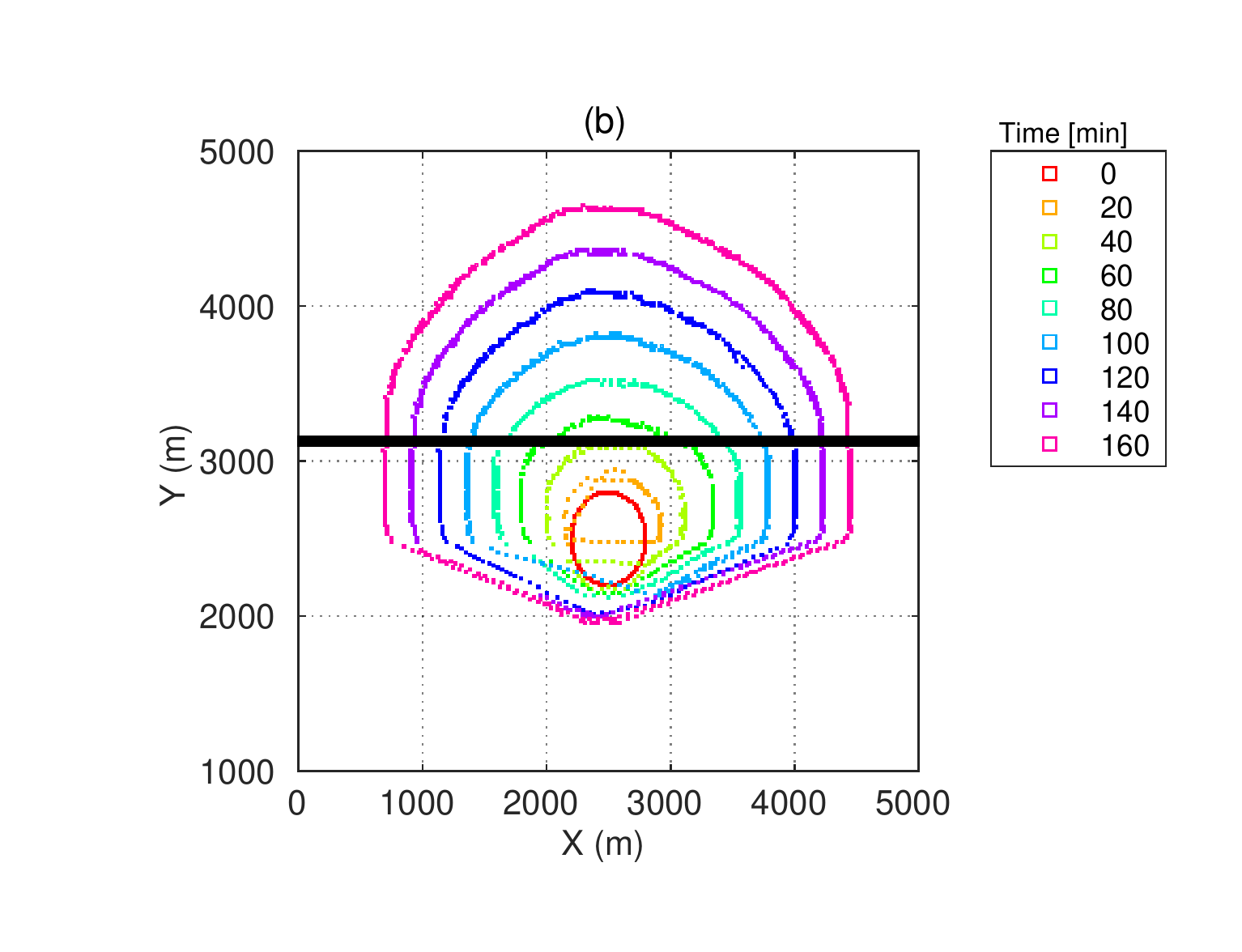}
    \end{subfigure}
     \caption{Same as Fig.~\ref{fig:FF_LSM_no_wind_turb_100}, but when both turbulence and fire-spotting are included
with $D=0.15\,\rm m^{2} s^{-1}$, $\mu=2.69$ and $\sigma=1.25$.}
     \label{fig:FF_LSM_with_wind_firespot_100}
\end{figure}

\smallskip
Another aspect contributing towards the increase in the fire spread due to new fire ignitions generated by the firebrands 
is presented in Fig.~\ref{fig:FF_LSM_with_wind_firespot_100}. With the inclusion of fire-spotting along with turbulence, the evolution of the fire-front is faster in comparison to the effect of turbulence alone (Fig.~\ref{fig:FF_LSM_with_wind_turb_100}). 
The flank fire and the head fire are also well simulated in both the approaches, and again a larger spread out the flanking fires is observed in the DEVS scheme. 
The proposed mathematical formulation of spot-fires along with turbulence is effective in mimicking the rapid advancement of the fire-line due to the generation of new ignition points by the firebrands. It is also effective in simulating the new ignition points developed across the no fuel zones. Within the current parametrisation, the firebrand landing distances lie close to the main fire and are not long enough to develop a new secondary fire, but in the next section, a sensitivity study of the spatial pattern of the firebrand landing distances is also presented. 
 
\begin{figure}[p]
   \begin{subfigure}{0.5\textwidth}
	\centering
    \includegraphics[trim = 2cm 8cm 4cm 8cm, clip=true, width = 8cm, height =6cm]{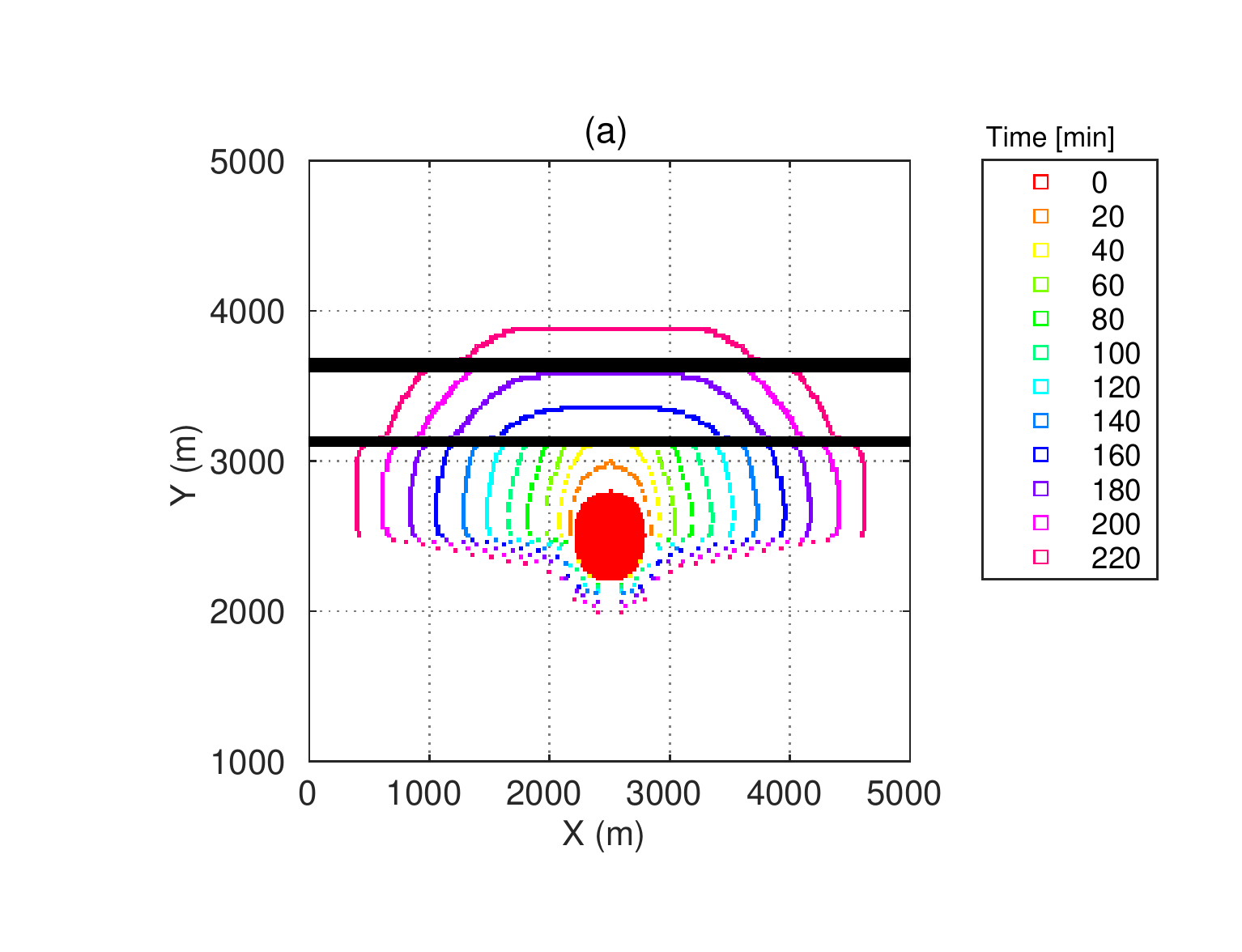}
       \end{subfigure}
    \begin{subfigure}{0.5\textwidth}
	\centering
    \includegraphics[trim = 2cm 8cm 4cm 8cm, clip=true, width = 8cm, height =6cm]{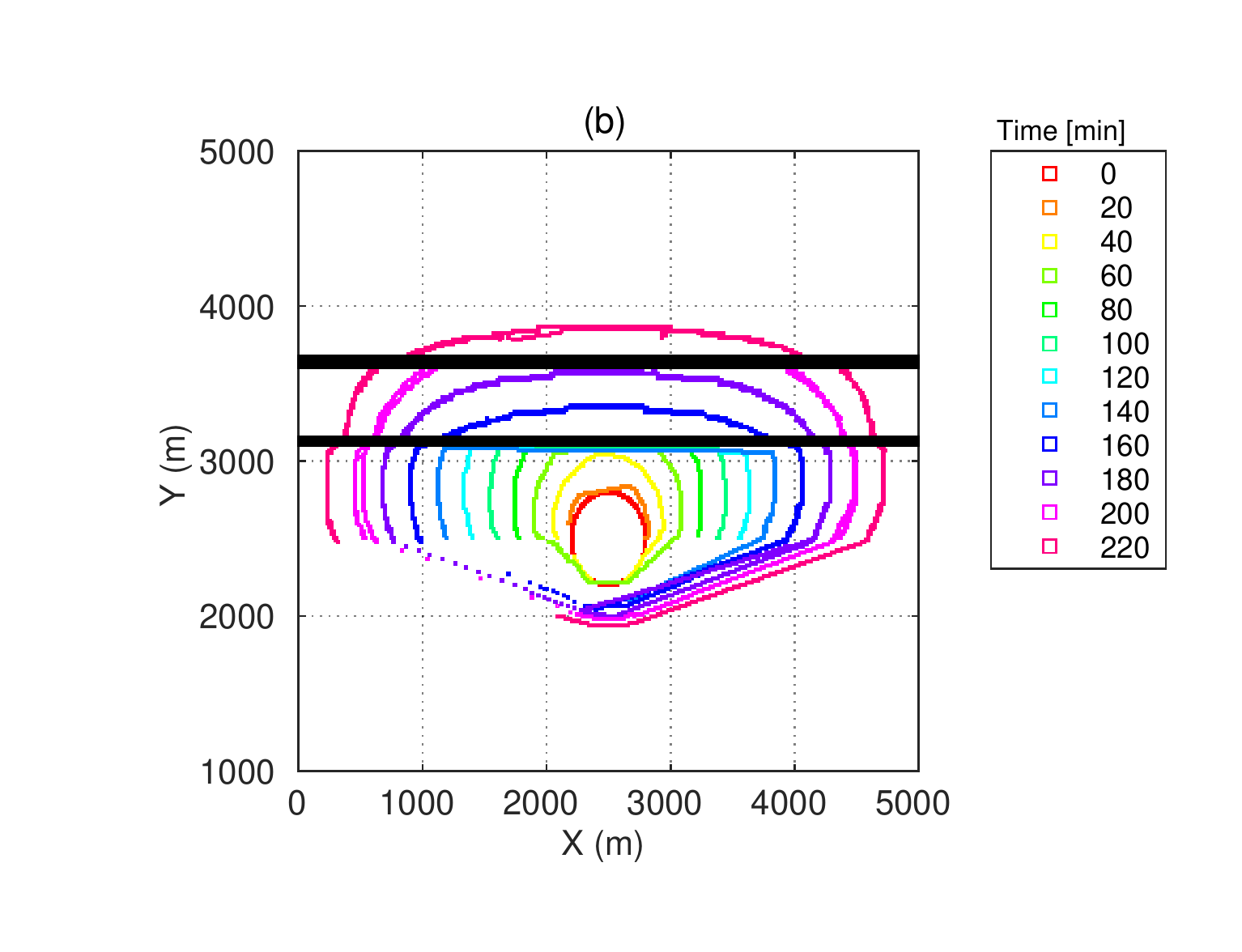}
    \end{subfigure}
     \caption{Evolution in time of the fire-line contour with turbulence, a north wind of $3 \,\rm m s^{-1}$ and two fire-break zones 
for a) LSM and b) DEVS based simulators. The initial fire-line is a circle of radius $300\,\rm m$. The turbulent diffusion coefficient is $D=0.075\,\rm m^{2} s^{-1}$.}
     \label{fig:FF_LSM_with_wind_turb_two_break}
\end{figure}

\begin{figure}[p]
   \begin{subfigure}{0.5\textwidth}
	\centering
    \includegraphics[trim = 2cm 8cm 4cm 8cm, clip=true, width = 8cm, height =6cm]{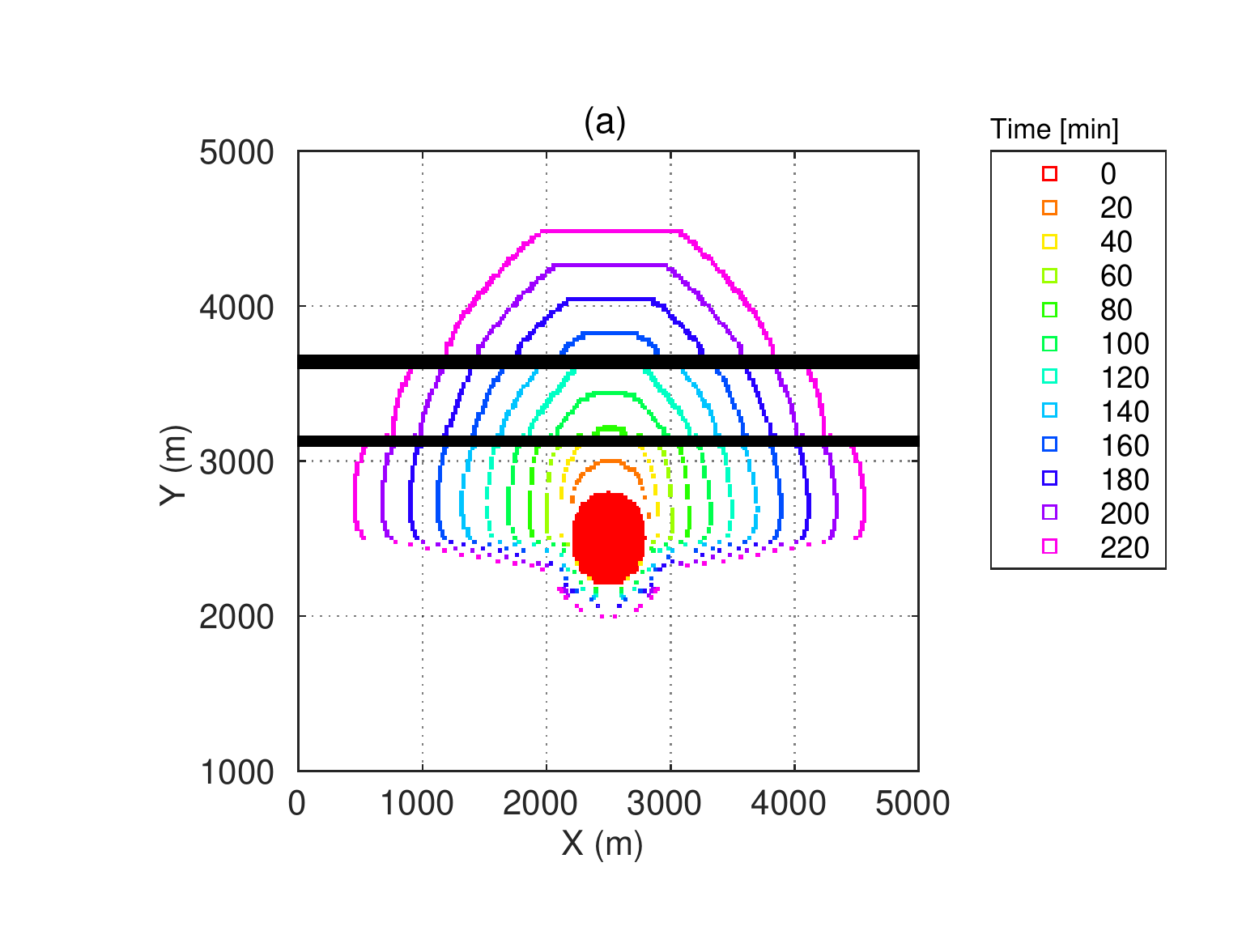}
       \end{subfigure}
    \begin{subfigure}{0.5\textwidth}
	\centering
    \includegraphics[trim = 2cm 8cm 4cm 8cm, clip=true, width = 8cm, height =6cm]{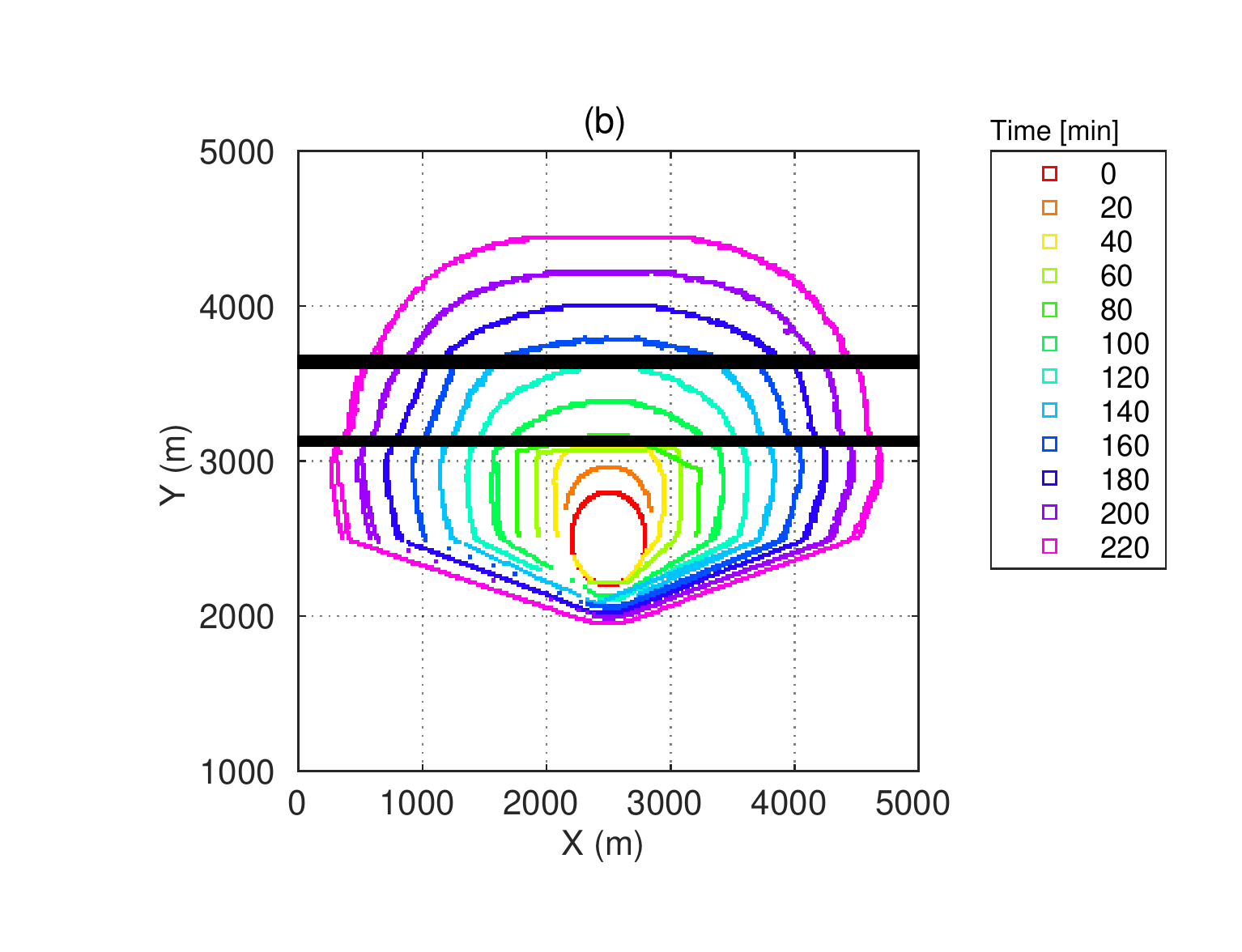}
    \end{subfigure}
     \caption{Same as Fig.~\ref{fig:FF_LSM_with_wind_turb_two_break}, but 
when both turbulence and fire-spotting are included with $D=0.075\,\rm m^{2} s^{-1}$, $\mu=2.69$ and $\sigma=1.25$.}
     \label{fig:FF_LSM_with_wind_firespot_two_break}
\end{figure}

\smallskip
Fig.~\ref{fig:FF_LSM_with_wind_turb_two_break} shows the time evolution of the fire-front in presence of a mean wind velocity of $3 \,\rm m s^{-1}$ 
and two fire-break zones. The results are obtained when only the effects of turbulence are included in the formulation and the diffusion coefficient $D = 0.075 \rm\, m^2s^{-1}$. 
Similarly, Fig.~\ref{fig:FF_LSM_with_wind_firespot_two_break} shows the time evolution of fire-front but when both turbulence and fire-spotting are included in the formulation. Both approaches follow an identical evolution pattern as observed in other cases. 
As expected, the spread of the fire-front is faster and is able to overcome both the fire-break zones when turbulence is included; 
but inclusion of fire-spotting along with turbulence causes the fire to propagate at a much faster rate and results in a larger spread. Again, the major differences which arise in the evolution of flank fires between the two methods are due to the dissimilarity in the construction of the direction of the propagation of the active burning points.

\section{Insight on firebrand distribution}
\label{firebrand}
In the simulations discussed previously, 
the phenomenon of fire-spotting contributes towards an increase in the burning area and a faster propagation of fire,
but the landing distances are not sufficiently long enough to develop new secondary fires detached from the primary fire. 

\smallskip
Generally, the firebrands generated in low intensity fires fall close to the primary fire and although they contribute 
towards a faster fire spread, the separation from the primary fire is not large enough for the new ignitions to develop a new secondary fire. 
But high intensity fires and extreme meteorological conditions such as strong wind/extremely high temperatures/low humidity can facilitate 
the transport of embers over longer distances and the area under firebrand attack can be sufficiently far away from the primary fire 
\cite{wilson-1962,permin-twmh-1971, pagni-fsj-1993, trelles_etal-1997, quintiere-1998}. 
In Ref. \cite{koo_etal-ijwf-2010} and references therein, a brief account of the impact of various historical forest fires is provided emphasising the importance of strong winds, and extreme meteorological conditions 
on the fire-spotting phenomenon. They describe different historical fires where low intensity fires coupled with strong winds and low relative humidity caused a large scale fire-spotting, while on the other hand, fires with high intensities in calm meteorological conditions had a negligible contribution from the firebrands.

\smallskip
In lognormal distribution (\ref{lognormal}), 
a suitable choice of the parameters $\mu$ and $\sigma$ can be utilized to modify the firebrand landing distributions to represent different fire intensities and wind conditions. In this study, the main interest lies in the mathematical modelling of the firebrands landing distributions, hence constant meteorological conditions are assumed for all the simulations presented here. 
It is also assumed that each active ignition point is capable of generating firebrands and all embers are controlled by identical transport forces.  

\begin{figure}[p]
   \begin{subfigure}{0.5\textwidth}
	\centering
    \includegraphics[ scale = 0.40]{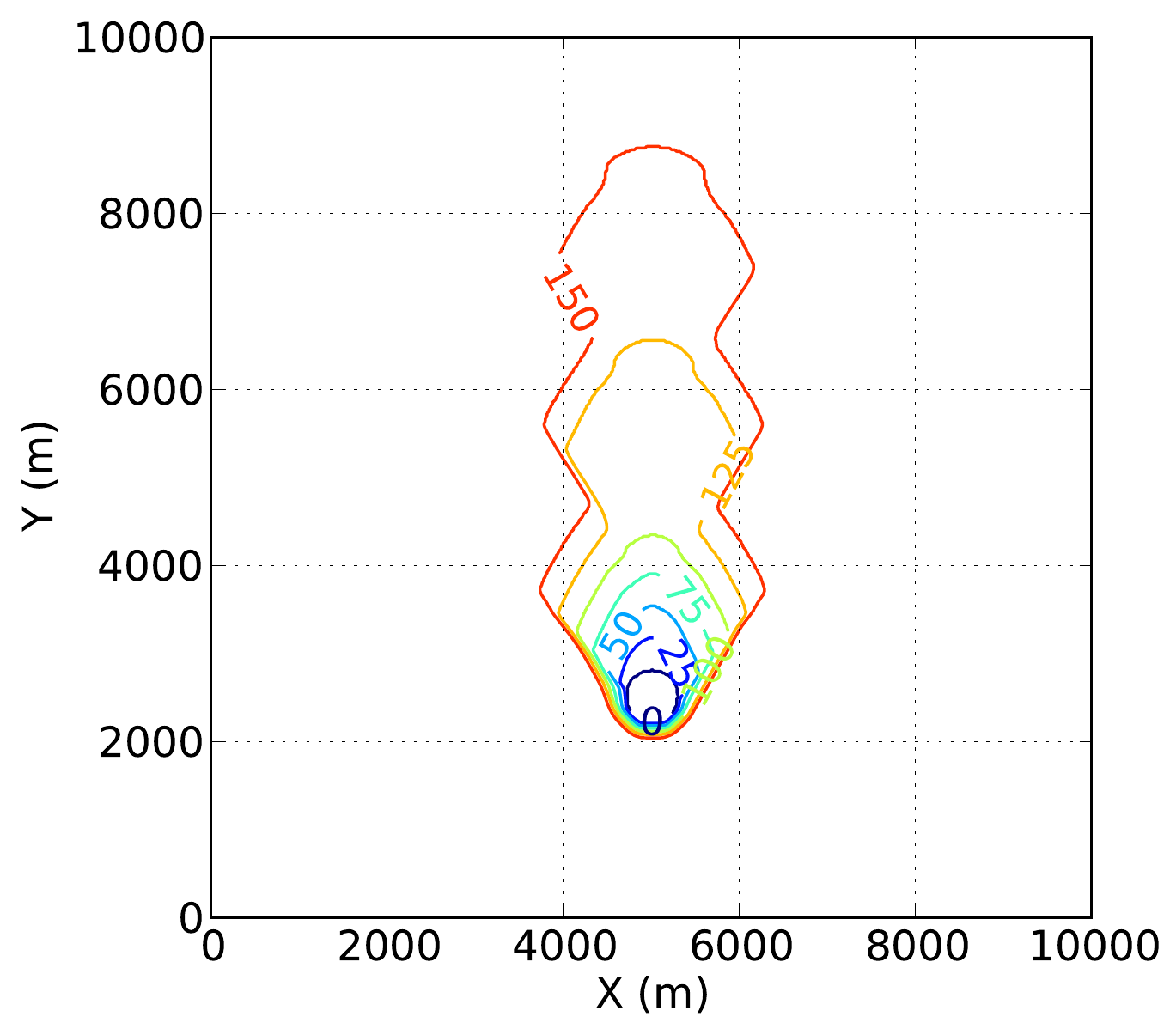}
    \caption{}
       \end{subfigure}
    \begin{subfigure}{0.5\textwidth}
	\centering
    \includegraphics[ scale = 0.40]{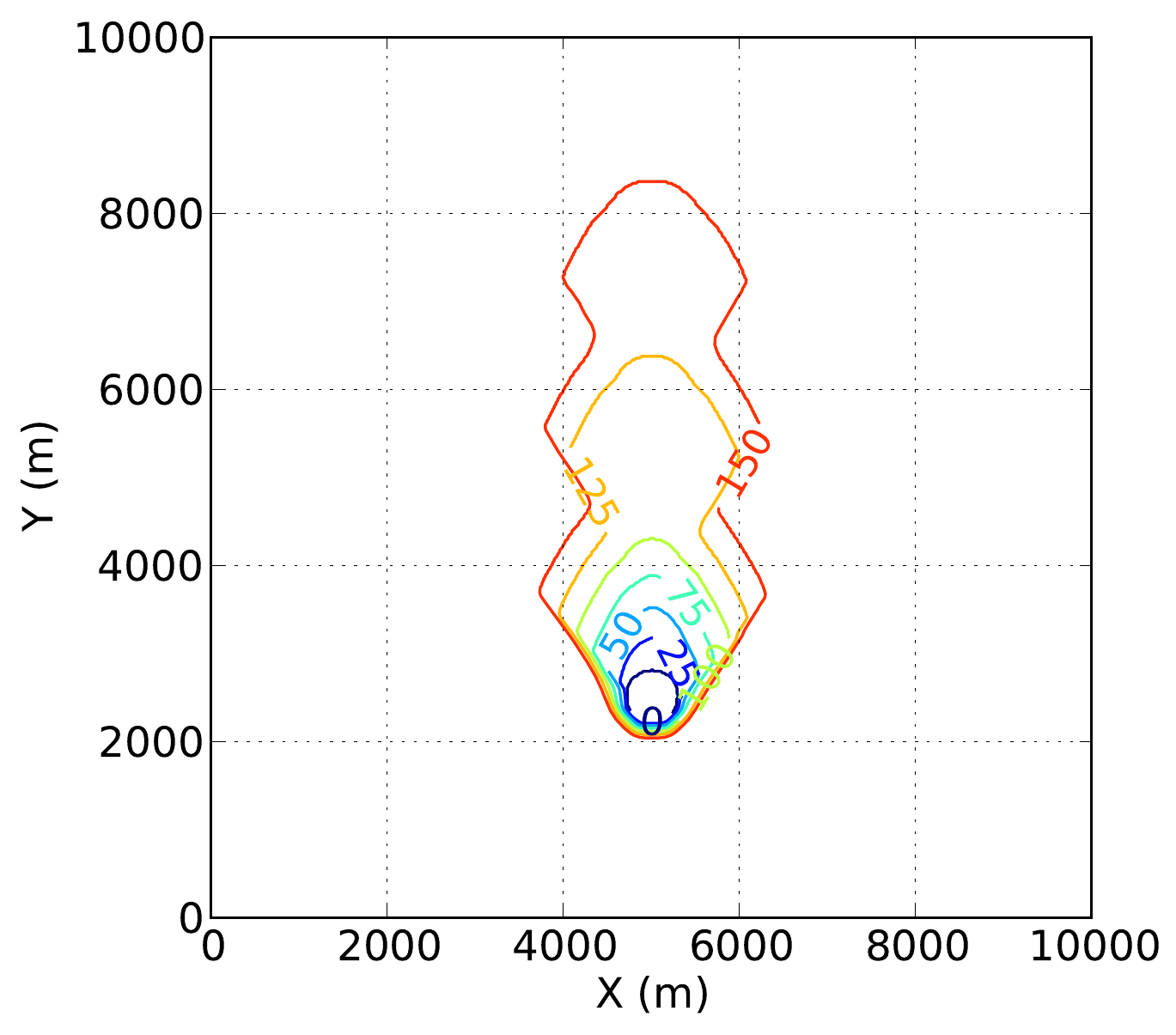}
    \caption{}
    \end{subfigure}\\
       \begin{subfigure}{0.5\textwidth}
	\centering
    \includegraphics[scale = 0.40]{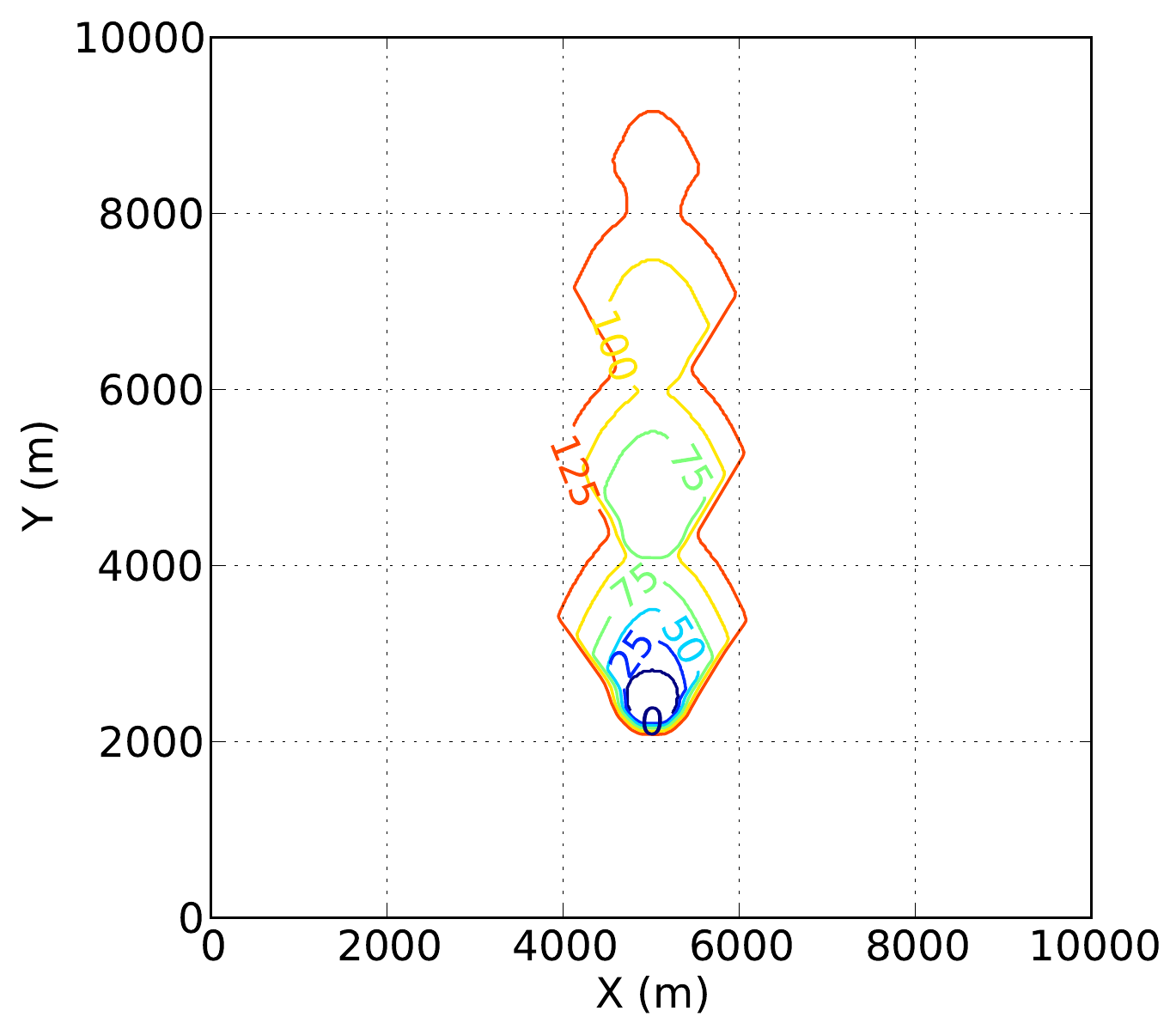}
    \caption{}
       \end{subfigure}
    \begin{subfigure}{0.5\textwidth}
	\centering
    \includegraphics[ scale = 0.40]{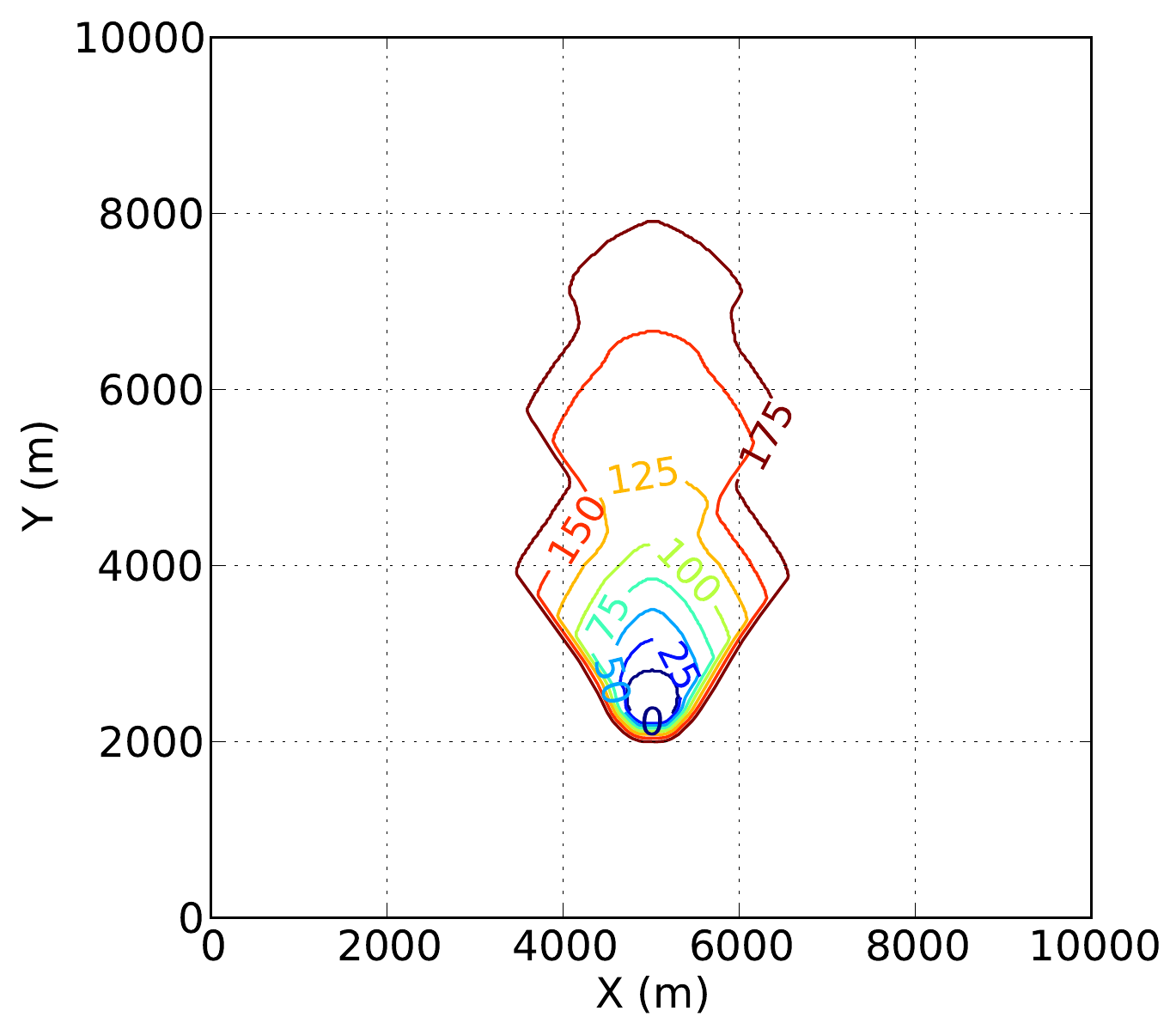}
    \caption{}
    \end{subfigure}\\
       \begin{subfigure}{0.5\textwidth}
	\centering
    \includegraphics[scale = 0.40]{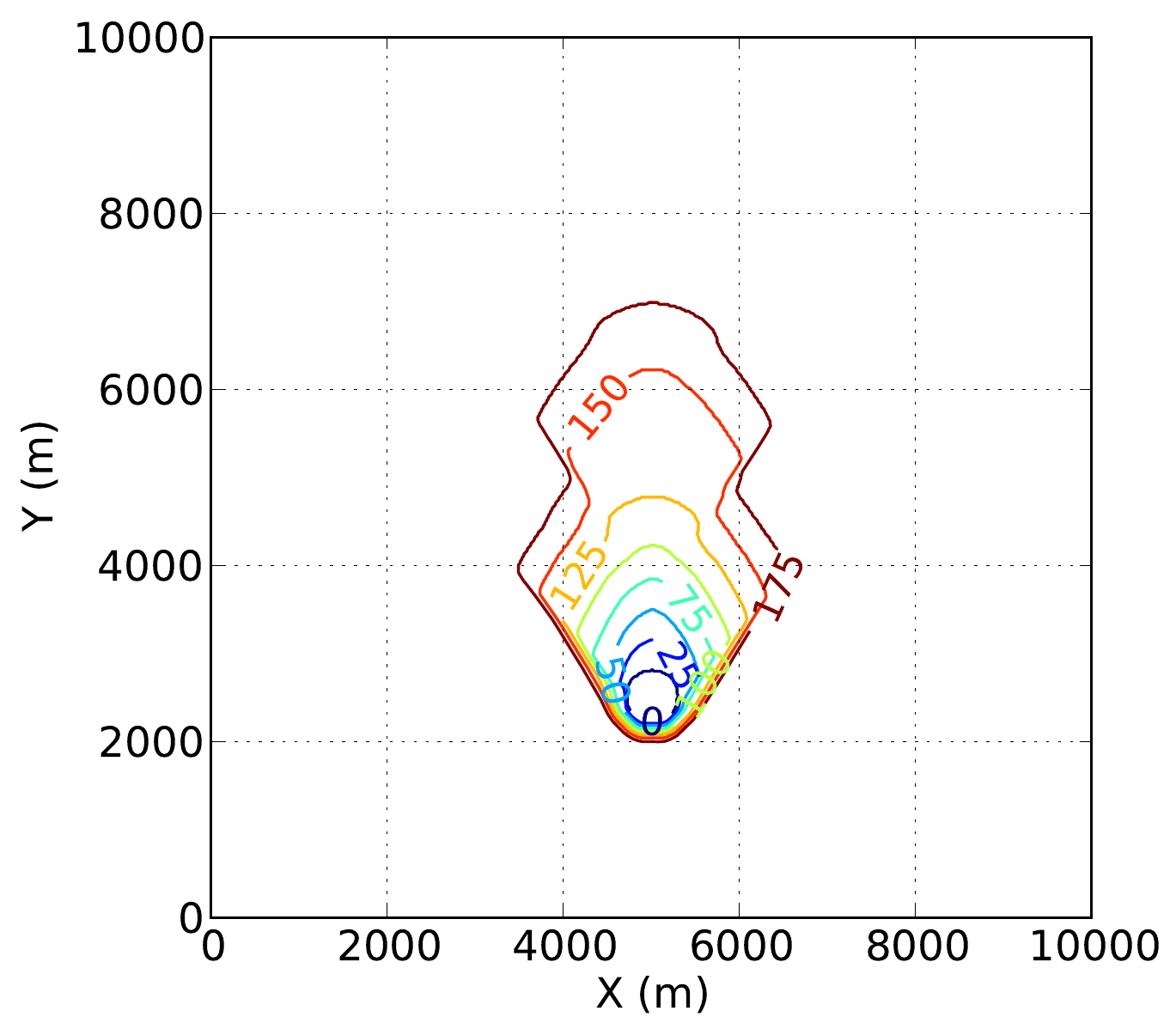}
    \caption{}
       \end{subfigure}
    \begin{subfigure}{0.5\textwidth}
	\centering
    \includegraphics[scale = 0.40]{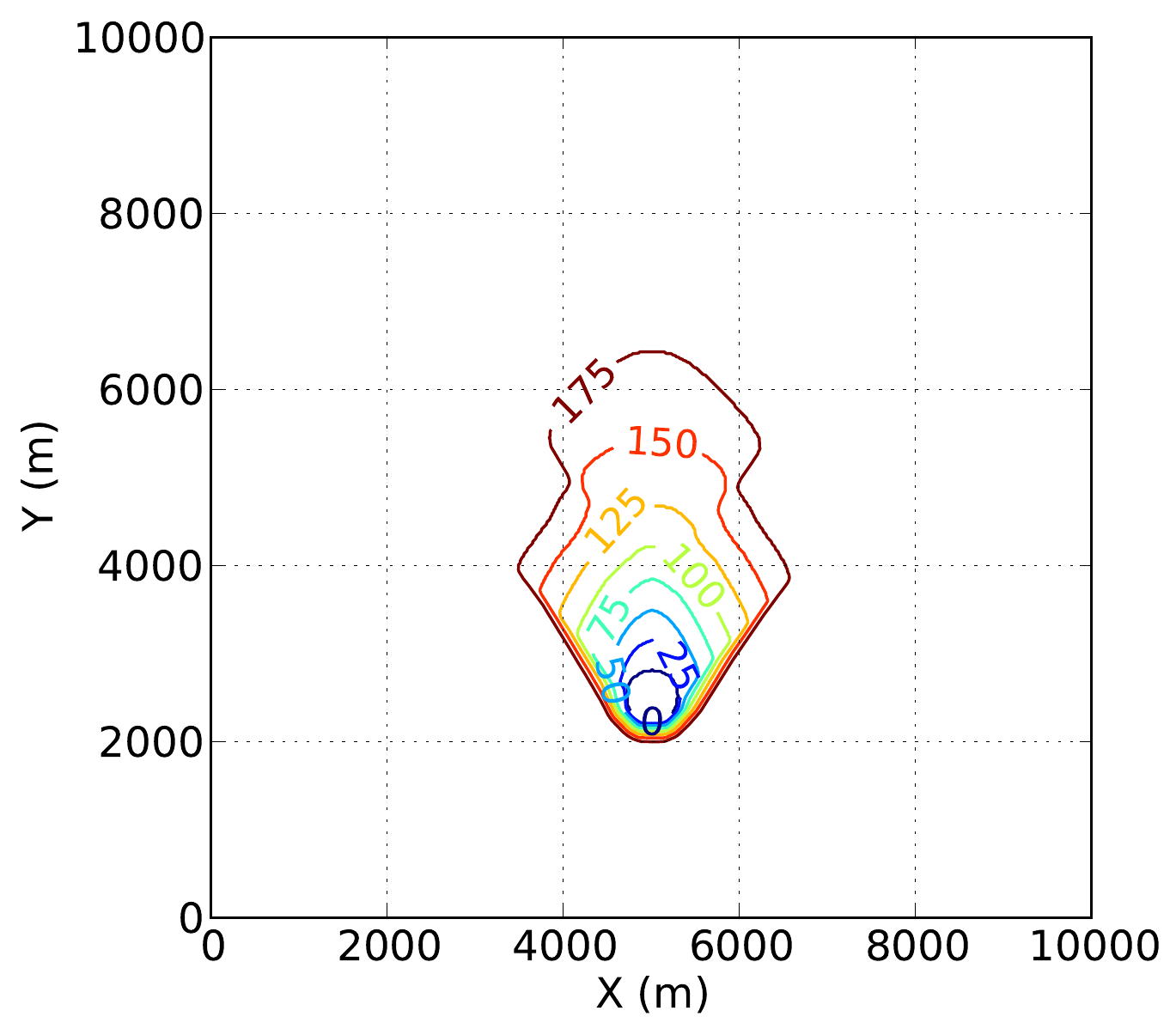}
    \caption{}
    \end{subfigure}
     \caption{Series of figures showing the variation in the fire propagation due to increasing value of $\mu$. From a-f, $\mu$ varies from $13$ to $18$, 
in increments of $1$, while $\sigma = 5$.}
     \label{fig:Firebrand_increasing_mu}
\end{figure}

\begin{figure}[p]
   \begin{subfigure}{0.5\textwidth}
	\centering
    \includegraphics[ scale = 0.40]{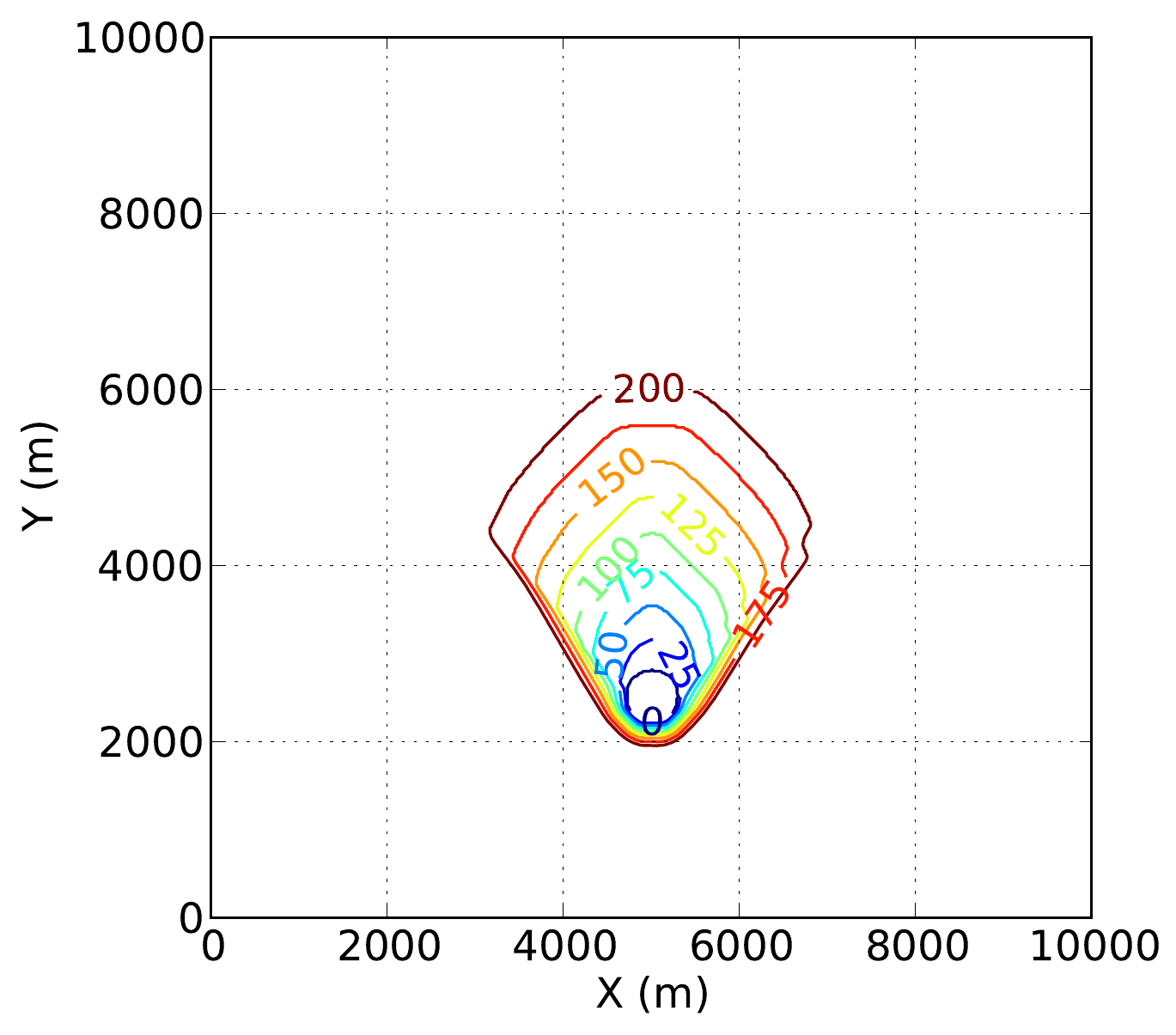}
    \caption{}
       \end{subfigure}
    \begin{subfigure}{0.5\textwidth}
	\centering
    \includegraphics[ scale = 0.40]{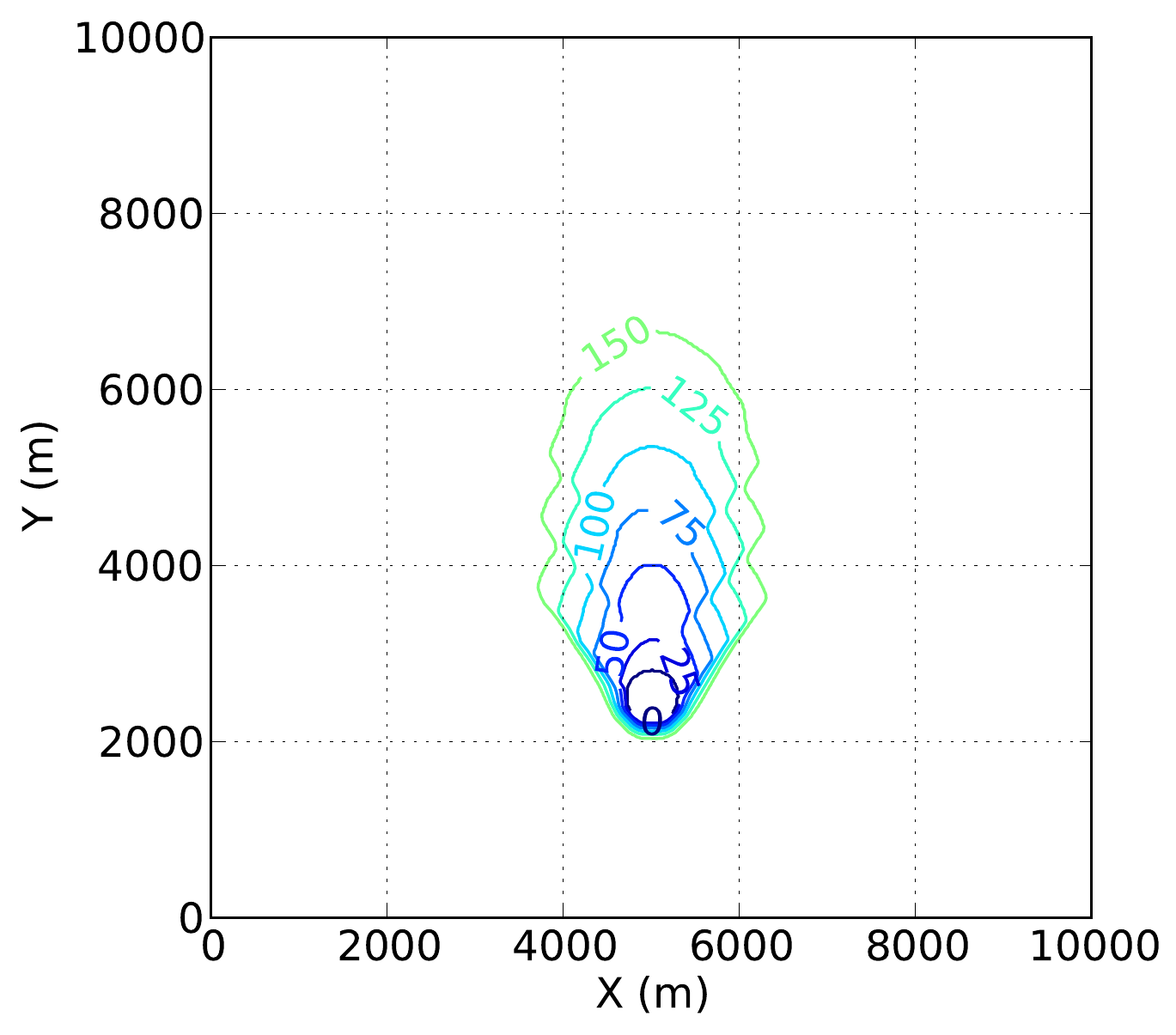}
    \caption{}
    \end{subfigure}\\
       \begin{subfigure}{0.5\textwidth}
	\centering
    \includegraphics[scale = 0.40]{LSM_15_5.pdf}
    \caption{}
       \end{subfigure}
    \begin{subfigure}{0.5\textwidth}
	\centering
    \includegraphics[ scale = 0.40]{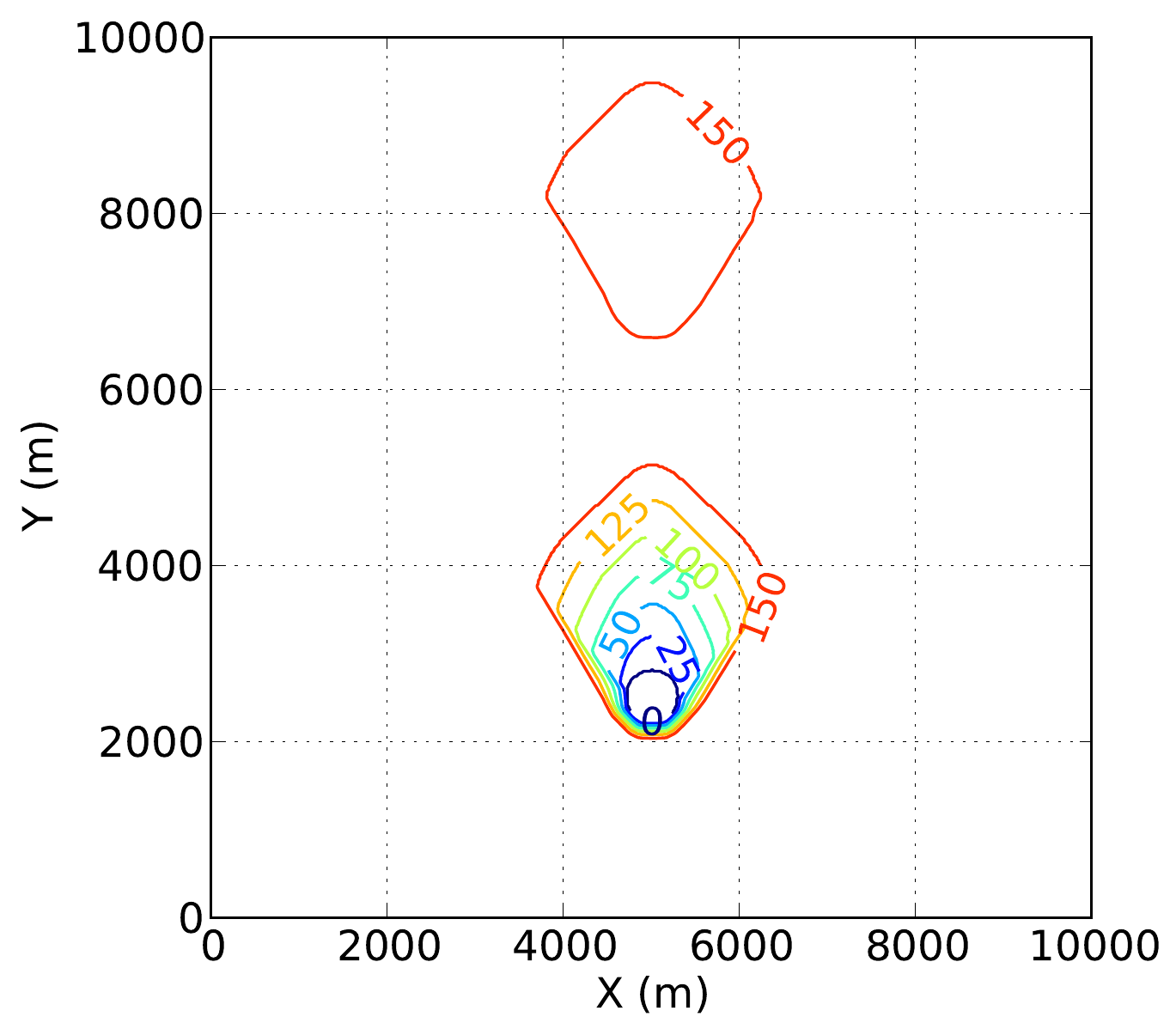}
    \caption{}
    \end{subfigure}\\
       \begin{subfigure}{0.5\textwidth}
	\centering
    \includegraphics[width = 5.3cm,height = 5cm]{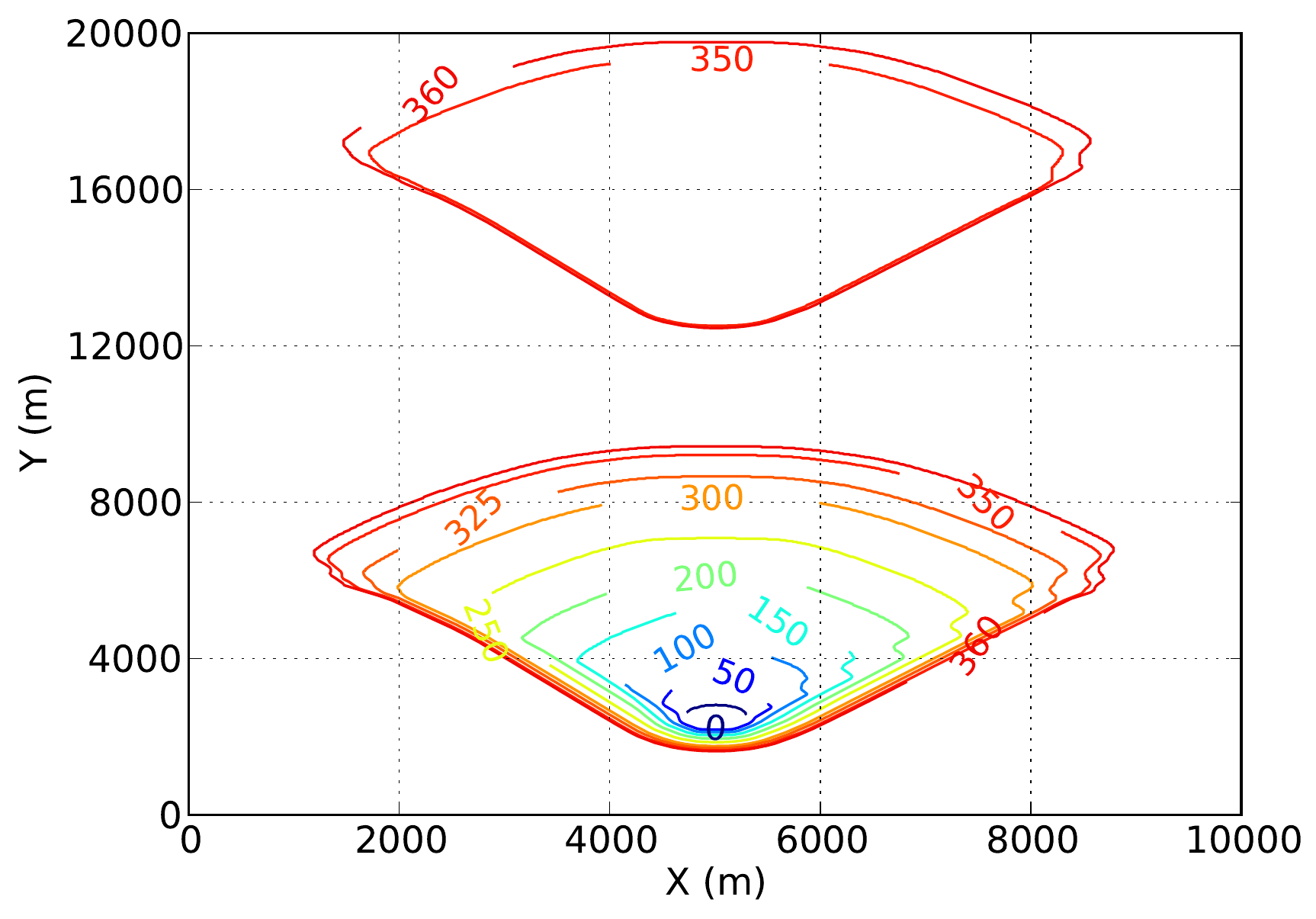}
    \caption{}
       \end{subfigure}
    \begin{subfigure}{0.5\textwidth}
	\centering
    \includegraphics[width = 5.3cm,height = 5cm]{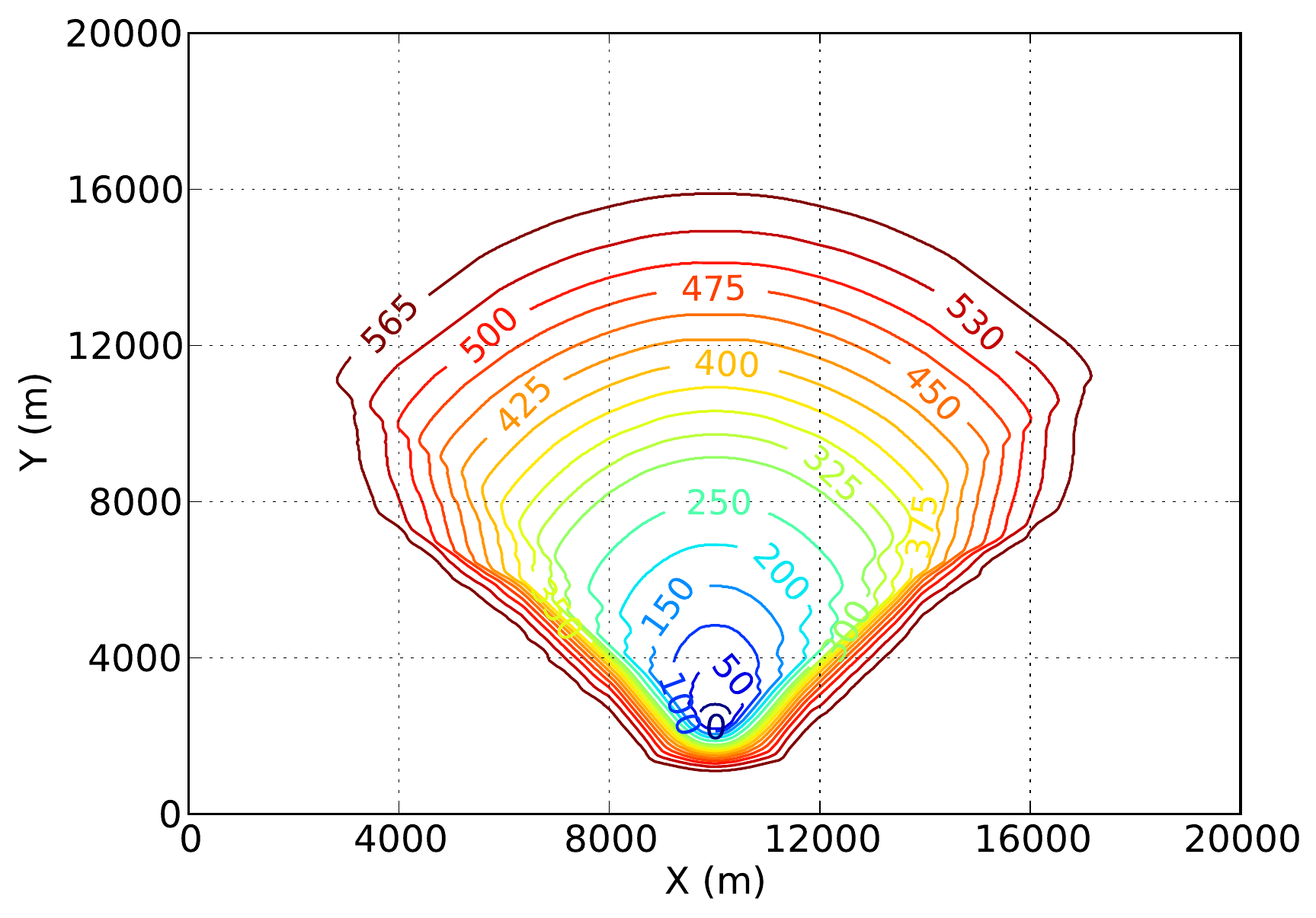}
    \caption{}
    \end{subfigure}
     \caption{Series of figures showing the variation in the fire propagation due to increasing value of $\sigma$. 
From a-f, $\sigma$ varies from $4$ to $6.5$, in increments of $0.5$, while $\mu = 15$.}
     \label{fig:Firebrand_increasing_sigma}
\end{figure}

\smallskip
Fig.~\ref{fig:Firebrand_increasing_mu} and Fig.~\ref{fig:Firebrand_increasing_sigma} show the variation of the area under the firebrand attack 
with different values of $\mu$ and $\sigma$. A constant $\sigma$ and an increasing $\mu$ points towards a decrease in the effectiveness of the firebrands, 
and slows down the fire propagation. A slight deviation from the pattern is observed for $\mu = 15$ and $\sigma = 5$, 
when the firebrands travel far enough to create a new secondary fire. 

\smallskip
For an increasing value of $\sigma$ (Fig.~\ref{fig:Firebrand_increasing_sigma}a-c) with constant $\mu$, 
the fire evolution increases as larger number of firebrands land away from the source, but the landing distances are still not long enough to delineate the new ignitions from the main fire. With further increase in $\sigma$, a secondary fire emerges away from the main source. 
Equal contribution to the firebrands by each flaming point leads to a regular distribution pattern of the secondary fire. As time progresses, the primary fire also advances and covers up the gap with the secondary fire. 
With further increase in $\sigma$, the probability of the firebrands landing at farther distances increases, and this can lead to a potentially explosive firebrands attack. Fig.~\ref{fig:Firebrand_increasing_sigma}f shows the effect of fire-spotting at shorter scales contributing to a faster rate of spread, but the development of new secondary fires is not observed. For this particular value of $\sigma$, the statistical distribution of the long range landing distances lies outside the domain and is impossible to resolve them in the limited time and spatial scales considered in this simulation.  

\begin{figure}
   \begin{subfigure}[t]{0.3\textwidth}
	\centering
    \includegraphics[scale = 0.4]{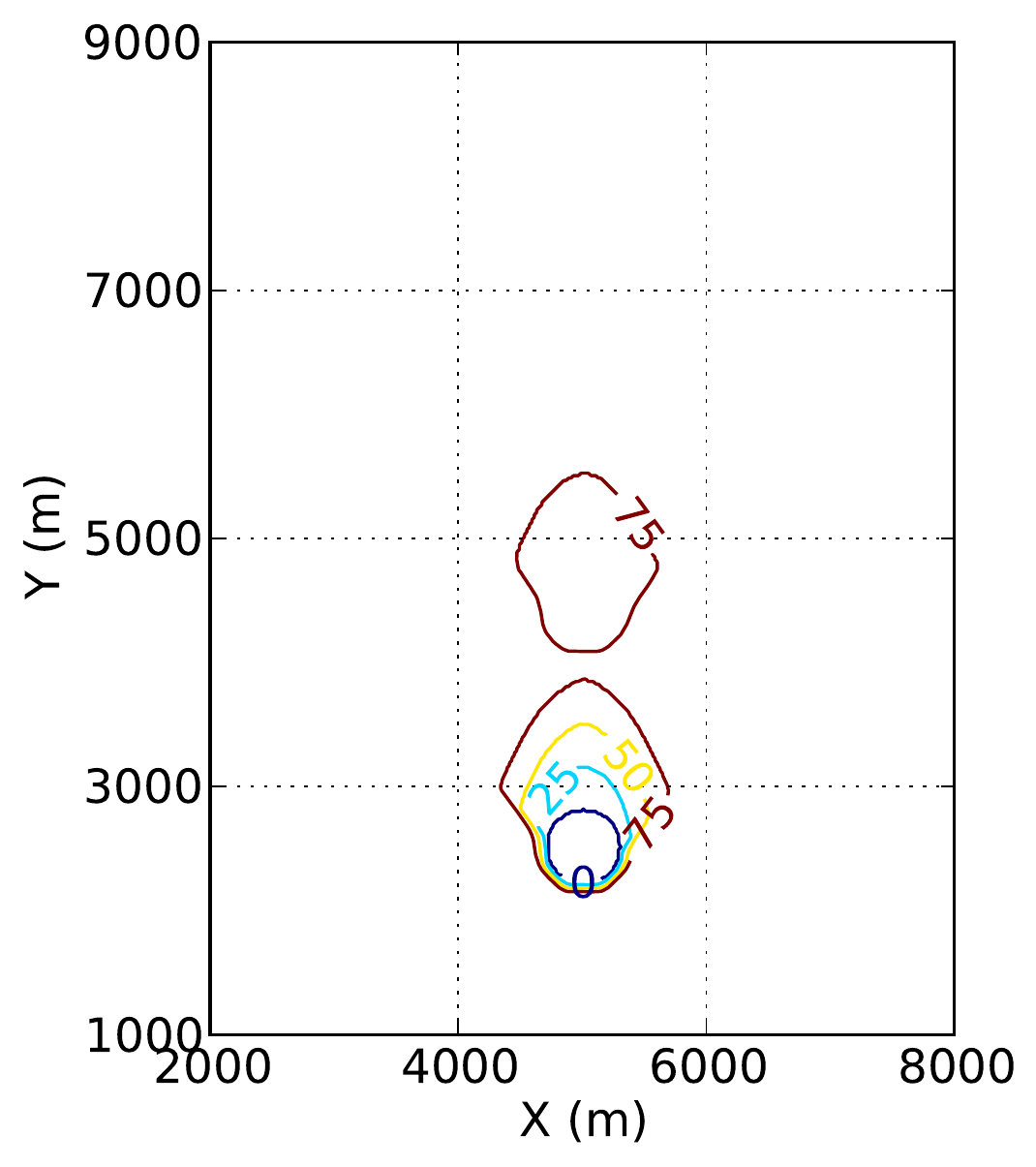}
    \caption{}
       \end{subfigure}
    \begin{subfigure}[t]{0.3\textwidth}
	\centering
    \includegraphics[scale = 0.4]{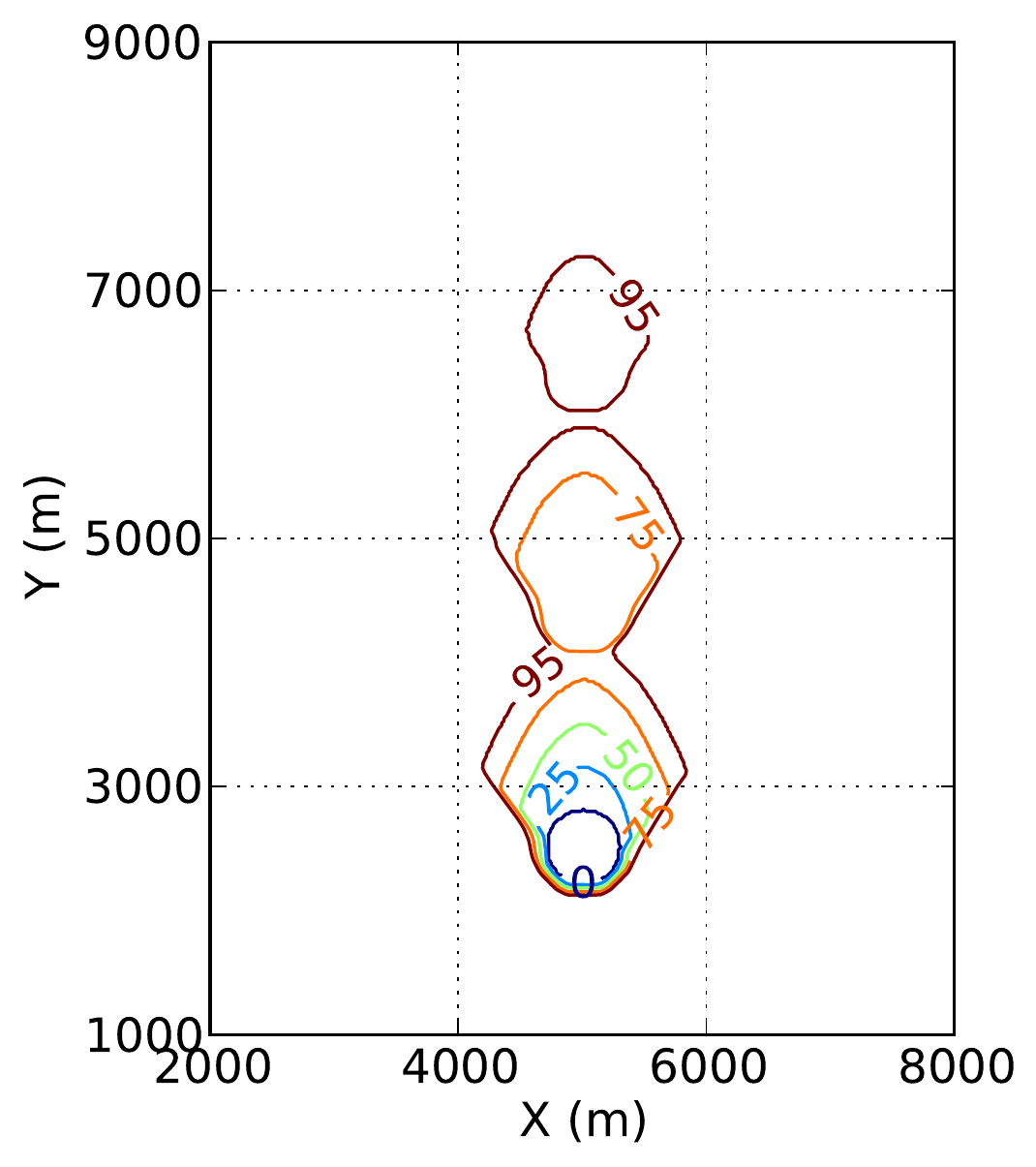}
    \caption{}
    \end{subfigure}
       \begin{subfigure}[t]{0.3\textwidth}
	\centering
    \includegraphics[scale = 0.4]{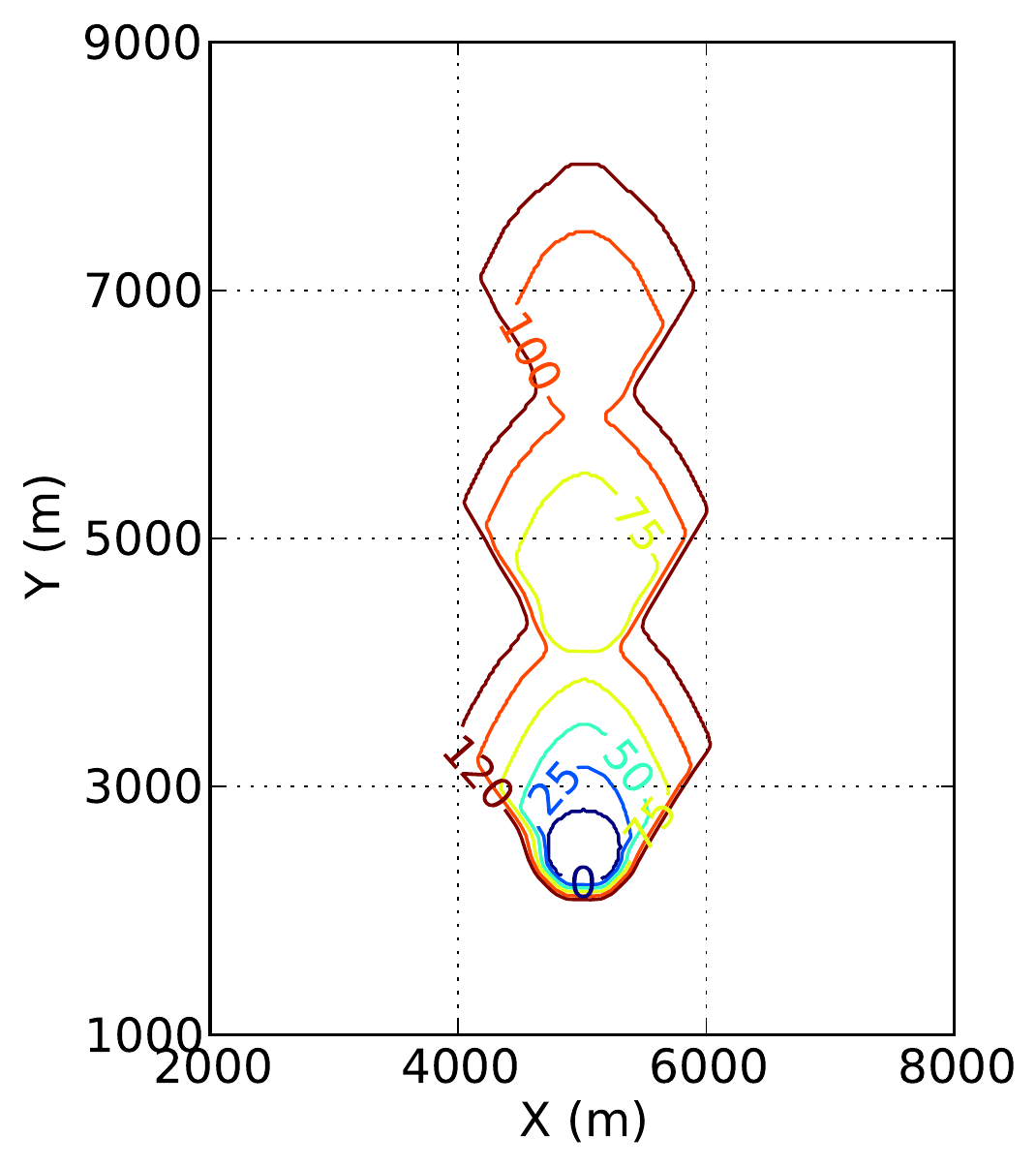}
    \caption{}
       \end{subfigure}
     \caption{Series of figures showing the generation of secondary fires due to fire-spotting and their subsequent merging with the primary fire  
when $\mu = 15$ and $\sigma = 5$.}
     \label{fig:Firebrand_secondary_fires}
\end{figure}

\smallskip
As a consequence of the idealised set-up for simulations, the new secondary fires evolve with the same intensity as the primary fire and  also serve as a source of generation of newer fire brands and henceforth the tertiary fires. Fig.~\ref{fig:Firebrand_secondary_fires}a-f demonstrates the generation of new secondary fires by the primary fire, and tertiary fires by the secondary fires. As time progresses, the each of the fire-line evolves independently along with a small contribution from its parent fire-line, and eventually they merge among each other.

\smallskip  
In a lognormal distribution, an increasing value of the $\mu$ causes a right shift of the maximum, 
indicating towards a larger jump length, but the simulations suggest a decreasing trend in the landing distance. 
This can be explained by the fact that the ignition of the unburned fuel is not an instantaneous process, but rather an accumulative process. 
The fuel starts burning only when the temperature is high enough to allow spontaneous combustion. 
Though a larger $\mu$ allows the firebrands to fall at larger distances but it limits the frequency of landing. 
Insufficient number of firebrands landing at farther distances diminishes the efficacy of their attack to cause combustion; hence leads to a smaller contribution from fire-spotting. 
For an effective combustion, sufficient number of firebrands should be available for an adequate heat exchange with the unburned fuel. 
On the contrary, an increasing $\sigma$ causes the lognormal distribution to have a lower maximum 
and an elongated right tail of the distribution, but with a high probability of landing far away from the main source.
The increase in the number of firebrands landing away from the maximum leads to a rapid ignition of the unburned fuel and a faster propagation of the fire. 

\smallskip
According to the increasing and decreasing trend of the landing distance with respect to increasing $\sigma$ and decreasing $\mu$ respectively, 
within the lognormal characterisation assumed here for the firebrand landing distribution, the reference length scale $\mathcal{L}$ of the landing distance can be stated as:
\be
\mathcal{L} \propto \frac{\sigma^{\alpha}}{\mu^{\beta}} \,,
\label{L}
\ee
where $\alpha >0$ and $\beta >0$ are exponents to be established.

\smallskip
From relation (\ref{L}) it can be deduced that neither the maximum nor the mean of the lognormal distribution can be considered as an estimation of the landing distance length scale $\mathcal{L}$. Physically, the landing distance can be described in terms of the dimensions of the convective column, wind conditions, fire intensity and fuel characteristics. Further analysis towards a physical parametrisation of fire-spotting is postponed for a future paper, where firebrand distributions different from the lognormal will be also considered.

\section{Conclusions}
\label{conclusions}
This article describes a mathematical formulation to model the effects of turbulence and fire-spotting in wild-land fire propagation models. Previously, this formulation has been applied to an Eulerian LSM based wildfire propagation method, and this article describes the implementation of the mathematical formulation into a Lagrangian DEVS based wildfire propagation method. This formulation splits the motion of the front into a drifting part and a fluctuating part. The drifting part is independent of the fluctuating part and is given by the fire perimeter determined from one of the various wildfire propagation methods that exist in literature (DEVS and LSM in the present study); while the fluctuating part is generated by a comprehensive statistical description of the system and includes the effects of random processes in agreement with the physical properties of the process. This study also provides a unique opportunity for the comparison of two simple wildfire simulators based on the Eulerian LSM and on the Lagrangian DEVS scheme. Numerical simulations based on identical set-ups are used to compare the performance of the two approaches.

\smallskip
A series of idealised numerical simulations show that the proposed approach for random effects emerges to be suitable for both LSM and DEVS based simulators to manage the real world situations related to random character of fire, e.g., increase in fire spread due to pre-heating of the fuel by hot air,  vertical lofting and transporting of firebrands, fire overcoming no fuel zones, and development of new isolated secondary fires. The simulations from both  LSM and DEVS techniques follow nearly identical patterns, 
and the only difference which distinguishes the outputs from the two techniques is due to the distinction in the treatment of the direction of propagation. 
In DEVS, the active burning points propagate in a direction given by the bisector of the angle made by each marker with its immediate left and right markers, in contrast to the normal direction of propagation used in LSM. Such differences result in ``spurious" flanking fires in DEVS, which 
are anomalous with respect to the definition of the ROS, but qualitatively provide a more ``realistic" representation of the fire contour. Due to this fact, the two methods can be considered complementary to each other for simple situations, but with increasing complexity and the introduction of random processes a validation exercise is necessary to single out the best representation of the fire propagation.  

\smallskip
A brief study on the sensitivity of the firebrand landing distribution to the changes in the shape of the lognormal distribution indicates that 
the shape parameters ($\mu$ and $\sigma$) can control the spread of the firebrand landing and the lognormal distribution is capable of simulating new secondary fires separated from the main fire perimeter.  
In future, the formulation for the firebrand landing distance would be modified to include information about the wind speed, fire intensity, radius of the firebrands and fuel characteristics.

\smallskip
Though without a validation test case, these simulations do provide an insight into the flexibility of this formulation to incorporate 
effects of random processes such as the turbulent heat diffusion and fire-spotting, specially while characterising the landing distance of firebrands. It can be deduced that the mathematical formulation is independent of the choice of the method used to ascertain the drifting part and can serve as a versatile addition to the existing fire spread simulators to include the effects of random spread. The results from this study provide a support to the implementation of the proposed approach for the effects of random phenomena into operational softwares such as WRF-SFIRE and ForeFire. The source code utilized for all the simulations in this article is hosted at \url{https://github.com/ikaur17/firefronts}.

\section*{Acknowledgement}
This research is supported by MINECO under Grant MTM2013-40824-P, 
by Bizkaia Talent and European Commission through COFUND programme under Grant AYD-000-226, 
and also by the Basque Government through the BERC 2014-2017 program and 
by the Spanish Ministry of Economy and Competitiveness MINECO: BCAM Severo Ochoa accreditation SEV-2013-0323. The authors would also like to thank the two anonymous reviewers for their insightful comments which helped in improving the manuscript both in presentation and scientific content.


\clearpage
\clearpage


%

%
%
%
%
%

%

%
%
%
%

%
\clearpage

\end{document}